\documentclass{aa}  

\usepackage{graphicx}
\usepackage{bbm}
\usepackage{placeins}
\usepackage{txfonts}
\newcommand{\bs}[1]{{\boldsymbol #1}} 
\newcommand{\vct}[1]{{\rm vec}\left( #1 \right)} 
\newcommand{\sinc}[1]{{\rm sinc}\left( #1 \right)} 
\newcommand{\D}[1]{{\rm d} #1}
\newcommand{\m}{$^{\rm m}$}

\newcommand{\gaia}{{\it Gaia}}

\begin{document}

   \title{Spectrophotometric calibration of low-resolution spectra}

   \author{M. Weiler, J.~M. Carrasco, C. Fabricius, and C. Jordi
             }

\authorrunning{M. Weiler et al.}

   \institute{Departament de F{\'i}sica Qu{\`a}ntica i Astrof{\'i}sica, Institut de Ci{\`e}ncies del Cosmos (ICCUB), Universitat de Barcelona (IEEC-UB), Mart{\'i} i Franqu{\`e}s 1, E 08028 Barcelona, Spain\\
              \email{mweiler@fqa.ub.edu}
             }

   \date{Received 12 October 2019 / Accepted 17 March 2020}

 
  \abstract
   {Low-resolution spectroscopy is a frequently used technique. Aperture prism spectroscopy in particular is an important tool for large-scale survey observations. The ongoing ESA space mission \gaia~is the currently most relevant example.} 
   {In this work we analyse the fundamental limitations of the calibration of low-resolution spectrophotometric observations and introduce a calibration method that avoids simplifying assumptions on the smearing effects of the line spread functions.}
   {To this aim, we developed a functional analytic mathematical formulation of the problem of spectrophotometric calibration. In this formulation, the calibration process can be described as a linear mapping between two suitably constructed Hilbert spaces, independently of the resolution of the spectrophotometric instrument.}
   {The presented calibration method can provide a formally unusual but precise calibration of low-resolution spectrophotometry with non-negligible widths of line spread functions. We used the \gaia~spectrophotometric instruments to demonstrate that the calibration method of this work can potentially provide a significantly better calibration than methods neglecting the smearing effects of the line spread functions.}
{}
 

   \maketitle
%

\section{Introduction}

Spectroscopy at low spectral resolution is a frequently used observational technique. In particular, aperture prism spectroscopy can be employed to obtain a large number of spectra with a single exposure. If the dispersion is low enough to avoid major problems with blending spectra from different sources, aperture prism
spectroscopy is a suitable approach for large-scale astronomical surveys. As such, it has been applied repeatedly, from the generation of the Henry Draper catalogue in the early 20th century \citep{Pickering1890} to the ongoing European space mission \gaia~\citep{Gaia2016a}. The use of slitless spectroscopy with low resolution will continue in the future, for example with space missions such as EUCLID \citep{Costille2016} and WFIRST \citep{Akeson2019}. However, the spectrophotometric calibration of data coming from such missions, that is, the transformation of the observational data to a well-defined approximation of the spectral energy distribution (SED) of the observed object, can be difficult. Attempts at a stringent spectrophotometric calibration of such data have led to mixed results, from reliable and frequently used calibrations (e.g. for slitless infrared spectroscopy with HST/WFC3) to considerable complications in parts of the spectra due to the effects of low resolution (e.g. near-ultraviolet objective prism spectroscopy with HST/STIS at wavelengths longer than 300~nm; \citealt{JMA2005}).\par
As described in detail in Sect.~\ref{sec:theory} of this work, complications in calibration arise from the observational spectrum being the result of a convolution, or rather variable-kernel convolution, of the SED. In particular, at very low resolution and dispersion, this may turn the reconstruction of the SED from the observational spectrum into an ill-posed problem, making the numerical solution highly unstable. To address this difficulty, in this work we approach the problem of spectrophotometric calibration of low-resolution spectra from a new perspective, making use of a functional analytic formulation of the problem; that is, we exploit the vector nature of functions in vector spaces of square-integrable functions. By doing so, the calibration problem is essentially turned into a problem of linear algebra. Such a mathematical approach was introduced in photometry by \cite{Young1994} for the problem of transformations between different photometric passbands. The use of this formalism in photometry was expanded by \cite{WeilerEtAl2018} to the problem of passband reconstruction and applied to the \gaia~data release 2 passbands \citep{Weiler2018,MAW2018}. Here, we expand this approach from photometry to spectrophotometry, where it can provide a better understanding of the inherent problems of calibrating low-resolution spectra. The approach results in a different concept for the calibration as compared to commonly applied calibration methods and provides a stringent quantitative relationship between the calibration result and the true SED of the observed astronomical object. We put particular emphasis on the spectroscopic instruments of the \gaia~mission,  which is an obvious example on which to test the application of the approach outlined here. Moreover, we demonstrate that it can potentially increase the scientific outcome of the \gaia~spectroscopic data.\par

The European Space Agency's (ESA) ongoing mission \gaia~\citep{Gaia2016a} provides a repeating all-sky survey of all point-like astronomical sources in the magnitude range of approximately 6\m~ to 20\m. Designed as an astrometric mission, its primary purpose is high-precision astrometry. The required observations are being performed in a wide photometric band, the $G$ band, ranging from approximately 330~nm to 1100~nm. The spacecraft is spinning and precessing, resulting in a repeating scan of the whole sky. During the scanning, the images of the astronomical objects are moving over {\it Gaia}'s focal plane, where their light is recorded by an array of charge-coupled devices (CCDs) operating in time delay and integration (TDI) mode. In this operation mode of the CCDs, the charges generated by the light from a point-like or nearly point-like astronomical source (e.g. a star, asteroid, Quasi-Stellar Object (QSO)) are shifted through the CCD as the source image moves over the CCD due to the scanning motion of the spacecraft. The apparent velocity of the source's image with respect to the CCD and the shifting velocity of the charges are synchronised. The direction of movement of the source image in the focal plane is referred to as the along scan (AL) direction, and both the CCD lines and the rectangular aperture of the \gaia~telescopes are aligned with respect to this direction. Each passage of an image of an astronomical object over the focal plane is referred to as a 'transit'.\par
Besides its astrometric capabilities, \gaia~is equipped with spectroscopic capabilities. One of the spectrometers on board, referred to as the Radial Velocity Spectrometer (RVS), performs medium resolution spectroscopy on a relatively narrow wavelength range (845~nm to 872~nm), its primary objective being to determine spectroscopic radial velocities. The other two spectrometers are two very low-resolution slit-less spectrographs, one covering the short-wavelength part of the $G$ band, from about 330~nm to 680~nm, and the second covering the long-wavelength part, from about 630~nm to 1100~nm. The two instruments are referred to as the blue and red photometers, or BP and RP, respectively. When referring to both, we may use the abbreviation XP. The spectral resolving powers of the BP and RP instruments range between 20 and 90, depending on wavelength and instrument.\par
The \gaia~mission has already made two data releases (DRs; \citealt{Gaia2016b,Gaia2018a}). The latest, \gaia~DR2, provides information on almost 1.7 billion sources. For about 1.3 billion of them, photometry in the $G$ band and integrated photometry derived from the XP spectra have been published. The enormous size of the data set makes \gaia~one of the most interesting observational data sets, and the low-resolution XP spectra will result in an additional large increase in the scientific value of the data. We therefore chose the \gaia~XP instruments as a test case for the calibration schemes derived in this work for the case of low-resolution spectrophotometric data. \gaia \ XP spectra are expected to be published for the first time in the second half of 2021\footnote{\url{https://www.cosmos.esa.int/web/gaia/release}}, so this work relies entirely on simulations of the \gaia~spectrophotometric instruments for the time being. However, this is not a disadvantage as the aim of this work is to address problems related to low-resolution spectroscopy. For this purpose, a comparison with the 'true' values and functions is necessary, which requires the use of simulated data anyway. In Sect. \ref{sec:example} we describe our simulation of the XP instruments in more detail.\par
This work uses the \gaia~spectroscopic instruments as a detailed example. However, it does not provide a description of the default {\it Gaia Data Processing and Analysis Consortium} (DPAC) spectrophotometric calibrations for DR3. Discussion of the possible implementation of the method developed in this work within DPAC processing pipelines is beyond the scope of this work.\par

Throughout this work we assume the use of a detector that is recording photons, rather than photon energy such as a CCD. That is, we assume a detector that is recording the same signal at two different wavelengths if it detects the same number of photons at the two wavelengths. Due to the wavelength-dependency of the photon energy, this is not equivalent to the same energies entering the detector at the two different wavelengths. Recorded observational spectra therefore depend on the spectral photon distribution (SPD) observed, for which we use the symbol $s(\lambda)$ (with indices where SPDs of different astrophysical sources are to be considered). The SPD is related to the SED $S(\lambda)$ via
\begin{equation}
s(\lambda) = \frac{\lambda}{h\, c} \, S(\lambda) \quad ,
\end{equation}
with $h$ and  $c$ being the Planck constant and the vacuum speed of light, respectively. By using the SPD instead of the SED, we omit the frequent occurrence of the wavelength-dependent conversion factor $\lambda/(h\, c)$ in the equations of this work.\par
We define the response function $R(\lambda)$  as the ratio between the number of detected photons within a wavelength interval, $n(\lambda)\, \D{\lambda}$,  and the number of photons per wavelength interval entering the telescope, $n_0(\lambda)\, \D{\lambda}$:
\begin{equation}
R(\lambda) \coloneqq \frac{n(\lambda)\, \D{\lambda}}{n_0(\lambda)\, \D{\lambda}} \quad . \label{eq:response}
\end{equation}
Since the energy of a number of photons for a given wavelength $\lambda$ is proportional to the number of photons, the photon energy being $E = h\, c / \lambda$, Eq.~(\ref{eq:response}) is equivalent to the ratio of detected energy to energy entering into the telescope. Equation~(\ref{eq:response}) is therefore an unambiguous definition of response, both for photon-detecting and energy-detecting instruments. This definition is equivalent to what is sometimes referred to as the 'photonic response function' or 'photonic passband' (e.g. \citealt{Bessell2012}), in distinction to the re-normalised function of $\lambda \, R(\lambda)$. The function $\lambda \, R(\lambda)$ is sometimes used to describe the response of a photon detector in terms of the SED instead of the SPD, by absorbing the conversion function $\lambda / (h \, c)$ in a re-definition of the response function (e.g. \citealt{Bessell2000}).\par
Furthermore, we restrict this work to spectroscopical observations of point-like astronomical sources. Extended objects (such as resolved galaxies) may introduce significant complications to spectroscopy, in particular when using wide or no apertures, and their inclusion is therefore beyond the scope of this work.\par
In this work we make use of a number of special functions. We use $g(x,\sigma,\mu)$ to denote the Gaussian with standard deviation $\sigma$ and mean $\mu$, and $\Lambda(x)$ to denote the triangular function, i.e.
\begin{equation}
\Lambda(x) \coloneqq
\begin{cases}
1-|x| & \text{ for } |x| < 1\\
0 & \text{else} \quad .
\end{cases} 
\end{equation}
Here, $\sinc{x}$ is used to denote the normalised cardinal sine function, i.e.
\begin{equation}
\sinc{x} \coloneqq
\begin{cases}
\frac{\sin{(\pi\, x)}}{\pi \, x} & \text{ for } x \ne 0 \\
1 & \text{ for } x = 0 \quad .
\end{cases}
\end{equation}
The symbol $\bar{\varphi}_n(x)$ denotes the $n$--th Hermite function, which is related to the $n$--th Hermite polynomial $H_n(x)$ via
\begin{equation}
\bar{\varphi}_n(x) = \sqrt{\frac{\sqrt{\pi}}{2^{n-1}\, n!}} \, g(x,1,0) \cdot H_n(x) \quad .
\end{equation}
For the actual computation of Hermite functions, we use the recursive relation
\begin{equation}
x \cdot \bar{\varphi}_n(x) = \sqrt{\frac{n}{2}}\, \bar{\varphi}_{n-1}(x) + \sqrt{\frac{n+1}{2}}\, \bar{\varphi}_{n+1}(x) \quad ,
\end{equation}
together with $H_0(x) = 1$ and $H_1(x) = 2\, x$. The Hermite functions represent a set of orthonormal functions on $\mathbb R$, i.e.
\begin{equation}
\int\limits_{-\infty}^{\infty}\, \bar{\varphi}_n(x) \cdot \bar{\varphi}_m(x)\, \D{x} = \delta_{n,m} \quad , \label{eq:HermiteOrthonormality}
\end{equation}
with $\delta_{n,m}$ being the Kronecker delta.\par
The function $\vct{\bf M}$ is the vectorisation of the matrix $\bf M$, that is, the column-wise re-arrangement of the elements of the $N \times M$ matrix $\bf M$ into an $N\cdot M \times 1$ vector. The vectorisation is related to the Kronecker product $\otimes$ via
\begin{equation}
\vct{{\bf A\, B \, C}} = \left({\bf C}^{\mathsf T} \otimes {\bf A}\right) \vct{{\bf B}} \quad . \label{eq:kronMatrix}
\end{equation}
As conventions throughout this work, we denote matrices by bold uppercase letters, vectors by bold lowercase letters, and operators by calligraphic uppercase letters. Table~\ref{tab:notations} provides an overview of the symbols and notations used throughout this work.\par
The outline of this work is as follows. In Sect.~\ref{sec:theory} we develop the principles for the calibration of low-resolution spectra. We start from basic considerations of the formation of spectra (Sect.~\ref{sec:formation}) and a brief overview of conventionally followed approaches to calibration (Sect.~\ref{sec:standard}), and then sketch the basic idea of the method developed in this work (Sect.~\ref{sec:sketch}). This idea is then put into a wider theoretical context (Sect.~\ref{sec:context}) and formulated stringently (Sect.~\ref{sec:matrix}). The remainder of Sect.~\ref{sec:theory} (Sect.~\ref{sec:expansion} to \ref{sec:uncertainties}) discusses different aspects of the developed calibration method, such as implementation issues, possible further generalisations, implications for the choice of calibration sources and the interpretation of the calibrated spectra, and the propagation of uncertainties.\par
In Sect.~\ref{sec:example}, we then consider \gaia's XP instruments in detail as an example for the application of the newly developed approach. We start with a description of \gaia's spectrophotometric instruments and how we simulate their behaviour (Sect.~\ref{sec:gaiaInstruments}), and provide an overview of the different scenarios under which we investigate the performance of the calibration approach (Sect.~\ref{sec:gaiaOutline}). After illustrating the difficulties described theoretically in Sect.~\ref{sec:theory} for the \gaia~instruments (Sect.~\ref{sec:problemIllustration}), we perform the different necessary calibration steps for the method developed in this work (Sect.~\ref{sec:exampleBases} to \ref{sec:GaiaSourceCalibration}). The remainder of Sect.~\ref{sec:example} then presents the analysis of the performance of the calibration approach for the different scenarios investigated (Sect.~\ref{sec:performanceFocusedTelescope} to \ref{sec:narrowFeatures}). Section~\ref{sec:summary} closes this work with a summary and discussion of the results.

\begin{table}
\renewcommand\arraystretch{1.2}
\caption{List of notations used throughout this work. \label{tab:notations}}
\resizebox{0.48\textwidth}{!}{
\begin{tabular}{l l}\hline
Symbol & Meaning \\ \hline
$\mathbbm{1}_N$ & $N \times N$ identity matrix\\
$\Delta(\lambda)$ & Dispersion function\\
$\mathcal F$ & Fourier operator \\
$f(u)$ & Sampled observational spectrum\\
$g(x,\sigma,\mu)$ & Gaussian with standard deviation $\sigma$ and mean $\mu$\\
$H(\nu)$ & Optical transfer function\\
$\mathcal I$ & Instrument operator\\
$I(u,\lambda)$ & Instrument kernel\\
${\bf I}$ & Instrument matrix\\
$\lambda$ & Wavelength \\
$\Lambda(x)$ & Triangular function \\
$L(u,\lambda)$ & Sampled line spread function\\
${\mathcal L}^2(X)$ & Space of square-integrable functions on $X$ over $\mathbb R$\\
$N$ & Dimension of $V$ and $W$\\
$N^\ast$ & Dimension of $V^\ast$ and $W^\ast$\\
$\nu$ & Frequency variable associated with pupil function\\
$\Pi(x)$ & Rectangular function \\
$\bar{\varphi}_n(x)$ & $n$--th Hermite function\\
$R(\lambda)$ & Response function\\
$s(\lambda)$ & Spectral photon distribution\\
$\bs{\Sigma}^y$ & Variance-covariance matrix of a random vector $\bf y$\\
${\rm sinc}(x)$ & Normalised cardinal sine function\\
$u$ & Focal plane variable \\
${\bf U}_{\rm M}\, \bs{\Sigma}_{\rm M}\, {\bf V}^{\mathsf T}_{\rm M}$ & Singular value decomposition of matrix $\bf M$\\
$V$ & Space spanned by calibration SPDs\\
$V^\ast$ & Finite-dimensional expansion of $V$\\
${\rm vec}({\bf M})$ & Vectorisation of the matrix $\bf M$\\
$W$ & Space spanned by observed calibration spectra\\
$W^\ast$ & Finite-dimensional expansion of $W$\\
$\square$ & Place holder for variables\\
$\star$ & Correlation operation\\
$\otimes$ & Kronecker product\\
\end{tabular}
}
\end{table}

\section{Theoretical framework \label{sec:theory}}

\subsection{Formation of spectra \label{sec:formation}}

Assuming a SPD $s(\lambda)$ entering into the telescope, an observational spectrum $f_o({\bf u})$ results in the focal plane (or, rather the plane of the light detector, if we include defocused instruments in the discussion). Here, $\bf u$ is a two-dimensional vector of focal plane variables, for which different choices can be made. The components of $\bf u$ might be the position in the focal plane, measured with respect to some reference point, or field angles, or pixel positions on a CCD detector deployed in the focal plane. The latter is the most natural choice for an observational spectrum recorded with a CCD, and we assume $\bf u$, without loss of generality, to be a vector of continuous real numbers, measured in units of CCD pixels. For convenience, we furthermore assume that the spectrum is aligned with the CCD rows or columns, and that the spectrum is integrated in the CCD direction perpendicular to the axis of alignment. In this case, the problem becomes one-dimensional, and we can assume the observational spectrum to depend only on a single real variable $u$, with non-integer values in $u$ corresponding to sub-pixel CCD positions. The value of $f_o(u)$ is the instrumental output to the input SPD at CCD position $u$, which is a number of photoelectrons per interval of time. The non-integer nature of $u$ reflects the possibility of different sub-pixel positioning of the source with respect to some reference position. The dispersion function $\Delta(\lambda)$ provides a relation between the focal plane coordinate $u$ and the corresponding wavelength. Finding $\Delta(\lambda)$ means performing the wavelength calibration of the observational spectrum, which then allows the transformation of $f_o(u)$ into $f_o(\lambda)$. The function $f_o(u)$ for a monochromatic SPD, i.e. $s(\lambda) = \delta(\lambda-\lambda_0)$, is the line spread function (LSF) for the wavelength $\lambda_0$, scaled with the instrumental response $R(\lambda_0)$, and transformed to the focal plane coordinate via the dispersion relation. The observed spectrum $f_o(u)$ for some general SPD is then obtained by integration over all wavelengths, i.e.
\begin{equation}
f_o(u) = \int\limits_{0}^\infty\, L_o(u(\lambda),\lambda) \, R(\lambda) \, s(\lambda) \, \D{\lambda} \quad . \label{eq:basicXP}
\end{equation}
In practice, instrumental functions such as the LSF, the dispersion, and the response may depend on the position within the focal plane, or on even more parameters (such as time in the case of instrumental ageing). Such additional dependencies are suppressed in Eq.~(\ref{eq:basicXP}) and the following equations for simplicity. We return to the problem of additional functional dependencies in Sect.~\ref{sec:multidimensional}. The relationship between the SPD entering into the telescope and the resulting observed spectrum can be expressed in different ways. The different formulations include using frequencies instead of wavelength, SEDs instead of SPDs, already assuming a wavelength calibration or not, separating the response into individual contributions from the detector, the optics, and the atmosphere, using the inverse of $R(\lambda)$, or already making particular assumptions (such as an LSF not changing with wavelength, resulting in a standard convolution problem). In practice, these different formulations do indeed occur (\cite{Bohlin1980,Lena1998,Cacciari2010}, only to name a few examples). Common to all these formulations is that the observed spectrum is expressed as an integral transformation of a 'true' spectrum, and so we may use a general form,
\begin{equation}
f_o(u) = \int\limits_{0}^\infty\, I_o(u,\lambda) \, s(\lambda) \, \D{\lambda} \label{eq:general}
,\end{equation}
to express the relationship between $f_o(u)$ and $s(\lambda)$. The kernel of this integral equation, $I_o(u,\lambda)$, fully describes the effect of the spectroscopic instrument and we refer to it as the 'instrument kernel' in this work. The manner by which this instrument kernel is separated into individual functions, the choice of variables, or the choice of assumptions imposed on it may be decided according to convenience and suitability. However, two aspects of the general formulation by Eq. (\ref{eq:general}) require special attention.\par
First, Eq. (\ref{eq:general}) is by its nature a linear equation. As a consequence, this formulation cannot take into account non-linear instrumental effects. In practice, non-linear effects can be introduced by the detector, as a relatively mild effect such as a non-linear CCD response, particularly at very low or very high count rates, or as potentially strong non-linearities resulting from charge transfer inefficiency. Non-linear effects are not taken into account in this work, and if they are relevant, they have to be corrected for before entering into the formalisms deployed here.\par
Second, the separation of the LSF from the instrument kernel is of particular interest, as the LSF is the only function contributing to $I_o(u,\lambda)$ that depends on both the wavelength and the focal plane variable. This property gives it a special role in the formation of the observed spectrum. Furthermore, the LSF can be conveniently described using the formalism of Fourier optics \citep{Goodman1996}. Given the important role that the LSF plays in particular for low-resolution spectrophotometry, we introduce some basic aspects of Fourier optics that we rely on in the course of this work.\par
We assume an instrument with a focal length $F$ and a point-like observed object at infinity. We describe the pupil function of the instrument as
\begin{equation}
P(x,y) = p(x,y) \cdot {\rm e}^{\mathfrak{i}\, 2\pi / \lambda \, W(x,y)} \quad .
\end{equation}
Here, $x$ and $y$ are coordinates in the pupil, and $p(x,y)$ describes the geometry of the pupil. In the exponential term, $\mathfrak{i}$ denotes the imaginary unit. This term describes phase shifts as a function of pupil coordinates, thus including the optical aberrations of the instrument. These aberrations may include many effects, from imperfections in the mirror shapes to a defocusing of the instrument. The effective path error due to non-perfect conditions is expressed by the function $W(x,y)$, with $W(x,y) \equiv 0$ representing the ideal, diffraction-limited instrument.\par
From this, we may express the optical transfer function (OTF) $H(x,y)$ as the autocorrelation of the pupil function,
\begin{equation}
H(x,y) = P(x,y) \star P(x,y) \quad ,
\end{equation}
and finally the point spread function (PSF) as the inverse Fourier transform of the optical transfer function,
\begin{equation}
PSF(\nu_x,\nu_y) = {\mathcal F}^{-1}\left\{ H(x,y) \right\} \quad .
\end{equation}
The LSF, $L_o(u,\lambda)$, is then obtained by integration of the PSF along the direction perpendicular to the dispersion direction. Assuming a suitable choice of coordinate system, the LSF can be expressed as a one-dimensional problem
\begin{equation}
L_o(u,\lambda) = {\mathcal F}^{-1}\left\{H(x)\right)\}\quad .
\end{equation}
With the size $d$ of a CCD pixel and the focal length $F$, we can write for the pixel scale $p_s$ (in radians per pixel) in a small angle approximation
\begin{equation}
p_s =  \frac{d}{F}\quad .
\end{equation}
Therefore, when choosing CCD pixels as the focal plane coordinates, we can use
\begin{equation}
u = \nu_x / p_s \quad .
\end{equation}
Recording the observational spectrum with a CCD detector requires that the CCD sampling be taken into account, that is, the fact that the CCD integrates the observed spectrum over one CCD pixel. The sampled observed spectrum $f(u)$ may therefore be written as
\begin{equation}
f(u) = \int\limits_{u-\Delta u/2}^{u + \Delta u/2} f_o(u^\prime) \, {\rm d} u^\prime = \left( f_o \ast \Pi \right) (u) \quad , \label{eq:pixel}
\end{equation}
where the last step is the convolution with a rectangular function. As the LSF is the only function in the formation of the observational spectrum depending on $u$, the convolution with the rectangular function can be included in the LSF. For convenience we introduce the sampled LSF,
\begin{equation}
L(u,\lambda) = L_o(u,\lambda) \ast \Pi(u) \label{eq:sampledLSF}
,\end{equation}
and the convolution is conveniently included by introducing a multiplication with a cardinal sine function to the OTF. Eventually, we obtain an expression analogous to Eq. (\ref{eq:basicXP}) for the relation between the sampled observed spectra and the SPD:
\begin{equation}
f(u) = \int\limits_{\lambda_0}^{\lambda_1} \, L(u,\lambda) \cdot R(\lambda) \cdot s(\lambda)\, {\rm d}\lambda \equiv  \int\limits_{\lambda_0}^{\lambda_1} \, I(u,\lambda) \cdot  s(\lambda)\, \D{\lambda} \quad . \label{eq:sampledXP}
\end{equation}

\subsection{The conventional approach and its limit \label{sec:standard}}

Calibration of spectrophotometric data usually involves a wavelength calibration, i.e. finding the dispersion relation, and transforming from the focal plane coordinate $u$ to a wavelength coordinate $\lambda^\prime$. Thus, Eq. (\ref{eq:sampledXP}) is brought to the form
\begin{equation}
f(\lambda^\prime) = \int\limits_{\lambda_0}^{\lambda_1}\, L(\lambda^\prime,\lambda)\cdot R(\lambda)\cdot s(\lambda)\, \D{\lambda} \quad . \label{eq:noLSF}
\end{equation}
If the width of the LSF is negligible, one may use the approximation $L(\lambda^\prime,\lambda) = \delta(\lambda-\lambda^\prime)$, and the integral disappears. Therefore, for a sufficiently narrow LSF, one obtains the approximation
\begin{equation}
R(\lambda^\prime) \approx \frac{f(\lambda^\prime)}{s(\lambda^\prime)} \quad . \label{eq:stdApprox}
\end{equation}
With knowledge of the observational spectrum and the SPD for some source, it is possible in this approximation to obtain an estimate for the response. Moreover, with knowledge of that response and the observational spectrum of another source, it is possible to obtain an estimate for the SPD of this other source. In the following, we generally refer to the first step, that is, deriving knowledge on the instrument from given observational spectra and the corresponding SPDs, as the 'instrument calibration'. We refer to the second step, deriving the SPD from the known instrument and observational spectrum, as the 'source calibration'. Recovering the SPD of a source from the observational spectrum is an inverse problem. We refer to proceeding in the opposite direction, that is, predicting the observational spectrum from a known SPD, as the 'forward calibration'.\par
Instead of  Eq. (\ref{eq:stdApprox}), some authors (e.g. \citealt{Vacca2003}) use an approximation in a different form, for example by assuming
\begin{equation}
\begin{split}
f(\lambda^\prime) = \int\limits_{\lambda_0}^{\lambda_1}\, L(\lambda^\prime,\lambda)\cdot R(\lambda)\cdot s(\lambda)\, \D{\lambda} \approx \\
\quad \int\limits_{\lambda_0}^{\lambda_1}\, L(\lambda^\prime,\lambda)\cdot R(\lambda)\, \D{\lambda} \cdot \int\limits_{\lambda_0}^{\lambda_1}\, L(\lambda^\prime,\lambda)\cdot  s(\lambda)\, \D{\lambda} \label{eq:vaccasLSF}
\end{split}
.\end{equation}
This latter approximation confers the advantage that, with some estimate of the LSF, a smoothing is already included in the division of the observational spectrum by the SPD. As the assumption of a negligible influence of the LSF may not be justified at wavelength ranges where the function $R(\lambda)\cdot s(\lambda)$ varies strongly, such as in the vicinity of absorption lines, this approach gives preference to smooth SPDs that are poor in spectral features for the determination of the response function. Typically, hot white dwarfs have SPDs with such properties, making them suitable calibration sources.\par
While the SPD used for the determination of the response function can be selected smoothly by choosing suitable calibration sources, the response function cannot be modified once the instrument design is fixed. The validity of the approximation (\ref{eq:stdApprox}) depends on the product of the response function and the SPD, which has to be smooth enough to render the effect of the LSF negligible. If the response function changes strongly on a wavelength scale which is not large compared to the width of the LSF, the approximation by Eq. (\ref{eq:stdApprox}) is insufficient for the determination of $R(\lambda)$, irrespective of the choice of the calibration source. If the effect of the LSF is not negligible, meaning the integral in Eq. (\ref{eq:sampledXP}) can be eliminated, then one is left with solving the integral equation (\ref{eq:sampledXP}), either for the instrument kernel $I(u,\lambda)$ (in the instrument calibration) or for the SPD $s(\lambda)$ (in the source calibration).\par
However, being left with the integral equation (\ref{eq:sampledXP}) results in serious difficulties. In mathematical terms, it is a Fredholm equation of the first kind, a type of integral equation notorious for being very ill-conditioned \citep{Press2007}. The problem with an integral equation of this type is nicely illustrated following \cite{Schmidt1987}. As follows from the Riemann-Lebesgue Lemma (see Theorem 10.6a in \citealt{Henrici1988}),
\begin{equation}
\lim_{\nu \to \infty} \int\limits_{\lambda_0}^{\lambda_1}\, I(u,\lambda) \cdot {\rm sin}(\nu\, \lambda)\, {\rm d}\lambda = 0 \quad . \label{eq:RiemannLebesgue}
\end{equation}
This result is graphically interpreted by the smoothing effect of the LSF, which fully smooths out the sine oscillation if only its frequency becomes large enough. A simple proof results directly from applying integration by parts to Eq.~(\ref{eq:RiemannLebesgue}). Thus, for any constant $\alpha \in {\mathbb R}$ we obtain
\begin{equation}
\lim_{\nu \to \infty} \int\limits_{\lambda_0}^{\lambda_1}\, I(u,\lambda) \cdot \left[s(\lambda) + \alpha\cdot {\rm sin}(\nu\, \lambda)\right]\, {\rm d}\lambda = f(u) \quad , \label{eq:problem1}
\end{equation}
meaning that the function sequence $s_\nu \coloneqq  s(\lambda) + \alpha\cdot {\rm sin}(\nu\, \lambda)$, with $f_\nu(u)$ being the corresponding observational spectrum, shows the phenomenon of weak convergence: $\lim_{\nu \to \infty} f_\nu(u) = f(u)$, while at the same time $\lim_{\nu \to \infty} s_\nu(\lambda) \ne s(\lambda)$. As a consequence, when allowing for an arbitrarily small uncertainty on the observational spectrum $f(u)$, arbitrarily large changes in the SPD $s(\lambda)$ become consistent with the observation. This is valid even if the instrument (i.e. $I(u,\lambda)$) is assumed to be perfectly known. Improving the calibration of the LSF, the response, and the dispersion, although highly desirable for reducing systematic errors, can therefore not mitigate the intrinsic degeneration in the calibration problem. For a non-zero error on the observational spectrum, which is the case for all practical cases, a meaningful inversion of Eq.~(\ref{eq:sampledXP}) is impossible without imposing additional constraints on the solution for $s(\lambda)$. When replacing $L(u,\lambda)$ by $L(\lambda-\lambda^\prime)$, the problem turns into a convolution problem, and this result corresponds to the notorious problem of deconvolution.\par
At this point we may formulate the objective of this work more clearly. We use 'low-resolution spectrophotometry' here to refer to cases in which the effect of the LSF cannot be neglected in the instrument calibration. If the LSF effects can be neglected, an approximation of the form of Eq. (\ref{eq:stdApprox}) --or similar-- holds, and we refer to this approximation as the 'conventional approach' in this work for simplicity. If the LSF effects cannot be neglected, the calibration problem becomes considerably more difficult by the nature of the integral equation one is left with for linking SPDs and observational spectra. Whether the smearing effects of the LSF can be neglected or not is not so much a question of absolute resolution, but depends on the instrument design itself. If the response function changes significantly over wavelength scales that are small or similar to the width of the LSF (expressed in wavelengths), the LSF effects become relevant, at least at parts of the wavelength interval covered by the spectroscopic instrument. As the response function  is usually a rather smooth, slowly changing function for most instruments, such problems rarely occur in practice. In some cases however, the response function can vary too abruptly with respect to the width of the LSF. This may be caused by strong changes in the response function itself, i.e. because of the use of wavelength cut-off filters or electronic resonances in the reflecting materials of the mirrors, or by an overly low dispersion. In the remainder of this work, we investigate the possibilities for calibrating such cases. Finally, we apply the results to the case of the \gaia~spectrophotometric instruments, one of the few but important cases where the conventional approach may be insufficient.

\subsection{Sketch of the idea \label{sec:sketch}}

As illustrated by Eq. (\ref{eq:problem1}), for any observational spectrum $f(u)$ there exists an infinite number of possible SPDs $s(\lambda)$ that are all in agreement with the error-effected observational spectrum. The inversion of Eq. (\ref{eq:sampledXP}) therefore requires additional information to decide which of the possible SPDs is the most likely and therefore  preferred solution. Usually, additional constraints are introduced to the problem to find the preferred solution among all possible solutions. Regularisation introduces a second parameter next to the agreement with the observational data, and then requires an optimisation with respect to a weighted combination of the two parameters. A common choice for the second parameter is for example the power of a higher derivative of the solution, thereby implying some smoothness on the solution \citep{Press2007}. However, for SPDs, such a smoothness approach may be inadequate, as SPDs by their very nature tend to have very different levels of smoothness on very different wavelength scales.\par
We may therefore rather rely on the simplest possible assumption we can make on the solution: suppose we have a calibration source with both known SPD and known observational spectrum, then as a reasonable approach, from all SPDs consistent with the observational spectrum, we may select  the one that is actually the known SPD. Expanding this simple proposal, we may assume that we have a set of $n$ calibration sources, with known SPDs $s_i(\lambda)$, $i=1,\ldots,n$, and known observational spectra $f_i(u)$. We may then require, for each of the $n$ observational spectra, the solution to the corresponding known SPD. We may expand this approach further to the case of some observational spectrum which is not a calibration source, that is, for which the SPD is unknown. Suppose that this observational spectrum $f(u)$ can be expressed uniquely as some linear combination of the observational spectra of the calibration sources, that is,
\begin{equation}
f(u) = \sum\limits_{i=1}^n\, a_i \cdot f_{i}(u) \quad ,
\end{equation}
with $a_i$ being the coefficients of the linear combination. As the preferred solutions for the SPDs of the calibration sources are the known SPDs, and the instrument is linear, this immediately implies that the preferred solution for the SPD corresponding to $f(u)$ is a linear combination of the SPDs of the calibration sources, with the same coefficients of development $a_i$ that hold for the observational spectra:
\begin{eqnarray}
f(u) & = & \sum\limits_{i=1}^n\, a_i \cdot \int\limits_{\lambda_0}^{\lambda_1}\, I(u,\lambda) \cdot s_i(\lambda)\, {\rm d}\lambda \\
 & = & \int\limits_{\lambda_0}^{\lambda_1}\, I(u,\lambda) \cdot \underbrace{\left(\sum\limits_{i=1}^n\, a_i \cdot s_i(\lambda) \right)}_{s(\lambda)} \, {\rm d}\lambda \quad .
\end{eqnarray}
This basic idea, though conceptually very simple, proves to be the starting point for a powerful alternative approach to the problem of calibrating low-resolution spectrophotometric data, which makes the intrinsically ill-posed inversion of the observational spectra manageable. In the following sections we develop this approach in detail and work out a practical formalism.

\subsection{The abstract context \label{sec:context}}

For developing the approach sketched above, we first take the problem to a more abstract level. Suppose some functions $s(\lambda)$ and $f(\lambda)$,  connected via a linear mapping. This mapping is described by a linear operator $\mathcal I$, which we refer to as the 'instrument operator'. We may write the relationship between $s$ and $f$  as
\begin{equation}
f = {\mathcal I} \, s \quad . \label{eq:basicLinear}
\end{equation}
The linearity of the operator $\mathcal I$ means that a relation
\begin{equation}
{\mathcal I} \left[\alpha \, s_1 + \beta \, s_2 \right] = \alpha \, {\mathcal I}\, s_1 + \beta  \,{\mathcal I} \, s_2 
\end{equation}
holds, for $\alpha, \beta \in {\mathbb R}$. For practical purposes, a suitable representation of $\mathcal I$ must be found. If we assume that the functions $s$ and $f$ are well behaved with respect to differentiability, in particular that they are infinitely often continuously differentiable functions with compact support, then there exists one and only one generalised function (distribution) $I$ such that the linear map is represented by the integral transform
\begin{equation}
f(u) = \int\limits_{\lambda_0}^{\lambda_1}\, I(u,\lambda) \cdot s(\lambda) \, {\rm d}\lambda \quad ; \label{eq:schwartz}
\end{equation}
(see Theorem 5.2.1 in \citealt{Hoermander2003}). This is then Eq. (\ref{eq:sampledXP}) used above, and motivated by physical considerations. The fact that $I$ is a generalised function is required to allow for a statement in the most generalised form. For example, the identity transform is a linear transformation, and expressing the identity transform by an integral transform requires the Dirac distribution in the integral kernel. For the purpose of this work, we can think of $I(u,\lambda)$ as a simple two-dimensional function. From here, we may continue as described above, ending with the problem that, if the integral cannot be removed by the assumption of a sufficiently narrow LSF, the calibration problem becomes ill-posed.\par
\begin{figure}
   \centering
   \includegraphics[width=0.47\textwidth]{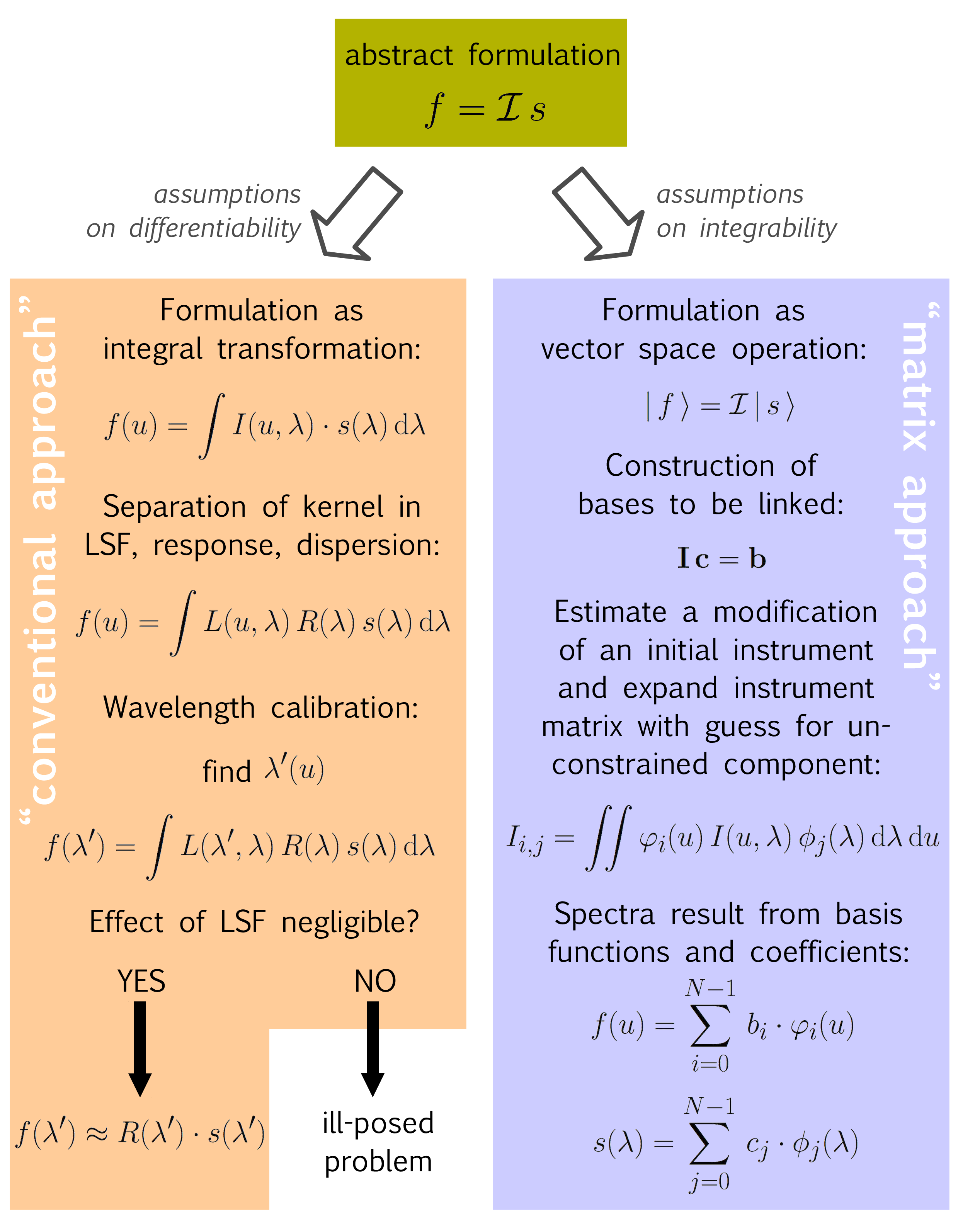}
   \caption{Overview of the conventional approach and the matrix approach from this work. The bra-ket notation $| \; \rangle$ is used to stress the vector nature of functions.}
              \label{fig:scheme}
    \end{figure}

However, the integral transform of Eq. (\ref{eq:schwartz})  is not the only possibility to represent the abstract relationship of Eq. (\ref{eq:basicLinear}). If we can assume that both $s$ and $f$ are square-integrable functions with compact support, then $\mathcal I$ represents a linear mapping between two Hilbert spaces of square integrable functions over the field of real numbers. We use ${\mathcal L}^2(X)$ to denote a vector space like this, with $X$ being the support of the square-integrable functions. A linear map between two such vector spaces may be expressed by a matrix transformation of the form
\begin{equation}
{\bf f} = {\bf I} \, {\bf s} \quad .
\end{equation}
Here, $\bf f$ and $\bf s$ are the representations of $f$ and $s$ with respect to some bases for the two Hilbert spaces, and $\bf I$ is a matrix which expresses the linear map with respect to the chosen bases. Here, $\bf I$, $\bf f$, and $\bf s$ are intrinsically countably infinite in dimension. In this representation, the instrument is therefore not represented by an integral kernel $I(u,\lambda)$, but by a matrix transforming between bases of two Hilbert spaces of square-integrable functions. We refer to this approach as the 'matrix approach'. The formal structure of the conventional approach and the matrix approach and how they are related may be summarised schematically as shown in Fig.~\ref{fig:scheme}. For practical applications, the bases, and with it the matrix representation of the instrument operator, has to be chosen in a suitable way. The method by which this is chosen is outlined in the following.

\subsection{The matrix approach \label{sec:matrix}}

Suppose we have a set of $n$ calibration sources available, that is, sources for which both the SPDs and the observational spectra are available. Then, the SPDs of these $n$ sources span an $N$-dimensional subspace of ${\mathcal L}^2[\lambda_0,\lambda_1]$, with $1 \le N \le n$. We use $V$ to denote this sub-space. Analogously, the observational spectra of the $n$ sources span an $N$-dimensional sub-space of ${\mathcal L}^2[{\mathbb R}]$, which we denote $W$. For practical purposes, we may assume that the observational spectra are identical to zero everywhere outside an finite interval $[u_0,u_1]$. Let $\{\varphi_i\}_{i=1,\ldots,N}$ and $\{\phi_j\}_{j=1,\ldots,N}$ be two orthonormal bases of $W$ and $V$, respectively. The orthonormality conditions are
\begin{eqnarray}
\int\limits_{u_0}^{u_1} \, \varphi_k(u) \cdot \varphi_l(u) \, \D{u} & = & \delta_{k,l} ,\\
\int\limits_{\lambda_0}^{\lambda_1} \, \phi_k(\lambda) \cdot \phi_l(\lambda) \, \D{\lambda} & = & \delta_{k,l} \quad .
\end{eqnarray}
Any observational spectrum in $W$ and SPD in $V$ can then be expressed by a development into these bases, that is,\begin{eqnarray}
f(u) & = & \sum\limits_{i=0}^{N-1}\, b_i \, \varphi_i(u) ,\\
s(\lambda) & = & \sum\limits_{j=0}^{N-1} \, c_j\, \phi_j(\lambda) \quad .
\end{eqnarray}
The coefficients of these developments are given by
\begin{eqnarray}
b_i & = & \int\limits_{u_0}^{u_1} \, f(u) \cdot \varphi_i(u) \, \D{u} ,\\
c_j & = & \int\limits_{\lambda_0}^{\lambda_1} \, s(\lambda) \cdot \phi_j(\lambda) \, \D{\lambda} \quad .
\end{eqnarray}
We arrange the coefficients of these developments in the $N\times 1$ vectors $\bf b$ and $\bf c$.\par
If the relationship between the SPDs and the corresponding observational spectra is linear, there exists a matrix $\bf I$ such that
\begin{equation}
{\bf I} \, {\bf c}= {\bf b} \quad . \label{eq:basicMatrix}
\end{equation}
The elements of the $N \times N$ matrix, the instrument matrix, are the images of the basis functions of $V$ expressed in the basis of $W$. The instrument matrix is thus linked to the instrument kernel via the relationships
\begin{equation}
I_{i,j} = \int\limits_{u_0}^{u_1} \int\limits_{\lambda_0}^{\lambda_1} \, \varphi_i(u) \cdot I(u,\lambda) \cdot \phi_j(\lambda) \, \D{\lambda}\, \D{u} \quad . \label{eq:instrumentMatrixElements}
\end{equation}
If we have chosen suitable bases for $V$ and $W$, we may find the coefficients of the $N$ SPDs and observational spectra in these bases, and arrange the $N$ coefficient vectors ${\bf b}_i$ and ${\bf c}_i$, $i=1,\ldots,N$, in the $N \times n$ matrices $\bf B$ and $\bf C$. These are then linked via the relation
\begin{equation}
{\bf I} \, {\bf C} = {\bf B} \quad .
\end{equation}
This equation can then be solved for $\bf I$ by multiplying the $N \times N$ identity matrix from the left and using Eq. (\ref{eq:kronMatrix}):
\begin{equation}
\left(\, {\bf C}^{\mathsf T} \otimes \mathbbm{1}_N \, \right) \, {\rm vec}({\bf I}) = {\rm vec}({\bf B}) \quad . \label{eq:instrumentMatrix1}
\end{equation}
Because of the unity matrix in the Kronecker product, this last system of equations decomposes into $N$ independent sets, one for each row of the instrument matrix.\par
Once the instrument matrix has been derived from the calibration sources, the coefficients $\bf c$ for any source in $W$ can be obtained by solving Eq. (\ref{eq:basicMatrix}), with $\bf I$ and $\bf b$ given. The elements of $\bf b$ have to be obtained from a set of $M$ error-affected discrete observations $f_k(u_k)$ at discrete focal plane locations $u_k$, $k=1,\ldots,M$. We may write the values of the $f_k(u_k)$  as the $M \times 1$ vector $\bf f$. Evaluating the $N$ basis functions $\varphi_i(u)$ at these discrete locations and writing the values in the $M \times N$ matrix $\bf H$, we obtain
\begin{equation}
{\bf H} \, {\bf b} = {\bf f} \quad .
\end{equation}
This equation may be combined with Eq. (\ref{eq:basicMatrix}) to obtain
\begin{equation}
{\bf H} \, {\bf I} \, {\bf c} = {\bf f} \label{eq:fullSolution}
,\end{equation}
to be solved for $\bf c$ in one step.

\subsection{Expanding the matrix approach \label{sec:expansion}}

Up to now, the calibration formalism only holds for sources whose observational spectra fall within the vector space $W$ because the instrument matrix is constrained by the calibration sources only for the transformation between $V$ and $W$. In practice, it may be necessary to also calibrate  sources for which this restriction is insufficient, that is, for sources whose observational spectra are not well represented by a linear combination of the observational spectra of the calibration sources. We may use Eq. (\ref{eq:instrumentMatrixElements}) to expand the instrument matrix to include any additional basis functions, provided we know the instrument kernel $I(u,\lambda)$. We are therefore left with the task of obtaining a guess for the instrument kernel which is in agreement with the instrument matrix that transforms between $V$ and $W$. Such an estimate for the instrument kernel may be based on the initial knowledge we have on the instrument. We then have to select relevant additional basis functions, both for the SPDs and the observational spectra, which are chosen to be orthonormal with respect to the already used basis function. This way we expand the sub-spaces $W$ and $V$ spanned by the calibration sources to larger sub-spaces of ${\mathcal L}^2[u_0,u_1]$ and ${\mathcal L}^2[\lambda_0,\lambda_1]$, which we denote $W^\ast$ and $V^\ast$, respectively. Finally, we can use Eq. (\ref{eq:instrumentMatrixElements}) to compute the additional elements of the instrument matrix, which we then denote ${\bf I}^\ast$.\par
The way in which the instrument kernel can be estimated depends on the particular instrument to be calibrated. We may however consider some basic problems that occur in this context. Usually we can expect to have some a priori knowledge on the approximate appearance of the instrument kernel. This a priori knowledge may come from our knowledge of how the instrument has been built, putting constraints on the shape of the dispersion relation, the LSF, and the response function. We may therefore start with some initial guess for the kernel and modify this initial guess in such a way that it meets the constraints set by the instrument matrix $\bf I$, and such that the modification remains physically meaningful, that is, fulfilling smoothness and boundary constraints. Although only the kernel $I(u,\lambda)$ is required for the calibration of the instrument, physical constraints apply to the individual components from which the kernel is built. A smoothness consideration for example may apply to the dispersion relation, the LSF, and the response function independently. When deriving only the kernel, it may therefore be possible that it cannot meaningfully be decomposed into its individual components. It is therefore desirable to modify each individual component.\par
For the calibration restricted to $V$ and $W$, no wavelength calibration is required. The mapping can be constrained directly between the wavelengths of the SPDs and the focal plane coordinates of the observational spectra. As the initial guesses for the components depend on the wavelength as well, an explicit wavelength calibration is required for estimation of the instrument kernel. Solving for a particular instrument component, that is, either the dispersion relation, the LSF, or the response, by assuming that the remaining components are known, again leaves us with ill-posed problems. To solve such problems requires that the possible solutions be restricted to sufficiently smooth and bound ones. This can be achieved by applying a modification model to the initial guess for the instrument component under consideration. To be able to reduce the estimation of the instrument kernel to the problem of solving a system of linear equations, a linear modification model is preferable. One may consider the use of either an additive or a multiplicative modifications model, i.e. modifying the initial guess by either adding a function linear in its parameters, or by multiplying the initial guess with some function linear in its parameters. For positive functions that are smoothly approaching zero at the boundaries of their support, such as the response curve or the LSF, multiplicative models may be more advantageous because multiplication with a smooth function preserves such a behaviour more easily than adding functions. This approach is analogous to the one introduced by \cite{WeilerEtAl2018} for the reconstruction of the response curves from photometric observations.\par
However, modelling the LSF with a multiplicative modification model is complicated by the normalisation requirement. The LSF has to satisfy the normalisation condition
\begin{equation}
\int\limits_{u_0}^{u_1} \, L(u,\lambda)\, \D{u} = 1 \quad .
\end{equation}
Choosing a multiplicative modification model that by design preserves this normalisation condition is difficult. We may therefore  consider the modification of the LSF by convolving it with some normalised modification function instead of multiplying it with a modification function. The convolution of the LSF with a modification model may conveniently be implemented by including a parameterised multiplicative modification term in the OTF.\par
We consider using the following steps in the calibration:
First we estimate the dispersion relation. Such a calibration may rely on the initial guess for the dispersion, refined by observations. Emission line spectra can be useful for this purpose. For low-resolution spectrophotometry however, the emission lines have to be well separated in wavelength to avoid blending in the observational spectra. Furthermore, they have to be much stronger than the underlying continuum, or the continuum has to be known in the observational spectrum to allow for a separation from the response. If no sufficient information from emission lines is available, the global shape of the observational spectra of calibration sources can be used, together with some initial guess for the response and the LSF. As the dispersion relation should not allow for strong deviations from the initial expectation, such a simple approach may work sufficiently well.\par
Second, the LSF must be estimated, starting from an optimistic guess and adjusting the parameters of the multiplicative term in the OTF until the widths of the features in the observational spectra of the calibrators are well matched. As the convolution process renders the observational spectra rather insensitive to the fine structure of the LSF, simple modification terms in the OTF may already be sufficient, such as a Gaussian.\par
Finally, the response function needs to be estimated. Assuming that the dispersed LSF  is  known, finding the response function requires the solution of a Fredholm equation of the first kind again. Finding a smooth solution for the response function therefore requires smoothness conditions to be introduced. One may use a similar approach to the one by \cite{WeilerEtAl2018}, expressing the response functions as a multiplicative modification of an initial guess, with B-splines for the modification model. Equation (\ref{eq:stdApprox}) can allow the initial guess to be improved in wavelength regions where the smoothing effect of the LSF is expected to be small, and the knot sequence defining the B-spline basis functions may be adjusted to allow for more localised variations at wavelengths where the LSF effects are expected to be strong. Assuming that  $L(u,\lambda)$ is known, and with $B_k(\lambda)$, $k=1,\ldots,K,$ the B-spline basis functions, and $B_0(\lambda) \equiv 1$ to allow for the identity transformation, we obtain a system of $n$ integral equations,
\begin{equation}
f_l(u) = \int\limits_{\lambda_0}^{\lambda_1} L(u,\lambda) \left[ \sum\limits_{k=0}^K\, \alpha_k\, B_k(\lambda) \right] \, R_{ini}(\lambda) \, s_l(\lambda) \, \D{\lambda} \; , \; l=1,\ldots,n
,\end{equation}
which have to be fulfilled for the $n$ calibration sources when solving for the coefficients $\alpha_k$ of the modification model. This system of integral equations is brought into a more convenient form for solution when using Eq. (\ref{eq:instrumentMatrixElements}):
\begin{eqnarray}
I_{i,j} & = & \int\limits_{u_0}^{u_1} \int\limits_{\lambda_0}^{\lambda_1} \varphi_i(u) \, L(u,\lambda) \left[\sum\limits_{k=0}^K\, \alpha_k\, B_k(\lambda)\right]\, R_{ini}(\lambda) \, \phi_j(\lambda) \, \D{\lambda} \, \D{u} \nonumber \\
 &= & \sum\limits_{k=0}^K\, \alpha_k \, M_{i,j,k} \quad ,
\end{eqnarray}
with
\begin{equation}
M_{i,j,k} = \int\limits_{u_0}^{u_1} \int\limits_{\lambda_0}^{\lambda_1} \varphi_i(u) \, L(u,\lambda) \, B_k(\lambda)\, R_{ini}(\lambda) \, \phi_j(\lambda) \, \D{\lambda} \, \D{u} \quad .
\end{equation}
We may apply the vectorisation operation with respect to the indices $i,j$, take all elements of the instrument matrix into account, and write the $K$ elements $\alpha_k$ into a vector $\bs{\alpha}$, to obtain
\begin{equation}
{\bf M}\, \bs{\alpha} = {\rm vec}\left({\bf I}\right) \quad . \label{eq:newresponse}
\end{equation}
Here, $\bf M$ is the $N\cdot N \times K$ matrix containing the elements $M_{i,j,k}$ vectorised with respect to $i,j$. As the elements $M_{i,j,k}$ only depend on functions that are  known or are assumed to be known, the coefficients of the modification model can be obtained by simply solving a set of linear equations.\par
For the expansion from $V$ and $W$ to $V^\ast$ and $W^\ast$, suitable additional basis functions need to be chosen, and the additional elements of the instrument matrix ${\bf I}^\ast$ are computed using Eq.~(\ref{eq:instrumentMatrixElements}).

\subsection{Basis function construction \label{sec:implementation}}

By now we have not specified how the functions that constitute the bases and $V$ and $W$, and their expansions to $V^\ast$ and $W^\ast$, are constructed, and what the dimensionalities $N$ and $N^\ast$ are. Here we discuss possible approaches to these problems, starting with the construction of $V$ and $W$.\par
We assume $n$ observational spectra $f_i(u)$ and the corresponding SPDs $s_i(\lambda)$, $i=1,\ldots,n$, given. Strictly speaking, the sets of these $n$ spectra, both observational and the SPDs, will span $n-$dimensional vector spaces, as already random noise will prevent two measurements from being exactly identical. However, in practice, a much smaller dimension $N$ may be sufficient to describe the spectra sufficiently well. For low-resolution spectroscopy, spectral features, and with it differences between spectra, are blurred more in the observational spectra as compared to the SPDs. We therefore start the construction of the bases with $W$, seeking a set of basis functions that is able to reproduce all $n$ observational spectra within the limits set by the noise, with a number of basis functions $N$ being as small as possible. In principle functional data analysis provides tools to construct such a basis from functional principal components. However, this process can be complicated by strong variation in the absolute noise, both within a single observational spectrum and between different observational spectra. In practice, some noise suppression might be required for the construction of the basis functions. If this is the case, we may proceed as follows.\par
First we represent all observational spectra in a generic orthonormal basis (such as properly scaled Legendre polynomials, Hermite functions, etc.). The number of these generic functions is chosen such that all spectra can be described accurately within the noise. The goodness of fit for all sources, or a cross-validation, can serve as an indicator as to the number of basis functions for which this point has been reached. As the generic functions will be inefficient in describing the observational spectra, a reduction of the number of basis functions should be possible. This can be achieved by finding a suitable set of linear combinations of the original generic functions. As we require the more efficient basis to be orthonormal as well, the transformation from the generic functions to a new linear re-combination needs to be an orthogonal one. Furthermore, we wish for the more efficient basis that its basis functions are sorted according to their relevance in describing the entire set of observational spectra. These requirements are met if the transformation from the generic functions to a new basis is done by applying singular value decomposition (SVD) to the matrix of coefficients of the observational spectra in the generic basis. For the $n$ observational spectra, if we put the $m$ coefficients in the generic basis into an $n \times m$ matrix $\bf C$, the matrix ${\bf V}_{\rm C}$ provides the required orthogonal transformation to a more efficient linear combination of the generic basis functions. The first $N$ of these orthogonally re-combined basis functions serve as the basis that spans $W$, $\{\varphi_i(u)\}_{i=1,\ldots,N}$. Here, $N$ again is determined by testing how many basis functions are required to describe all observational spectra within the limits imposed by the noise.\par
For the construction of $V$, the first $N$ most relevant basis functions also need to be constructed. A similar approach to that for $W$ can be taken. However, should the noise in the SPDs be small enough, and the wavelength sampling sufficiently dense, one may simply compute the basis functions by omitting the step of an intermediate generic basis, and apply the SVD directly to the matrix containing the sampled SPDs, interpolated to a common wavelength grid if necessary. The result represents the basis functions $\{\phi_j(\lambda)\}_{j=1,\ldots,N}$ on a discrete wavelength grid.\par
Additional basis functions need to be chosen for the expansion of $V$ and $W$ to $V^\ast$ and $W^\ast$. We may consider three different strategies for finding such basis functions. First, we could make use of model spectra for astronomical objects. By doing so, we would be able to include basis functions containing features expected to occur in real spectra. However, extensive model libraries would be required to cover all the possible kinds of spectra that may be found. Second, we could use some generic functions of wavelengths to start with. This approach would be independent of what one expects to find in real spectra, while it might be possible to end up with poor matches with real spectral features. Finally, one could combine both approaches and use both model spectra and generic functions as a compromise between realistic spectral features and generic flexibility. This last option may be most promising, and we compare all three approaches for the \gaia~XP example in Sect.~\ref{sec:GaiaExpansion}, finding a combined approach indeed to be the best solution for an expansion of the vector spaces between which the calibration is performed.

\subsection{Generalisation to higher dimensions \label{sec:multidimensional}}
Up to now we have treated an observed spectrum as a function depending on one focal plane variable $u$ only. However in many cases, such as spectroscopy with CCD detectors, the observed spectra are intrinsically two dimensional, depending on a focal plane variable along the dispersion direction, $u$, as well as on a variable $v$ perpendicular to this direction, i.e. $f = f(u,v)$. When performing slit-less spectroscopy it might be desirable to treat spectra in two dimensions, in order to separate overlapping spectra from more than one source, a process which is referred to as 'de-blending'. We may also consider further variables on which a spectrum depends, such as the position in the focal plane, or time. The latter cases are caused by instrumental variations across the focal plane or with time, and are usually treated independently from the step of linking an observational spectrum with the SPD. However, there is no reason not to include these 'internal' instrumental calibration steps  in the matrix approach developed in the previous section, at least formally. We therefore consider a calibration which  directly links observational spectra for each individual observational circumstance (e.g. position in the focal plane or time of observation) with the SPD, instead of correcting for instrumental variations in a first step and then linking the thus-corrected observational spectra with the SPD. For the approach of this work, this means we extend the function $f(u)$ to a function $f(u,\square)$, where the $\square$ serves as a placeholder for any additional variables the spectrum may depend on. Analogously, the instrument kernel $I$ is also to be treated as dependent on these variables, resulting in an extended version of Eq.~(\ref{eq:sampledXP}):
\begin{equation}
f(u,\square) = \int\limits_{\lambda_0}^{\lambda_1} I(u,\lambda,\square) \cdot s(\lambda)\, \D{\lambda} \quad .
\end{equation}
The only change in the formalism is thus the need to find not a one-dimensional basis $\{\varphi_i(u)\}_{i=1,\ldots,N}$ that can represent the observational spectra $f(u)$, but rather a higher-dimensional basis $\{\varphi_i(u,\square)\}_{i=1,\ldots,N}$ that can describe the observational spectra as a function of several parameters $f(u,\square)$. The principal approach to finding such bases is no different from the one-dimensional case discussed above. Functional data analysis provides tools to construct such bases from observational data (e.g. \citealt{Chen2017}), or a generic basis might  also be used in this case. The main difference is the strong increase in the computational cost of such a generalisation to a multi-dimensional problem, making implementation demanding. Discussion of such an implementation of a matrix approach in the multi-dimensional case is therefore beyond the scope of this work.

\subsection{Implications of the matrix approach \label{sec:implications}}

The nature of the matrix approach differs very much from  that of the conventional approach, and consequently has a number of implications that are noteworthy.\par
First, the matrix approach affects the optimal selection of calibration sources. For the conventional approach, in principle a single calibration spectrum is sufficient to derive the response function. The division between the observational spectrum and the SPD makes a smooth, feature-poor calibration spectrum preferable. For the matrix approach however, the instrument is constrained on the space spanned by the calibration sources. This means that a large set of calibration sources, which includes many different shapes of the SPDs, is preferable, making the matrix approach more demanding in terms of calibration data than the conventional approach. However, this requirement meets with the requirement on calibration sources for the problem of reconstructing the response function from photometric instead of spectrophotometric data \citep{WeilerEtAl2018}.\par
Second, the scientific interpretation of the calibration result obtained with the matrix approach also differs from that obtained when using the conventional approach. While in the conventional approach the calibration result is the function $s(\lambda)$ itself, or a low-resolution approximation of it, the matrix approach solves for the representation of the function $s(\lambda)$ in some suitably chosen basis. Though the function $s(\lambda)$ can in principle be constructed from the representation, an important difference arises where the interpretation of the calibration errors is concerned. The ill-posed nature of the calibration of low-resolution spectroscopy makes the concept of error on the SPD not a well-defined concept. As expressed by Eq. (\ref{eq:problem1}), any function with sufficiently high frequency can be added to the solution for the SPD without leaving the error bound on the observational spectrum $f(u)$. As a consequence, solutions $s(\lambda)$ to the calibration problem of Eq. (\ref{eq:sampledXP}) can be found that can take any value at any wavelength. Therefore, there does not exist a meaningful error or interval of confidence around the preferred solution for the SPD. This problem is handled by the matrix approach by providing errors on the coefficients of the development in the chosen basis. Poorly constrained basis functions in this representation, such as highly oscillating
functions, have large, and sometimes excessively large, error on the corresponding coefficients of the development. However, this does not affect the interpretation of the development, as the large errors only indicate the poor constraints on the coefficients, and therefore a low weight is given to them in the interpretation, for example when comparing coefficients for different sources. Nevertheless, including the poorly constrained coefficients with very large errors in the reconstruction of the function $s(\lambda)$  would result in a formally correct but uninterpretable SPD. It is therefore preferable to analyse the calibration results obtained with the matrix approach in terms of coefficients in the development in a basis rather than in the form of the function $s(\lambda)$ itself. This kind of interpretation is unusual as compared to conventional spectrophotometry, but is by no means more difficult or less conclusive. We discuss a method for reducing the effects of very uncertain development coefficients in Sect.~\ref{sec:GaiaSourceCalibration}.\par
A third difference in the interpretation of the results obtained with the matrix approach, as compared to the conventional approach, arises from the fact that the solution for the SPD does not reflect the true spectral resolving power of the instrument, but rather the resolving power of the spectra used for the calibration of the instrument. As a consequence, it is not permissible to interpret individual spectral features in the reconstructed function $s(\lambda)$, as they are not constrained by the observational spectrum itself, but are introduced by the calibration process. However, when interpreting the coefficients of the SPD rather than the reconstructed SPD, such uncertainties are omitted immediately, as in the case of handling errors.\par
While at first glance the change of resolution may appear to be  a problematic effect, it confers an advantage in the scientific use of the SPDs. It is of interest to perform synthetic photometry based on the SPDs in order to link the calibration result to other photometric systems \citep{Carrasco2017}. Considering the filter passband $r(\lambda)$ and some 'true' SPD $s_0(\lambda)$, i.e. an SPD as it is received from some astronomical source, following \cite{WeilerEtAl2018}, the observed flux ${\mathfrak f}_0$ in the filter passband is
\begin{equation}
{\mathfrak f}_0 = \int\limits_{0}^{\infty}\, r(\lambda) \cdot s_0(\lambda) \, \D{\lambda} \quad .
\end{equation}
Now we may perform synthetic photometry using an approximation to $s_0(\lambda)$ with a lower resolution, denoted $s(\lambda)$. Here, $s(\lambda)$ is related to $s_0(\lambda)$ via
\begin{equation}
s(\lambda) = \int\limits_0^\infty \, L(\lambda,\lambda^\prime) \cdot s_0(\lambda^\prime) \, \D{\lambda^\prime} \quad ,
\end{equation}
with $L(\lambda,\lambda^\prime)$ being the LSF. For the synthetic flux $\mathfrak f$ in the filter passband $r(\lambda)$ with the low-resolution approximation of the SPD, we obtain
\begin{equation}
{\mathfrak f} = \int\limits_{0}^{\infty}\, r(\lambda) \cdot  \int\limits_0^\infty \, L(\lambda,\lambda^\prime) \cdot s_0(\lambda^\prime) \, \D{\lambda^\prime} \, \D{\lambda} \quad , \label{eq:integratedPhotometry1}
\end{equation}
and thus ${\mathfrak f} \ne {\mathfrak f}_0$. The synthetic flux in general does not agree with the observed flux if the influence of the LSF in the lower-resolution approximation to the true SPD is not negligible. Inverting the order of integration in Eq.~(\ref{eq:integratedPhotometry1}) results in a formulation in which the influence of the LSF in synthetic photometry can be visualised in an intuitive way. One obtains
\begin{equation}
{\mathfrak f} = \int\limits_{0}^{\infty} \underbrace{\int\limits_{0}^{\infty} L(\lambda,\lambda^\prime) \cdot r(\lambda) \, \D{\lambda}}_{r^\prime(\lambda^\prime)} \, s_0(\lambda^\prime) \, \D{\lambda^\prime} \quad .
\end{equation}
The inner integral, denoted $r^\prime(\lambda^\prime)$, is the filter passband with the effects of the LSF applied to it. Instead of applying the LSF to the SPD, it is therefore permissible to keep the SPD unmodified and apply the smoothing effect of the LSF to the passband $r(\lambda)$, exchanging $\lambda$ and $\lambda^\prime$ in the LSF. If the smoothing effect of the LSF in the low-resolution approximation of the SPD used in synthetic photometry is sufficiently large that it would significantly alter the shape of the passband if applied to it instead of the SPD, then systematic differences become apparent between the synthetic flux and the observed flux for some source in the given filter passband. The higher spectral resolution of the basis functions used in the matrix approach is therefore a prerequisite for reliable synthetic photometry using spectrophotometrically calibrated low-resolution spectra.

\subsection{Uncertainties and their propagation \label{sec:uncertainties}}

We conclude the formal discussion of the matrix method by considering the effects of uncertainties on the calibration results, both for random noise and systematic errors.\par
Random noise appears on the observational spectra, and may result in an inexact solution for the instrument matrix. If the bases that span $V$ and $W$ are sorted by their relevance in the representation of the observational spectra of the calibration sources, as they do in the approach described in Sect.~\ref{sec:implementation}, the higher-order columns of the instrument matrix will be more affected by the random noise on the observational spectra used in the calibration than the low-order columns. It may therefore be of advantage to exclude high-order columns when estimating the instrument kernel by applying Eq.~(\ref{eq:newresponse}), and then to recompute the instrument matrix from the estimated instrument kernel. This approach may help to improve the high-order columns of the instrument matrix, as the noise can be suppressed by using the smoothness conditions on the response function and the LSF to obtain a better guess.\par
However, the remaining error on the elements of the instrument matrix represents a source of systematic error in the calibration process, as this error affects the results of the source calibration for all calibrated sources. A strict treatment of this error would require the use of total least squares methods in the source calibration step, while solving for all sources simultaneously, or at least a large number of them. However, such methods  require a reliable statistical model for the errors on the elements of the instrument matrix, which in practice may not be available. Furthermore, the correlations of errors between the elements of the instrument matrix can be complex, which makes total least squares methods very computationally expensive even for rather small matrices such as the instrument matrices. In practice, the systematic errors may therefore remain and may be suppressed by improving the calibration data, i.e. using more calibration sources with better signal-to-noise ratios. In general, systematic errors due to an imperfect instrument matrix can be expected to mostly affect the spectra of objects with a signal-to-noise ratio that is similar to or higher than those of the calibration spectra.\par
A second kind of systematic error, which is of a different nature, may occur from the violation of the fundamental assumptions on which the matrix approach rests. The relationship between the observational spectrum and the SPD is established based on the assumption that every observational spectrum that can be described by a certain linear combination of basis functions has a SPD that can be described by a corresponding linear combination of basis functions. This assumption may be reliable as long as the variation in SPDs is not too large. If one allows for very complex SPDs however, this assumption may no longer be satisfied. We refer to this situation as the 'breakdown'. While the error on the instrument matrix refers to incorrect elements of the instrument matrix, the breakdown refers to effects caused by the truncation of the instrument matrix to a finite size, therefore not representing a transformation between basis functions that may play a role in describing the SPDs. When allowing for very complex SPDs, it might be possible to obtain a good representation of the observational spectra with a linear combination of basis functions and from it a physically meaningful linear combination of basis functions describing the SPDs, while the true SPD is a different one. If the signal-to-noise ratio is high enough, the true SPD may be found outside of the error intervals, which manifests itself as a systematic error. This error is difficult to detect, and we show in the example of Sect. \ref{sec:performanceSet} that it can become relevant for bright and cool M-type stars, which may present very complex SPDs. Another situation in which the breakdown may become relevant is the presence of narrow spectral features. We discuss this case in more detail in examples in Sect.~\ref{sec:narrowFeatures}.

\section{Worked example \label{sec:example}}

In the following, we apply the matrix approach as outlined theoretically to the case of simulated \gaia~spectrophotometers. We first introduce the instruments and how we simulate them (Sect.~\ref{sec:gaiaInstruments}). The different astronomical objects and instrumental setups used as test cases are then introduced (Sect.~\ref{sec:gaiaOutline}). After illustrating the ill-posed nature of the calibration problem with an example for the \gaia~ RP spectra (Sect.~\ref{sec:problemIllustration}), we go through all the steps of the instrument calibration (Sect.~\ref{sec:exampleBases} to \ref{sec:GaiaExpansion}), including some calibration aspects that are \gaia-specific (Sect.~\ref{sec:GaiaCombination}). We describe the source calibration for the test objects (Sect.~\ref{sec:GaiaSourceCalibration}) and evaluate the performance of the calibration for a large number of test sources, both stars and QSOs (Sect.~\ref{sec:performanceFocusedTelescope} and \ref{sec:performanceDefocusedTelescope}). Finally, we analyse the performance for narrow spectral features in detail (Sect.~\ref{sec:narrowFeatures}).\par
In this example, we stay with the assumption that the observational spectra depend on a single variable only. All dependencies on other parameters (such as time, position in the focal plane, etc.) are therefore assumed to be corrected by a previous calibration step that transformed all observational spectra to a common reference instrument.

\subsection{Gaia spectrophotometric instruments \label{sec:gaiaInstruments}}
The fact that neglecting the influence of the LSF in the calibration of \gaia, that is, using the approximation of Eq.~(\ref{eq:stdApprox}), is not good for the \gaia~spectrophotometric instruments has been noted before \citep{Pancino2010,Cacciari2010}. However, for deriving the SPD, a plain numerical inversion of the discretised integral equation has been proposed, which suffers strongly from the ill-posed nature of the problem. Not being able to employ the simple calibration scheme offered by Eq.~(\ref{eq:stdApprox}), one is forced to return to Eq.~(\ref{eq:sampledXP}), and to solve the integral equation for the SPD. In the following we do just that, making use of the matrix approach as outlined.\par
The approach is applied to both spectrophotometric instruments, BP and RP. For the computations we assume idealised instruments, with parameters taken from \cite{Gaia2016a}. In particular, we use the values of $D =$~1.45~m for width of the rectangular aperture in AL direction, $F =$~35~m for the focal length, and $d = $10~$\mu$m for the pixel size in AL direction, divided into four TDI phases. Instead of referring to 'pixels' we use the \gaia~related term 'samples' in the following. With these values, we can derive the LSF, for which we neglect all optical aberrations except for a possible defocusing of the instrument. Detector effects on the LSF are taken into account in the form of the TDI integration and the sample integration. The OTF is thus composed of a number of multiplicative terms, a triangular function resulting from autocorrelation of the rectangular aperture of \gaia, two cardinal sine functions resulting from the pixel integration and blurring when moving the accumulated electrons in the quarter-pixel steps during TDI integration, and a cardinal sine function resulting from a possible defocusing of the instrument \citep{Goodman1996}. The assumption for the OTF is therefore
\begin{equation}
H(\nu) = \Lambda(a\,\nu) \, \sinc{\nu} \, \sinc{\nu / 4} \, \sinc{b\, a\, \nu \left[1-a\, |\nu|\right]} \quad , \label{eq:OTF}
\end{equation}
with
\begin{equation}
a = \frac{\lambda}{D}\frac{ F}{d} \; , \quad b = \frac{D^2}{\lambda} \, \left(\frac{1}{F} - \frac{1}{F+\Delta F} \right) \quad .
\end{equation}
However, the assumptions on the LSF are still idealised  as it neglects a number of effects, resulting from the optics (e.g. mirror imperfections, resulting in aberrations), the detector (e.g. charge diffusion in the field free regions of the CCD detector, resulting in a smearing of the signal), and the spacecraft (e.g. variations in the spacecraft attitude, resulting in a smearing of the source image while integrating over one transit over the CCD). All these neglected effects lead to further broadening of the LSF, which makes the simplified assumptions of this work optimistic. Nevertheless, for illustrating the complications of low-resolution spectroscopy, the optimistic assumptions on the width of the LSF are fully sufficient. A more detailed account of the simulation of the \gaia~PSFs is provided by \cite{Babusiaux2004}. For computing Fourier integrals, we use the integration scheme by \cite{Bailey1994}.\par
 We introduce a convolution with a Gaussian to the LSF in order to include an empirical way of correcting the width of the LSF in case
of a poor initial guess. The standard deviation of this Gaussian, denoted $\sigma_u(\lambda)$,  may be chosen to be wavelength dependent. The empirically corrected OTF, $H^\prime(\nu)$, then becomes
\begin{equation}
H^\prime(\nu) = H(\nu) \cdot {\rm exp}\left\{-\frac{1}{2}\left[2\,\pi\, \sigma_u(\lambda)\right]^2\, \nu^2\right\} \quad . \label{eq:OTF2}
\end{equation}
We make use of this modified OTF when testing the robustness of the matrix approach.\par
For simulating the \gaia~XP instruments, further assumptions on the dispersion relations and response curves have to be made. The nominal dispersion relationships for \gaia's BP and RP instruments have been published by ESA, separate by field-of-view (FoV) and CCD row\footnote{\url{https://www.cosmos.esa.int/web/gaia/resolution}}. These relationships are similar for the different FoVs and rows, and for the purpose of this work we pick out the one for FoV~1, CCD row 4 as $\Delta(\lambda)$ for the simulations of the XP spectra.  We linearly extend the relations on both ends of the provided intervals in order to obtain a good simulation at
the boundaries of the wavelength intervals on which the dispersion relations
are provided.\par
\begin{figure}
   \centering
   \includegraphics[width=0.47\textwidth]{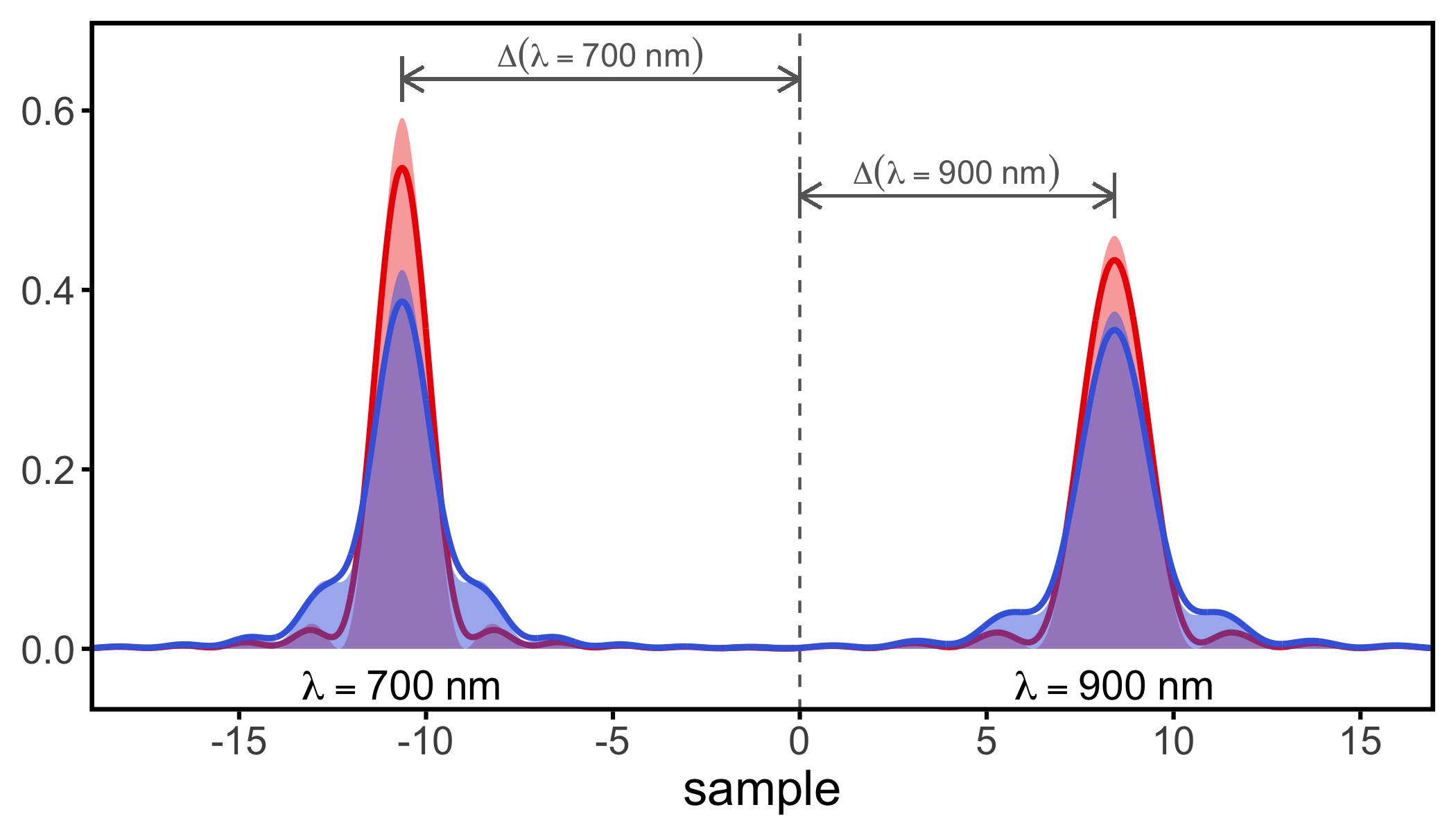}
   \caption{Simulated LSFs for wavelengths of 700~nm and 900~nm, including the dispersion relation $\Delta(\lambda)$ for RP. Red shaded areas show diffraction limited LSFs. Red lines show diffraction limited LSFs, with CCD effects included (pixel integration and TDI smearing). Blue shaded regions: optical LSFs for a defocusing of $\Delta F =$ 1~mm. Blue lines: LSFs for defocusing of 1~mm, with CCD effects included.}
              \label{fig:lsf}
    \end{figure}

An example for the simulated LSFs and dispersion relations is shown in Fig.~\ref{fig:lsf}. The LSFs for two different wavelengths, 700~nm and 900~nm, are shown using the RP dispersion relations. The shown LSFs include the diffraction limited case without and with the CCD effects, for the focussed instrument and for a defocusing of $\Delta F =$~1~mm displacement of the detector with respect to the true focal plane. The widening of the LSF with increasing wavelengths results from the optical terms in the OTF. The displacement $\Delta(\lambda)$ (with respect to the chosen reference wavelength of 800~nm) is the dispersion relation.\par
The nominal pre-launch response curves for BP and RP were published by \cite{Jordi2010}. However, these curves have been improved based on the published \gaia~DR2 integrated photometry for BP and RP since then by \cite{Evans2018}, \cite{Weiler2018}, and \cite{MAW2018}. The response curves by \cite{MAW2018} provide the best agreement between the synthetic photometry and the integrated XP photometry of \gaia~DR2, and we use these response curves for the simulations of this work. For BP, two response curves are provided by \cite{MAW2018}, which apply to different magnitude ranges. We select the curve for the bright magnitude range here, as this response curve appears to correspond to a better calibration of the integrated BP photometry.\par
We refer to the quantities used for the simulation of the \gaia~XP instruments, and quantities directly derived from them,  as the 'true' values when comparing with the calibration results.

\subsection{Outline of the calibration \label{sec:gaiaOutline}}

A set of   SPDs is required for the calibration and for this purpose  we selected the BaSeL WLBC99 spectral library \citep{Westera2002} as it provides spectra over a wide range of astrophysical parameters that cover the entire wavelength range of interest for the \gaia~spectrophotometric instruments. We select 108 spectra from this library with effective temperatures from 2500~K to 15,000~K, and surface gravities of $log(g)$ = 4.5, 2, and 1.5~dex.  We use this subset, for which we know both the SPD and the observational spectra, as calibration sources for the instrument calibration. Only knowledge of the observational spectra is assumed for  the remaining 2627 spectra in the BaSeL library, and we use these as a test set for the instrument calibration, deriving their SPDs. We expand this test set by applying interstellar extinction to the SPDs, using the extinction law by \cite{Cardelli1989}, and assuming a colour excess of $E(B-V) =$ 1\m.\par
For all spectra, we generate simulated observations for the 'true' instrument. The simulated observations consists of ten transits per source, with 60 samples per transit. Each transit has a random sub-pixel positioning. The random noise added to the observations consists of two components: the photon noise which depends on the assumed brightness of the source, and a normally distributed noise contribution with zero mean and a standard deviation of 10 electrons per transit. This noise component serves as a simple substitute for read-out noise and sky background noise on the data. The ten transits are combined in both the instrument and the source calibration. This combination requires that the SPDs of the sources be non-time dependent. For the instrument calibration, the absence of significant temporal variability may be a requirement for selecting calibration sources, and the combination is thus, in practice, justified. For the source calibration, the combination of different transits of the same source may not be justified for sources with temporal variability. The matrix approach itself is however not dependent on the number of transits, and may be applied to a smaller number of transits combined in the calibration.\par
The assumed brightnesses of the simulated sources are based on the $G$ band magnitudes, using the $G$ band response curve from \cite{MAW2018}. The choice for the $G$ band as the standard for the brightness in the two spectrophotometric instruments, BP and RP, is based on the use of the onboard estimate of the $G$ band magnitude to trigger gating in the XP CCDs to reduce the effective exposure time \citep{Gaia2016a}. For sources brighter than approximately 12\m~in $G$ band \citep{Carrasco2016}, the exposure time is reduced by integrating the spectra of the source only over a fraction of the CCD length in AL direction. This means that, for approximately 12\m~in $G$, the maximum signal-to-noise ratio is expected (ignoring possible saturation effects), independently of the SPD of the source. We therefore specify the magnitudes in $G$ band throughout, using magnitudes only down to 12\m, and the corresponding exposure time of 4.4167~s for a full CCD transit \citep{Gaia2016a}. Together with the aperture area of 0.7278~$\rm m^2$, the photon noise can be computed.\par
For the 108 calibration sources, we assume a $G$ band magnitude of 13\m~throughout. For the test sources, we consider three different magnitudes, namely $G=$ 13\m, 16\m, and 19\m. The brightest test sources thus have the same brightness, and with it the same signal-to-noise level as the calibration sources, while the other cases represent medium to low signal-to-noise levels. For the brightest case, we may expect to be sensitive to systematic errors in the calibration, which may become increasingly covered by random noise as the brightness is reduced to 16\m~ and 19\m.\par
In this work, we simulate two scenarios. In the first one, we assume a perfectly focused instrument. This corresponds to the most optimistic case as far as the effects of the resolution are concerned, as any neglected effects can only result in a broader LSF. However, the model of the LSF used in the calibration is the same as that used for producing the observations, expressed in Eq.~(\ref{eq:OTF}). This test case therefore allows in principle to perfectly reproduce the true LSF in the calibration process, and thus  also represents the most optimistic test case as far as systematic errors are concerned. 
However, in practice it may not be possible to exactly reproduce the true LSF with the LSF model in the calibration. We therefore also simulate a second scenario in which we introduce a systematic difference between the true LSF and the LSFs that can be represented by the LSF model used in this calibration.\par
In order to introduce a systematic mismatch between the true LSF and the modelled LSF in the second scenario, we introduce a mild defocusing of the telescope. This defocusing corresponds to a convolution of the focused LSF with a rectangular function. However, in the modelling of the LSF we  do not modify the corresponding cardinal sine term in Eq.~(\ref{eq:OTF}), which would allow for an exact compensation of the defocusing; instead we allow only to adapt the Gaussian term. The true LSF is given by Eq.~(\ref{eq:OTF}) for some defocusing $\Delta F$, while the model LSF is given by Eq.~(\ref{eq:OTF2}), with $b=0$. This way we introduce a systematic wavelength-dependent deviation between the true LSF and the modelled LSF, the degree of the deviation being adjustable by the choice of the defocusing $\Delta F$.\par
\begin{figure}
   \centering
   \includegraphics[width=0.43\textwidth]{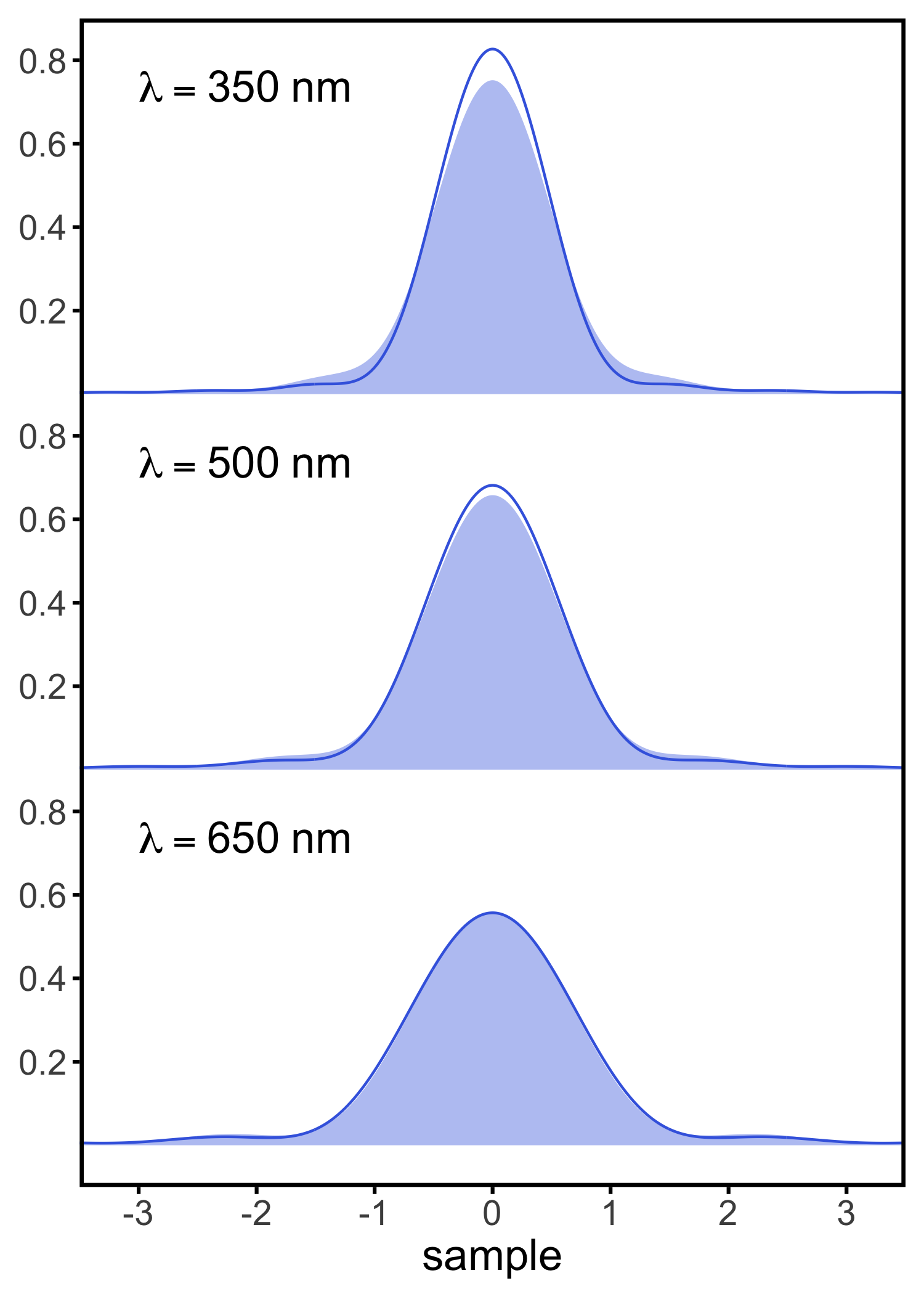}
   \caption{Line-spread functions for a defocusing of $\Delta F$ = 0.3~mm (shaded regions) and the adopted model LSFs (solid lines) for three different wavelengths: 350~nm, 500~nm, and 650~nm.}
              \label{fig:defocusedLSFs}
    \end{figure}

Here, we use a value of 0.3~mm for $\Delta F$. The true LSFs resulting from this model are shown as shaded regions in Fig.~\ref{fig:defocusedLSFs} for three different wavelengths: 350~nm, 500~nm, and 650~nm. The value of 0.3~mm for $\Delta F$ has been chosen because from this value onwards it becomes obvious in the calibration process that the LSF model in Eq.~(\ref{eq:OTF2}) no longer allows for a good representation of very hot calibration sources. Thus this value represents about the maximum deviation between the true LSF and the model LSF at short wavelengths that one may consider tolerable.

\subsection{Illustration of the problem \label{sec:problemIllustration}}

\begin{figure}
   \centering
   \includegraphics[width=0.49\textwidth]{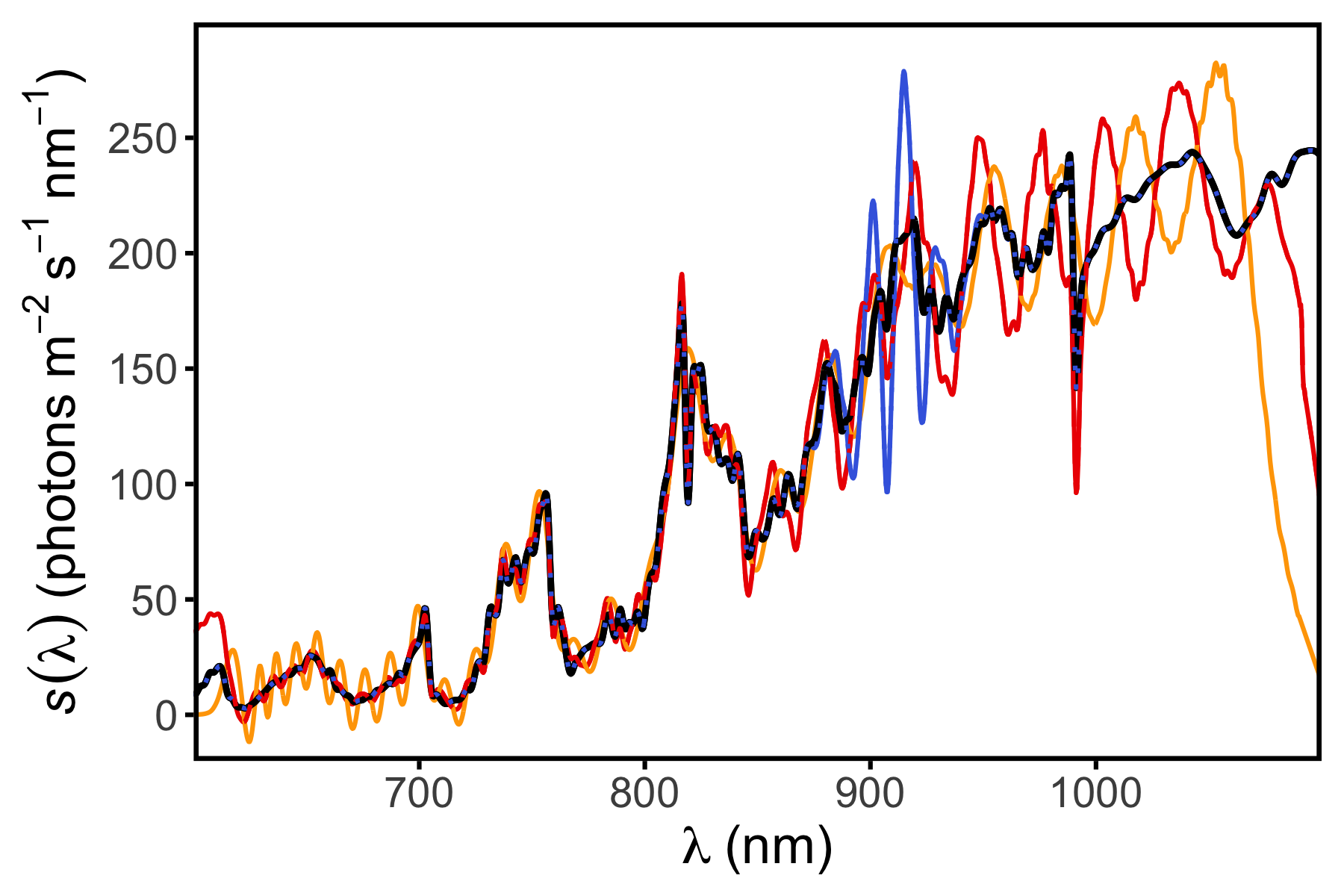}
   \includegraphics[width=0.49\textwidth]{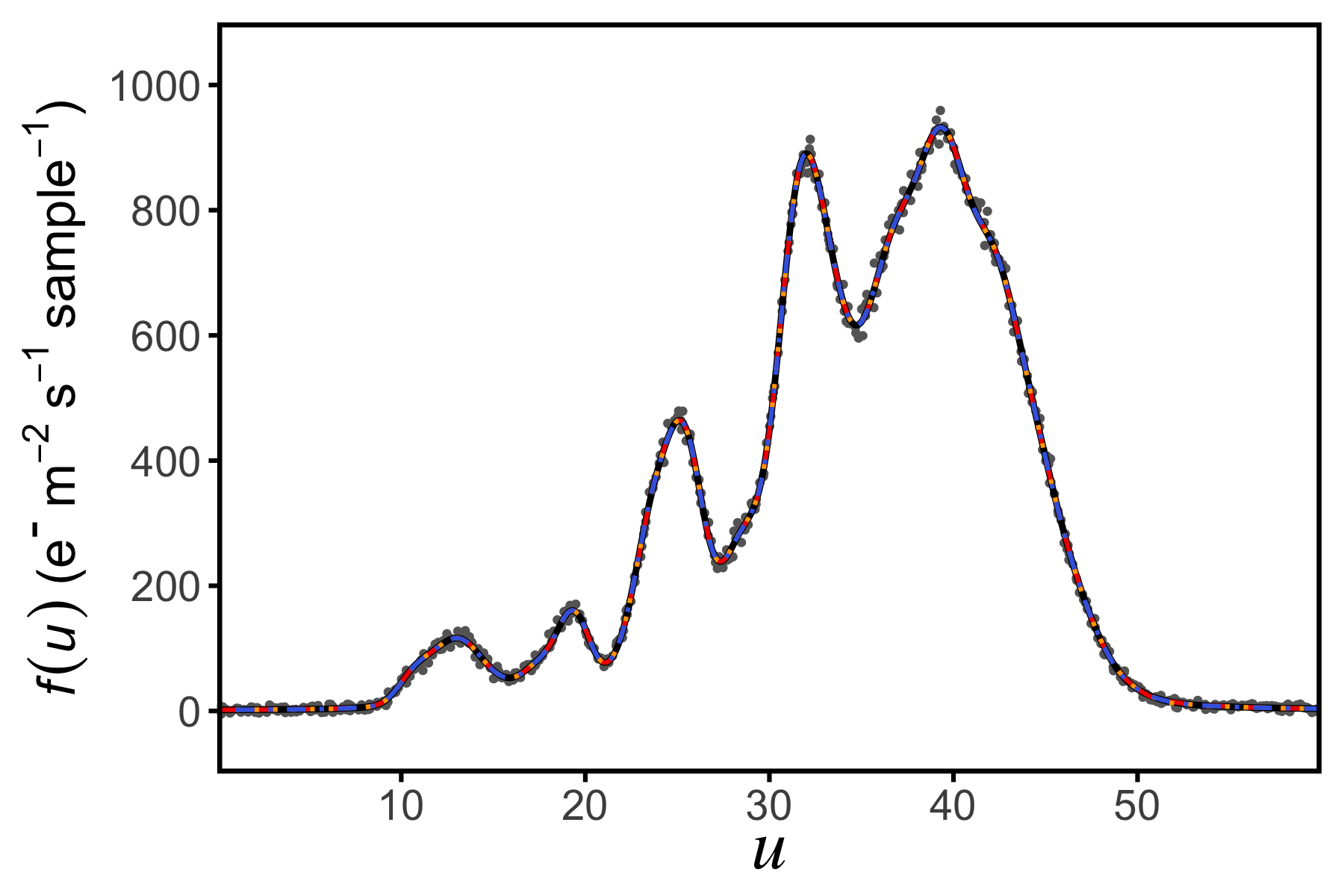}
   \caption{Top panel: SPD of an M-type BaSeL spectrum (black line) and three artificial SEDs (red, orange, and blue lines), scaled to a $G$ magnitude of 16\m. Bottom panel: Simulated continuous, noise-free observational RP spectra of the SPDs in the upper panel (solid lines; the differences between the lines are too small to be visible). The grey dotted symbols show simulated noisy observations.}
              \label{fig:degeneracyExample}
    \end{figure}

Before entering into the actual calibration problem, we first illustrate the ill-posed nature of the spectrophotometric calibration of the \gaia~BP and RP instruments with a graphical example. We take one of the BaSeL SPDs (scaled to $G$ = 16\m) over the wavelength range covered by the RP instrument. The SPD is shown as the black line in the top panel of Fig.~\ref{fig:degeneracyExample}; it belongs to an M-type source, which has a rather complex shape in its SPD. For this SPD, we compute the observational RP spectrum as a continuous function in the noise-free case. This observational spectrum is shown as the black line in the bottom panel of Fig.~\ref{fig:degeneracyExample}, together with a noisy simulated observational spectrum for ten transits, shown as grey dots.\par
Now we construct further SPDs different from the 'true' one, which result in observational RP spectra which are indistinguishable from the RP spectrum of the true SPD. The upper panel shows three examples for such SPDs. The orange line gives an example with a wavy pattern running through the whole SPD, illustrating the sensitivity of the inversion problem to wavy variations, indicated by Eq. (\ref{eq:problem1}) in Sect.~\ref{sec:standard}. The increasing width of the LSF with increasing wavelength allows for a lower frequency of the introduced wavy pattern at long wavelengths. The SPD shown as the red curve has been constructed such that the wavy pattern is suppressed at short wavelengths, whilst the curve remains similar to the orange curve at long wavelengths, and with a high phase shift in the pattern. The blue curve shows an SPD which is the true one, but with a rather localised wavy pattern placed around the wavelength of 910~nm. The blue curve is therefore difficult to distinguish from the black, true curve at wavelengths far from 910~nm.\par
The simulated RP spectra, using the same true instrument in all cases, are shown in the bottom panel of Fig.~\ref{fig:degeneracyExample}, plotted over the true one. The differences are nevertheless invisible to the eye in this kind of plot, as the differences between the continuous, noise-free RP spectra for the four SPDs in the top panel are much too small. Within the inner part of the RP spectra, where the signal is significant, the four RP spectra have relative differences of about $\rm 10^{-4}$. The significance of the differences between the simulated RP spectra must be judged by comparison with observational data. We therefore compute the $\chi^2$ values for the simulated noisy spectra shown in the bottom panel of Fig.~\ref{fig:degeneracyExample} for the RP spectra of the four different SPDs, and compute the $p$--value for the chi-squared distribution. In the example shown here, these $p$--values are between 0.941 and 0.943 for the four RP spectra, showing that the four SPDs are entirely indistinguishable from the observational data.\par
Although the observational RP spectra of the four different SPDs are very similar, one of them describes the given simulated observational data better than the others, even if the improvement over the other SPDs is highly insignificant. Nevertheless, as the differences in the goodness of fit are so small, changing the noise pattern on the observational data may easily make another SPD the best match with the observational data. This therefore introduces instability in the inversion process that derives the SPD from observational data.  The  SPD that provides the best agreement with the observational spectrum can change in response to minimal changes to the observational data. On the other hand, the improvement in agreement with the observational data may be entirely insignificant, and the best SPD may not be a good approximation of the true one. In the following sections we apply the matrix approach to the calibration problem, and see how it handles the ill-posed nature of the problem.

\subsection{Construction of $V$ and $W$ \label{sec:exampleBases}}

The two spectrophotometric instruments of \gaia, BP, and RP, covering different but overlapping wavelength ranges, require the vector spaces $V$ and $W$ to be constructed separately for BP and RP. We denote these $V_{\rm BP/RP}$ and $W_{\rm BP/RP}$, respectively. Although it would simplify the calibration if the vector space in which the SPDs are represented were constructed common to BP and RP,  in a first step we construct two separate spaces. As the spaces are constructed from the dominant spectral components, the dominant components over the entire wavelength range covered by BP and RP may not agree well with the dominant components on the individual wavelength ranges. We therefore obtain better results by keeping the calibration of BP and RP separated, and then merging the calibration results for BP and RP on the expanded vector spaces $V^\ast_{\rm BP/RP}$ to a single solution on a vector space $V^\ast$.\par
We begin the calibration by constructing bases from the observational spectra for BP and RP, respectively. In order to suppress the effects of strong heteroscedasticity in the observational spectra, we do not construct the basis functions directly from the data, but rather introduce a set of generic orthonormal basis functions for their representation, which we then use as a starting point for deriving a more suitable basis in coordinate space. For the orthonormal basis functions to start with, we use Hermite functions, $\bar{\varphi}_i(x)$. These functions are centred on the origin, and provide an orthonormal basis on $\mathbb R$. The choice is eventually driven by mere convenience, as Hermite functions converge to zero for sufficiently large values of $|x|$, which makes them less sensitive to outliers at the edges of observational spectra in comparison to polynomial bases. As is the case for other orthonormal bases, Hermite functions obey a recurrence relation with which they can be computed conveniently. Other bases, such as for example Legendre polynomials, may however be suitable as well. The use of orthogonal polynomials could bring the advantage of a lower sensitivity to errors in the subtraction of the sky background, should this be a practical issue. However, as far as the principles are concerned, which orthogonal functions are actually used is of no relevance. The Hermite functions used here have to be centred to the observational spectra and adjusted to their width. We therefore assume a relationship between the argument of the Hermite functions $x$ and the focal plane coordinate $u$ of the form
\begin{equation}
x = \frac{u-sft}{scl} \quad , \label{eq:substitution}
\end{equation}
with $sft$ being the shift in zero point, and $scl$ a scaling parameter. To maintain the orthonormality given by Eq.~(\ref{eq:HermiteOrthonormality}) under the substitution given by Eq. (\ref{eq:substitution}), we scale the Hermite functions with a factor $1/\sqrt{scl}$. For the shift we adopt a value of $sft=30$ samples, as this corresponds to the approximate centre of the spectra in both BP and RP. In order to obtain a good estimate for the scaling and the number of Hermite functions $N^\prime$ required for the representation of the observational spectra of the calibration sources, we use a cross-validation approach, dividing the calibration sources into two groups by transits. One group is fitted with Hermite functions using different numbers of Hermite functions and different values for $scl$. The goodness of fit for the data set, for which we use the normalised chi-square value, is then judged from the second group. We then select the number of Hermite functions and scaling parameter corresponding to the overall best goodness-of-fit parameter. For the test data generated and the focused telescope, these values were $N^\prime = 77$ Hermite functions and $scl = 2.52$ for BP, and $N^\prime = 67$ Hermite functions and $scl = 2.77$ for RP. For the defocused telescope, the numbers are $N^\prime = 69$ Hermite functions and $scl = 2.68$ for BP, and $N^\prime = 69$ Hermite functions and $scl = 2.76$ for RP. However, these numbers are not too accurately determined by the calibration sources, and the typical variations in $scl$ and $N^\prime$ are about 0.1 and 2, respectively, when using different sets of random observations.\par
In the following step, we optimise the bases by producing more efficient linear combinations of Hermite functions. This is done by applying a SVD to the matrix of coefficients, as outlined in Sect.~\ref{sec:implementation}. The number of linear combinations of Hermite functions is again fixed using the same cross-validation approach as before.\par
The basis functions for $V$ were produced by performing a spline interpolation of the BaSeL SPDs to a dense regular grid, and simply applying SVD to the resulting matrix of the tabulated fluxes for the 108 calibration sources. This was done on the wavelength intervals covered by BP and RP independently.\par
\begin{figure}
   \centering
   \includegraphics[width=0.49\textwidth]{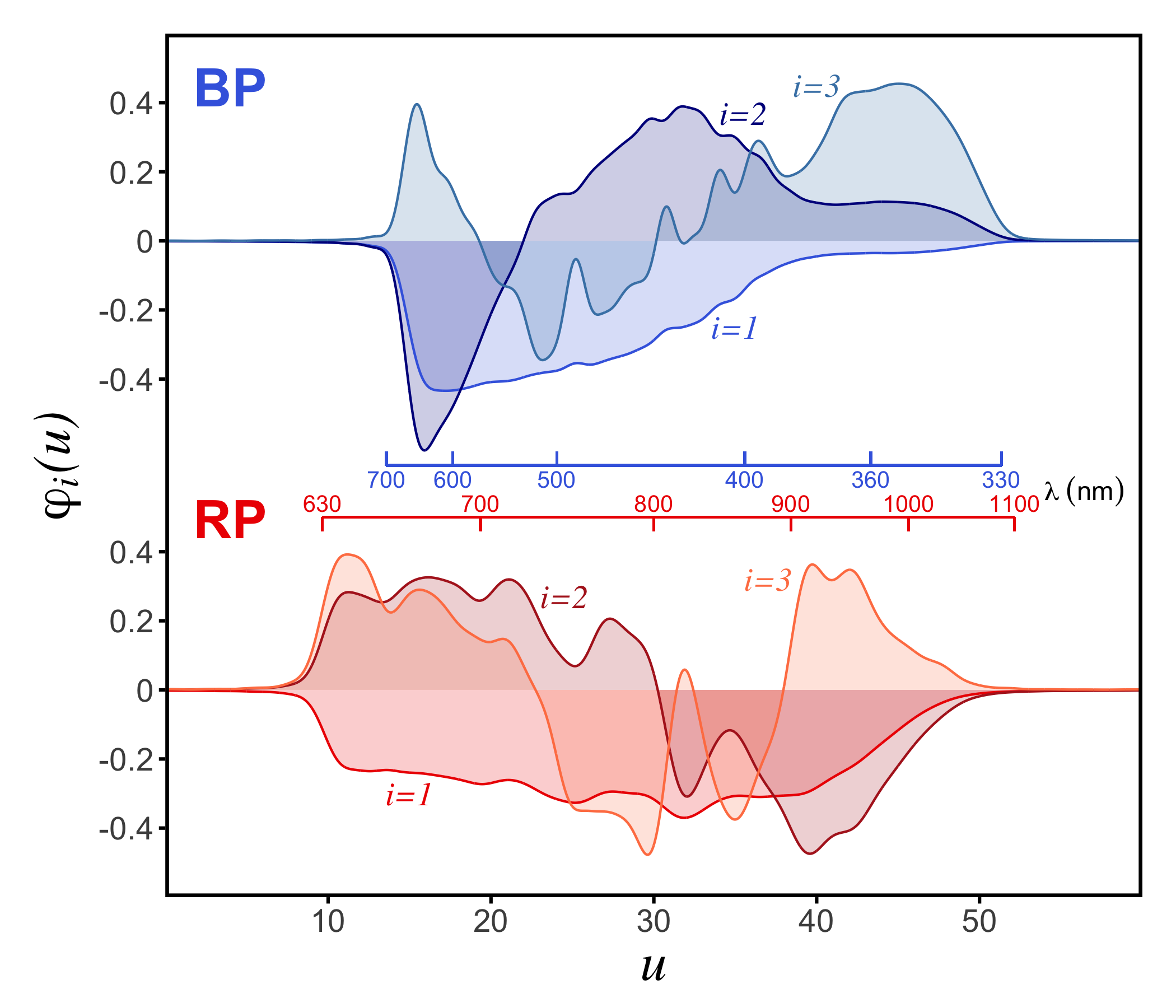}
   \includegraphics[width=0.49\textwidth]{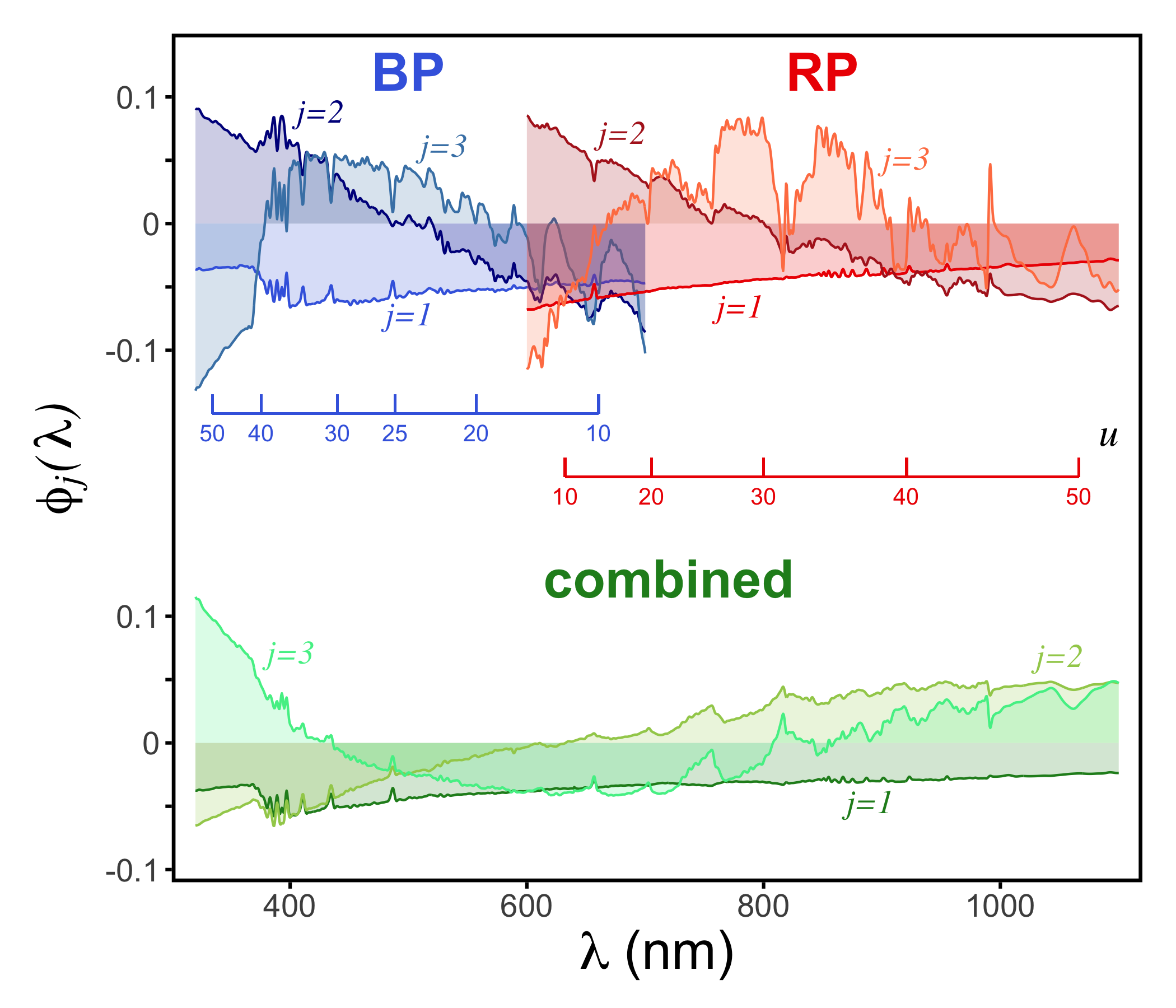}
   \caption{Examples for the basis functions for $W$ and $V$. Upper panel: First three basis functions $\varphi_i(u)$, $i=1,2,3$ for the observational spectra for BP and RP. The corresponding wavelength axes are indicated. Lower panel: First three basis functions $\phi_j(\lambda)$, $j=1,2,3$, for the SPDs for BP and RP, and the combined basis functions covering the entire wavelength range. The corresponding sample axes are indicated.}
              \label{fig:basisFunctions}
    \end{figure}

As examples, the first three basis functions for $V$ and $W$ are shown in Fig.~\ref{fig:basisFunctions}, both for BP and RP. For comparison, the wavelength and sample axis are shown as well. The non-linear dispersion relation results in the visible non-linear relationship between wavelength and sample.\par
Some common features of the bases can be seen. For increasing order of basis functions, the functions show more complex features, as they describe more and more detailed features of the data set as the order increases. The basis functions for $V$ are richer in small-scale features than the basis functions for $W$, which reflects the much lower spectral resolution of the observational spectra by \gaia~as compared to the calibration SPDs. The basis functions for $W$ show a strong decrease for very large and very small sample values $u$. This is caused by the response function dropping to zero well within the wavelength range covered by the presented sample range. As a consequence, very large and very small sample values contain only the contribution of the wings of the LSFs, which manifests itself in the basis functions as a strong and smooth decrease.

\subsection{Instrument calibration \label{sec:GaiaInstrumentCalibration}}

The instrument calibration starts by representing the observational spectra and SPDs of the 108 calibration sources in the bases $V_{\rm BP/RP}$ and $W_{\rm BP/RP}$ and solving Eq.~(\ref{eq:instrumentMatrix1}) for the instrument matrix $\bf I$, linking $V_{\rm BP/RP}$ and $W_{\rm BP/RP}$.\par
In preparation of the expansion of $V_{\rm BP/RP}$ and $W_{\rm BP/RP}$ we then estimate the instrument kernel $I(u,\lambda)$ from the instrument matrix $\bf I$. For this step we require initial guesses for the dispersed LSF and the response curve, as well as a modification model for the response.  The initial guess for $L(u,\lambda)$ can be adapted by manually modifying the free parameters introduced in Eq.~(\ref{eq:OTF2}). The response initial guess starts from a poor guess (using a scaled BP faint passband from \cite{MAW2018} instead of the correct one), which is improved by using a smooth calibration spectrum and Eq.~(\ref{eq:vaccasLSF}). The approximation using Eq.~(\ref{eq:vaccasLSF}) is then combined with the original initial guess at parts where the influence of the LSF is expected to be large, i.e. at the longest and shortest wavelengths on the interval covered by the response curves. A modification model based on B-splines is then chosen, with knot sequences such that the modification model is more dense at the long and short wavelength parts of the 
covered interval. Using this latter model, Eq.~(\ref{eq:newresponse}) can  be solved for the coefficients of the modification model. \par
\begin{figure}
   \centering
   \includegraphics[width=0.47\textwidth]{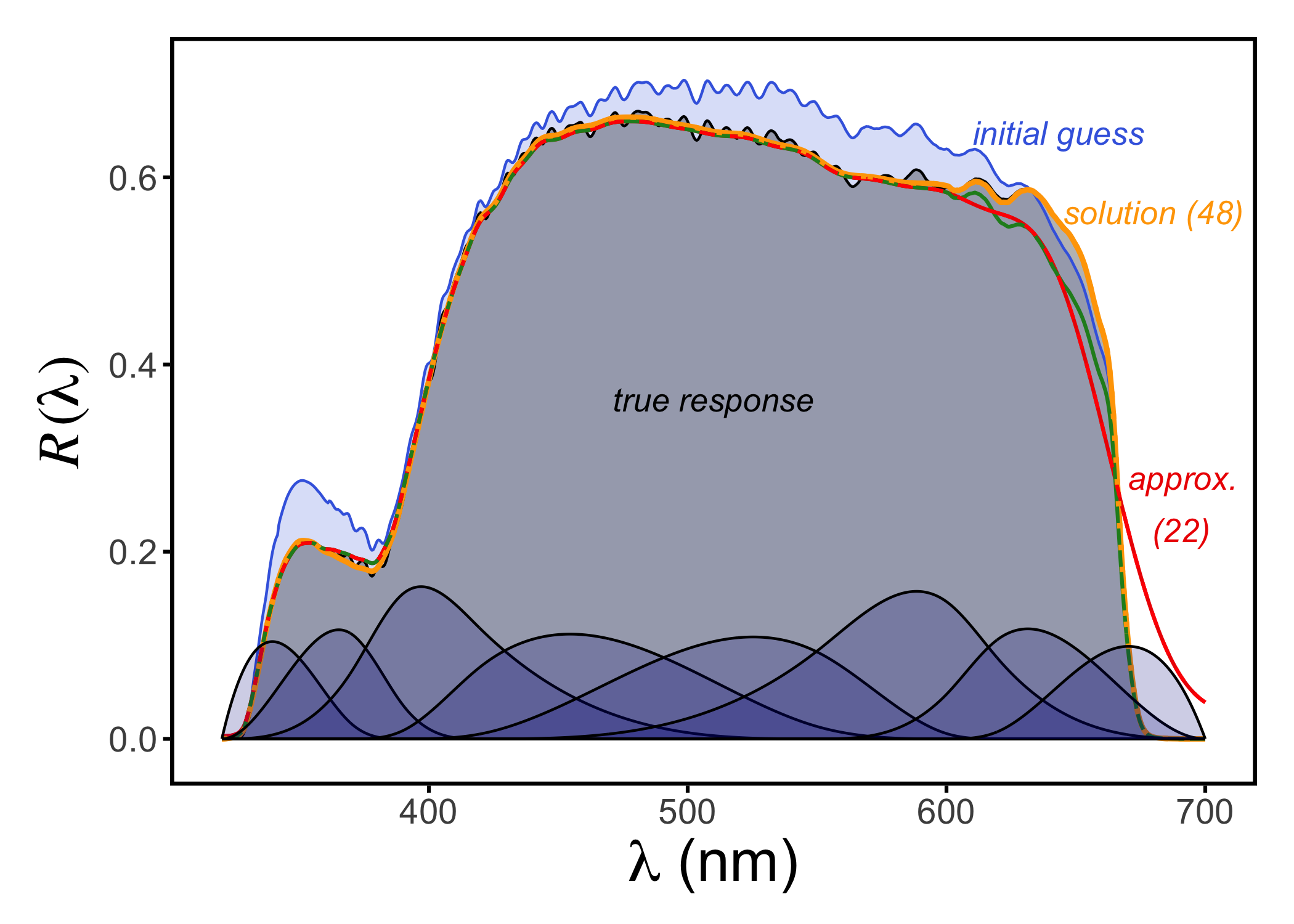}
   \caption{Illustration of the different steps of construction of the response function for BP. The blue line and shaded region is the initial guess, and the black line and grey shaded region is the true passband. The red line is the approximation using Eq. (\ref{eq:vaccasLSF}) and the green line the combination of the initial guess and the approximation with Eq. (\ref{eq:vaccasLSF}). The orange line shows the solution obtained with Eq. (\ref{eq:newresponse}). The B-spline basis functions $B_k(\lambda)$ are shown for comparison at the bottom of the plot.}
              \label{fig:response}
    \end{figure}

Figure~\ref{fig:response} illustrates the different steps in the response reconstruction for BP, starting from the initial guess, followed by the approximation by Eq.~(\ref{eq:vaccasLSF}), the combination with the initial guess, and the solution of Eq.~(\ref{eq:newresponse}). The B-spline basis functions are shown for comparison. A good reconstruction of the response curve is possible this way, and is barely dependent on the initial guess for the LSF. The differences between the focused and de-focused instrument are negligible in this calibration step. The LSF model can be refined using the new response curve if necessary.

\subsection{Expansion to $V^\ast$ and $W^\ast$ \label{sec:GaiaExpansion}}

For the expansion of the spaces $V$ and $W$ to $V^\ast$ and $W^\ast$ we test the three approaches described in Sect.~\ref{sec:expansion}, that is, using model spectra, generic functions without any particular relationship to astronomical spectra, and a combination of both.\par
For the first approach, we use the BaSeL SPDs of the test sources as a set of model spectra. We compute the orthogonal components of all these BaSeL SPDs with respect to $V_{\rm BP/RP}$, and then orthonormalise the orthogonal components using SVD. We then add as many of these new basis functions to $V$ as possible. The limit is given by the condition of keeping the covariance matrix of the solution well away from numerical singularity. This is required for the solution to remain interpretable. The conditioning of the covariance matrix is the square of the conditioning of the instrument matrix. With this approach to the expansion of $V_{\rm BP/RP}$, we expand $V_{\rm BP}$ with $N = 15$ to $V_{\rm BP}^\ast$ with $N^\ast = 55,$ and $V_{\rm RP}$ with $N = 18$ to $V_{\rm RP}^\ast$ with $N^\ast = 63$.\par
For the generic functions we use B-spline basis functions in wavelength, where we simply use a regular grid of spline knots over the wavelength interval covered by the BP and RP basis functions, respectively. We increase the density of the spline knots, and with it the density of B-spline basis functions, as the conditioning of the instrument matrix permits. By doing so, we obtain $N^\ast = 51$ for BP and $N^\ast = 58$ for RP.\par
Finally, we combine both approaches, using the first ten basis functions of the BaSeL model spectra, and then add B-spline basis functions as in the previous approach until the conditioning cannot be increased further. In this case, eventually $N^\ast = 61$ for BP and $N^\ast = 68$ for RP for the focused telescope case.\par
The consistently somewhat higher final numbers of basis functions for RP as compared to BP are caused by the M-type stars. The high complexity of their SPDs necessitates more basis functions to describe them, and their signal-to-noise ratio is much higher in RP than in BP for the case of fixed $G$ magnitudes.\par
In cases where $N^\ast$ is larger than 60, the number of coefficients to be determined in the source calibration is larger than the number of samples for a single transit. As an additional complication, corrupted samples (e.g. due to a cosmic ray hit) may further reduce the number of samples in a single transit that can be used for the determination of $\bf c$. If only a single transit can be used in the source calibration, for example,  for a source with temporal variation, Eq.~(\ref{eq:fullSolution}) may become underdetermined. Here,  SVD can be used to find a particular solution for the underdetermined system. With mild underdetermination, this approach provides good results. For the case of the expansion of $V$ and $W$ with a combination of model spectra and B-spline basis functions for RP, with $N^\ast =68$, the suppression of eight basis functions may result in overly large breakdown errors. In this case, the source calibration of single transits may require the construction of an expansion with less B-spline basis functions. Having constructed the additional basis functions that expand $V$ and $W$ to $V^\ast$ and $W^\ast$, the corresponding expanded instrument matrices for BP and RP, ${\bf I}_{\rm BP}^\ast$ and ${\bf I}_{\rm RP}^\ast$, can be computed using Eq.~(\ref{eq:instrumentMatrixElements}).

\subsection{Combination of BP and RP bases \label{sec:GaiaCombination}}
The two \gaia~instruments for two wavelength regions leaves us with the additional complication of combining the calibration results for BP and RP to a single spectrum. To approach this problem, we first combine the bases by expanding the domains of the BP and RP basis functions to the wavelength interval covered by both bases, continuing the basis functions with zero. The coefficient vectors for BP and RP we combine in a common vector ${\bf c} \coloneqq [{\bf c}_{BP}, {\bf c}_{RP}]$. If the wavelength intervals covered by BP and RP were found to not overlap, the combination would represent a fully orthonormal basis and the process of combination would already be completed. To take into account the overlap, we have to separate the bases at a transition wavelength $\lambda_{trans}$, by setting the basis functions for BP to zero for all wavelengths longer than or equal to $\lambda_{trans}$, and all basis functions for RP to zero for all wavelengths shorter than $\lambda_{trans}$. We use $\lambda_{trans} = 640$~nm. By doing so, we have constructed two sets of basis functions representing two orthogonal subspaces, while the basis functions for each subspace are not orthogonal with respect to each other. We fully orthonormalise the set of basis functions by first using an orthonormal matrix decomposition to the matrix $\bf B$ consisting of the sampled basis functions. Using the SVD, we use the columns of the matrix ${\bf U}_{\rm B}$ as the new sampled basis functions, and the matrix $\bs{\Sigma}_{\rm B} \, {\bf V}_{\rm B}^{\mathsf T}$ as the transformation matrix of the coefficient vector $\bf c$ to the new basis.\par
As the individual basis for $V^\ast_{\rm BP}$ and $V^\ast_{\rm RP}$ cover partly the same wavelength range, a few singular values could become very small, in particular if the expansion of the $V_{\rm BP}$ and $V_{\rm RP}$ has been done with basis functions mainly covering the overlap region between BP and RP. In the test cases of this work this situation occurs for the B-spline expansion to $V^\ast$ and $W^\ast$. We therefore introduce a cut-off value in the singular values, using less combined basis functions than the sum of the number of basis functions for BP and RP if appropriate. We use a cut-off value of 1\% to reduce the number of combined basis functions.\par
The resulting basis functions may appear rather artificial, so as an essentially cosmetic correction, we may arrange the coefficient vectors $\bf c$ for a large number of sources in a matrix $\bf C$, and again apply a SVD to sort the coefficients. For the final combined basis we use the columns of the matrix ${\bf U}_{\rm B} \, {\bf V}_{\rm C}$. The coefficients in the combined basis are then obtained from the coefficients for BP and RP via the transformation
\begin{equation}
{\bf c}^{comb} =  {\bf V}_{\rm C}^{\mathsf T} \, \bs{\Sigma}_{\rm B} \, {\bf V}_{\rm B}^{\mathsf T} \, {\bf c} \quad . \label{eq:combination}
\end{equation}
The first three basis functions for the combined basis for $V^\ast$ are also shown for illustration in the bottom panel of Fig.~\ref{fig:basisFunctions} in green.

\subsection{Source calibration \label{sec:GaiaSourceCalibration}}

The calibration for a given source is straight forward, solving Eq.~(\ref{eq:fullSolution}) for BP and RP, respectively, and then using Eq.~(\ref{eq:combination}) to obtain the coefficients in the combined basis. In order to compute the actual solution for the SPD from the coefficients in the combined basis, to get a visual impression of the shape of the SPD, or for further computations based on the SPD, it is of interest to suppress the coefficients with extremely large errors. We do so by following the scheme outlined below.\par
First we transform the coefficient vector in the combined basis into the reference system in which its covariance matrix diagonalises. The coefficient vector in this basis is obtained by multiplying the coefficient vector in the combined basis from the left by ${\bf V}^{\mathsf T}_{\Sigma}$, with ${\bf V}^{\mathsf T}_{\Sigma}$ resulting from the SVD of the covariance matrix of the coefficient vector, $\bs{\Sigma}^{\rm c}$. We then compute the standard deviation of the coefficients in a one-sided running window as a measure of the scatter of the coefficients with increasing order of basis function. We compare this scatter with the errors  of the coefficients, that is, the square roots of their variances. When the scatter of the coefficients becomes comparable to the errors in the coefficients, we consider them no longer well constrained.\par
\begin{figure}
   \centering
   \includegraphics[width=0.49\textwidth]{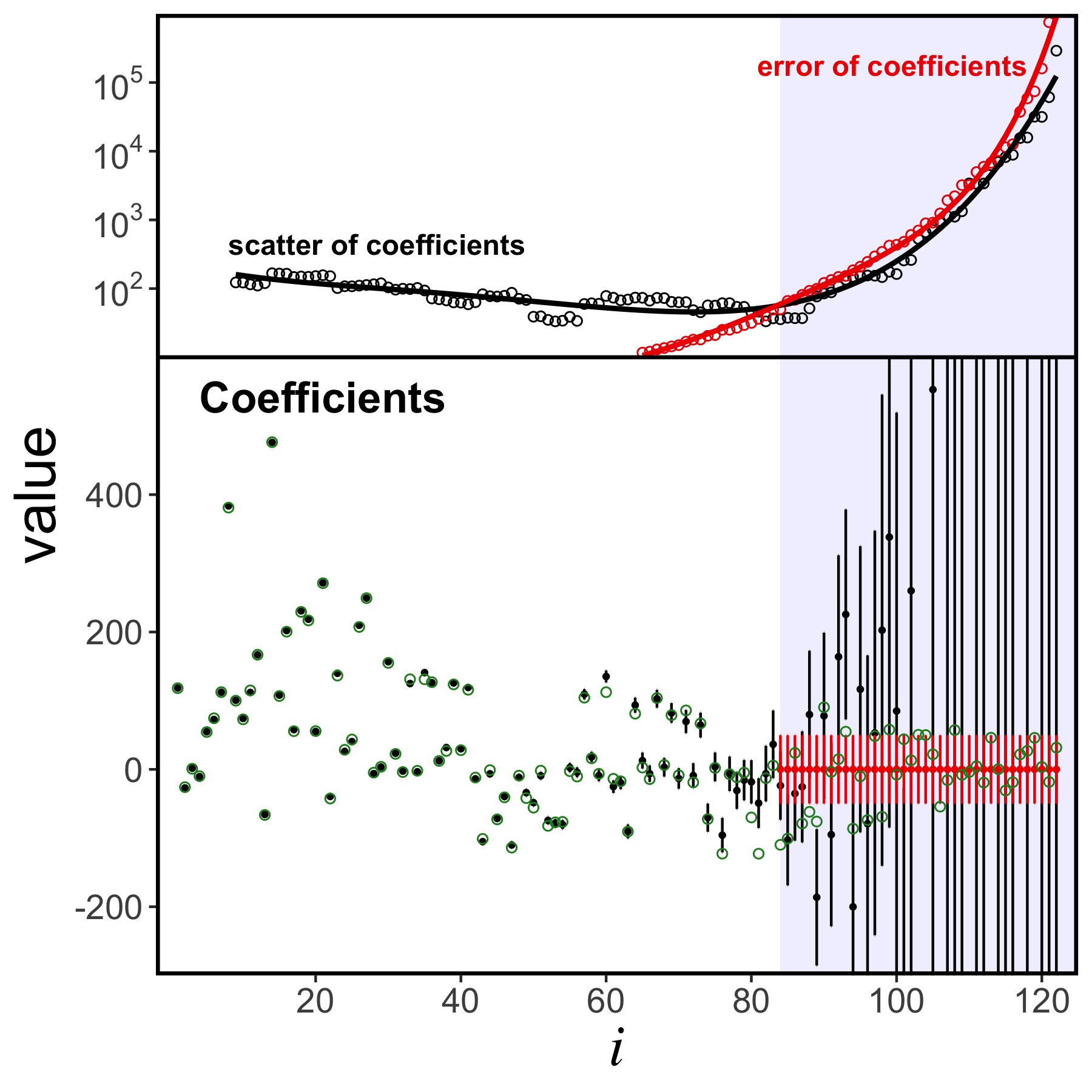}
   \caption{Example for the correction of poorly constrained coefficients for one star of the test set. Upper panel: Scatter of the coefficients in a window of eight coefficients (black symbols) and errors of the coefficients (red symbols). The black and red lines give polynomial approximations to the scatter and errors, respectively. Lower panel: Coefficients in the common XP basis, in the reference system diagonalising the covariance matrix. Black symbols: Result of the calibration. Red symbols: Corrected coefficients. Green symbols: True coefficients.}
              \label{fig:elimination}
    \end{figure}

Figure~\ref{fig:elimination} illustrates this process for one example star from the test set (with the effective temperature of 3750~K, $log(g)$ = 4, $[M/H] = -0.5$, and $E(B-V) = 0$) with $G = $ 16\m. In the top panel, the red symbols show the square root of the variance of the coefficients in the diagonal basis sorted in increasing order, together with the fit (red line). The black symbols give the standard deviation of the coefficients in a running window of  eight coefficients, together with a fit (black line). The scatter of the values of the coefficients, quantified by the standard deviation in the running window, strongly increases for high-order basis functions, and approaches the error of the coefficients, which we use as an indication for unconstrained coefficients. We set the coefficients to zero above the index of the basis function at which the fits to the two parameters first meet. However, the errors on these eliminated coefficients cannot be set to zero as this would result in a possible strong underestimation of the errors in further computations using the coefficients. In fact, the true coefficients are not zero, but show some variation. The errors attributed to the coefficients set to zero should therefore approximate the scatter of the true coefficients. Attributing the fitted error at the first coefficient that is set to zero to all higher-order coefficients provides a fairly good estimate for this variation of the true coefficients. We may refer to this process as high-order replacement for simplicity.\par
The result is shown in the bottom panel of Fig.~\ref{fig:elimination}. The black symbols show the coefficients with their errors. The excessive increase of the error for high-order basis functions is clearly visible. The red symbols show the coefficients set to zero, with their attributed errors. The green symbols give the true coefficients for comparison. For low-order basis functions, the calibration results in good agreement with the true coefficients, while the errors attributed to the high-order basis functions represent a reasonably good estimate for the actual scatter of the true coefficients.\par

\begin{figure}
   \centering
   \includegraphics[width=0.49\textwidth]{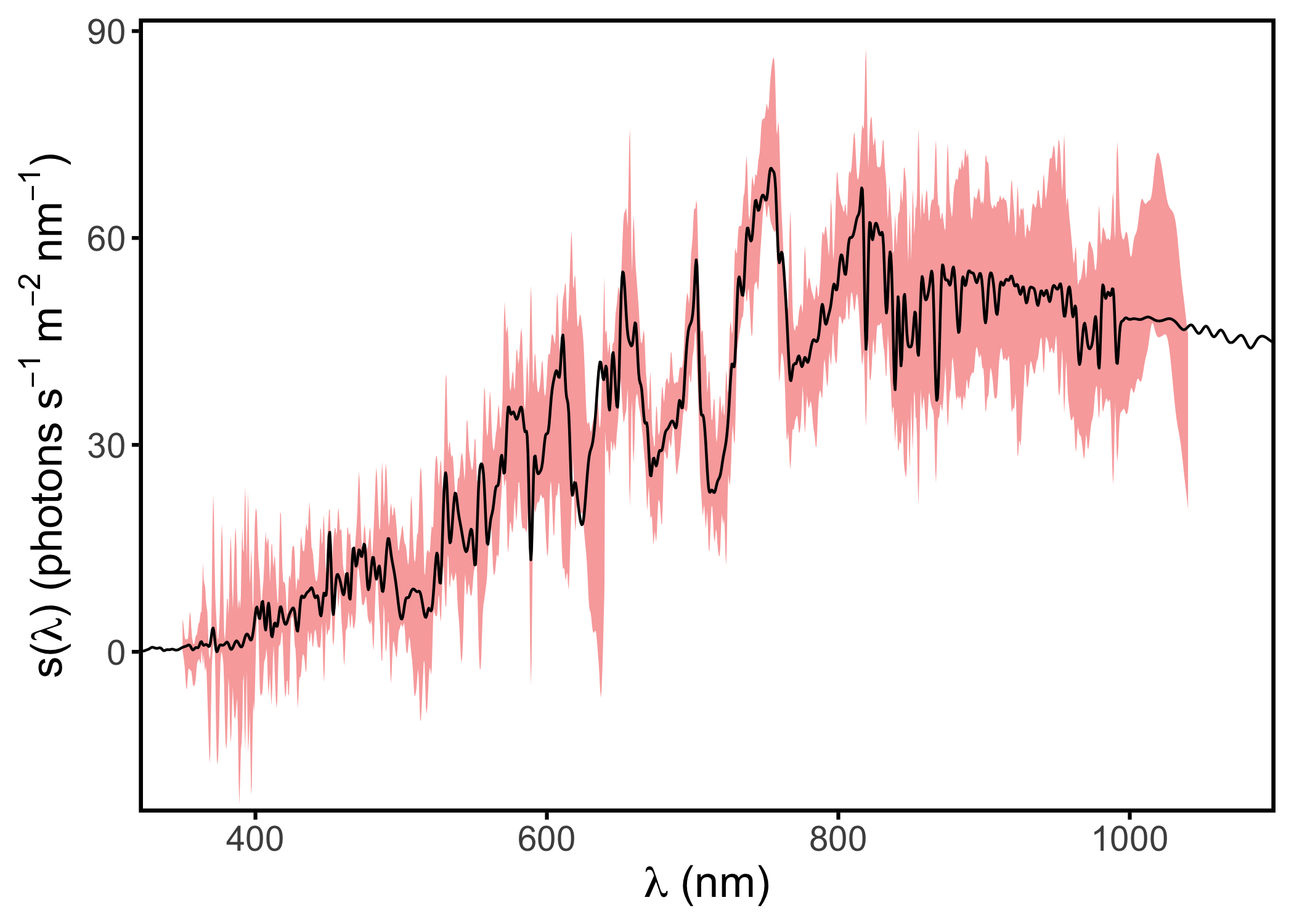}
   \caption{Spectral photon distribution for the test star from Fig.~\ref{fig:elimination} (black line), together with the 1-sigma error interval resulting from the coefficients with high-order replacement (red shaded region).}
              \label{fig:sampled}
    \end{figure}

The coefficients with  high-order coefficients replaced as described above can then be used to compute the SPD from a linear combination of the basis functions. This is illustrated for the same source in Fig.~\ref{fig:sampled}. The black line shows the true SPD of this source, while the red shaded region gives the 1-sigma uncertainty interval resulting from coefficients with high-order replacement.

\subsection{Calibration performance for a focused telescope \label{sec:performanceFocusedTelescope}}

In the following, we investigate the performance of the calibration of \gaia's spectrophotometry with the matrix approach in the case of a focused telescope, that is, in the absence of systematic errors due to an imperfect LSF model. We do this in three different steps. First, we give some examples of the calibration of individual stars to illustrate the peculiarities of the matrix approach. We then consider the statistical properties of the calibration of the overall set of test stars. Finally, we consider the case of QSOs with different redshifts as example cases for non-stellar SPDs.\par
In all investigations of the calibration performance, for each source we use the reference system in which the covariance matrix on the coefficients in the combined basis diagonalises. This way we can consider the coefficient vectors as vectors of uncorrelated random variables, with variances ${\rm diag}\left(\Sigma_{\Sigma}\right)$. We use all coefficients without high-order replacement.

\subsubsection{Performance for individual stars \label{sec:individualPerformance}}

In order to estimate the performance of the calibration for a given star, we compute the difference between the coefficients obtained from the calibration and the true coefficients. To make this difference interpretable, we divide it by the standard deviation (i.e. the square root of the variance) of the coefficients from the calibration. We do this in the combined basis and in the reference system in which the covariance matrix $\Sigma_{\bf c}$ diagonalises.\par
\begin{figure*}
   \centering
   \includegraphics[width=0.48\textwidth]{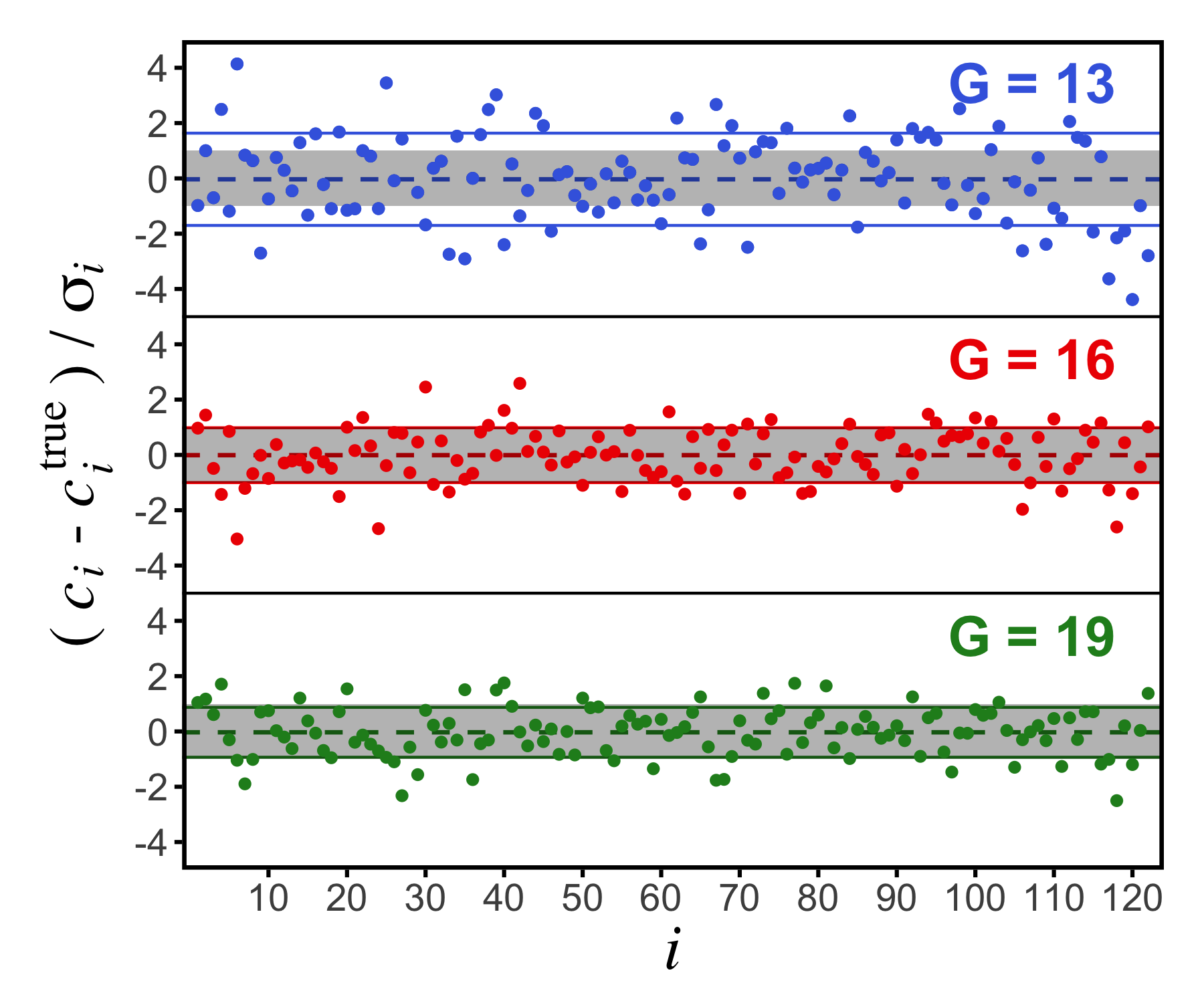}
   \includegraphics[width=0.48\textwidth]{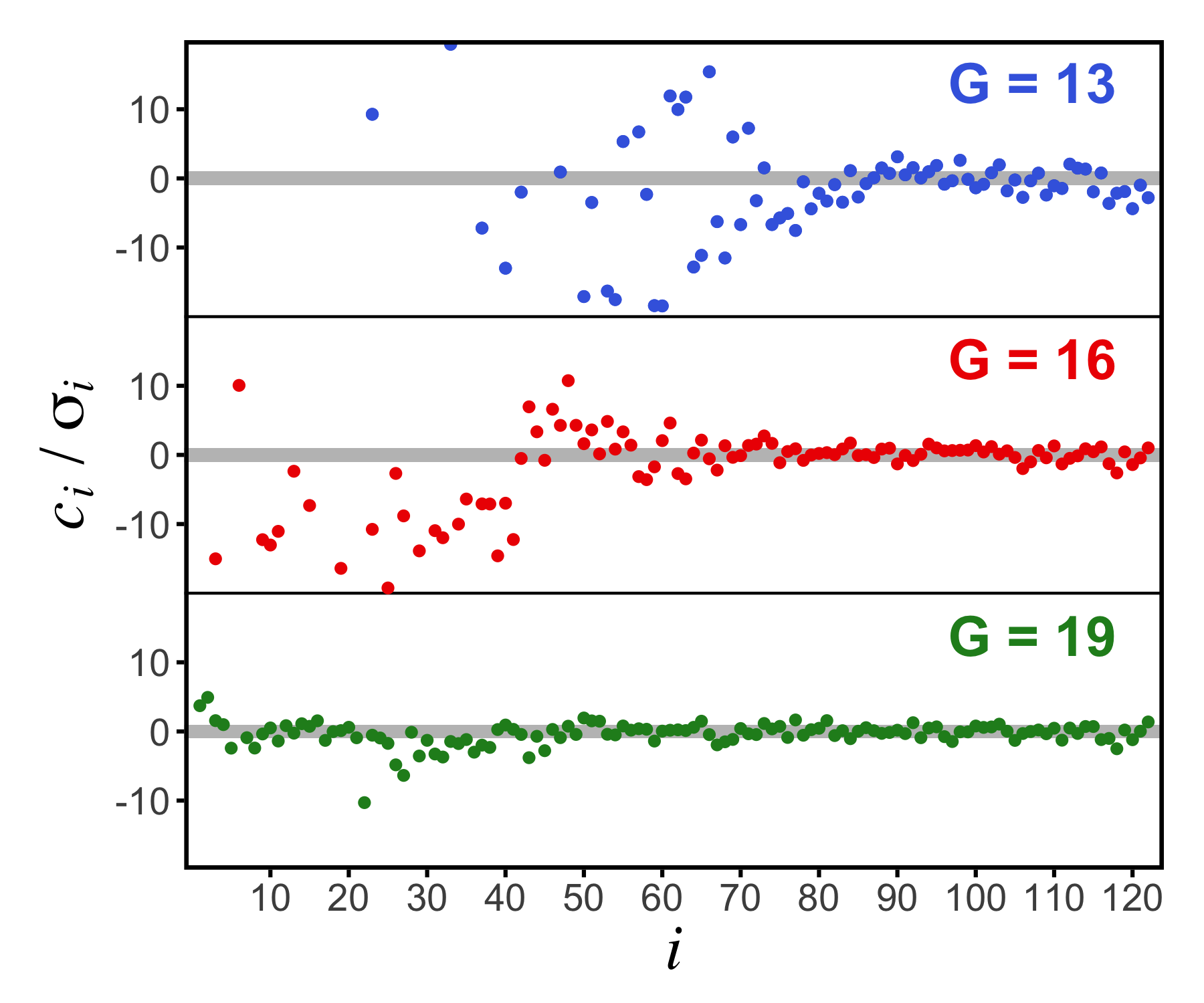}
   \caption{Example for the calibration results for one test SPD, and for three different $G$ band magnitudes, 13\m, 16\m, and 19\m. Left: Difference between the coefficients in the combined basis and true coefficients, normalised to the standard error. The grey shaded regions show the $\pm 1\sigma_i$ intervals, the dashed lines the sample mean, and the solid lines the sample standard deviation. Right:  Coefficients divided by the standard error. The grey shaded regions show the $\pm 1\sigma_i$ intervals.}
              \label{fig:performance1}
    \end{figure*}

The left-hand side of Fig.~\ref{fig:performance1} shows these normalised residuals, $(c_i - c_i^{true})/\sigma_i$, for one of the test sources  (with the parameters $T_{eff} = 7750$~K, $log(g) = 4$~dex, $[M/H] = -1$~dex, and $E(B-V)=0$), for three different magnitudes, $G =$ 13\m, 16\m, and 19\m, respectively. The result is obtained with the expansion of $V$ and $W$ using the combination of stellar model spectra and B-splines, which for this particular source is nevertheless irrelevant because its SPD is already contained in $V$, and is therefore not dependent on the expansion. For the bright source, the scatter of the normalised residuals is about 50\% larger than expected. This effect results mainly from systematic errors in the instrument matrix, which affect sources with signal-to-noise ratios that are  similar to or higher than those of the calibration sources (which are all of $G = $ 13\m). However, there is little indication for systematic deviations from the true coefficients, and so this uncertainty may be partially absorbed by artificially increasing the errors on the calibrated coefficients. For fainter sources of 16\m~ and 19\m, the scatter of the normalised residuals is in very good agreement with expectations, and the calibration can be considered successful.\par
The right-hand side of Fig.~\ref{fig:performance1} shows the coefficients for the same source obtained from the calibration, divided by the standard deviation. One can see that the coefficients for the lower order basis functions tend to have a higher signal-to-noise ratio than those for the higher order basis functions. With increasing magnitude, the signal-to-noise ratio decreases, with the coefficients for the higher order basis functions progressively approaching the noise level.\par

We also show the results for the forward calibration for the test star and $G=$ 16\m~in Fig.~\ref{fig:forwardCalibration}. In this case, we assume the SPD to be known, compute its projection onto the BP and RP basis functions, and use the instrument matrices to predict the BP and RP observational spectra. This approach is of little relevance in practice, as the use of the instrument kernel as derived in Sect.~{\ref{sec:GaiaInstrumentCalibration} provides a faster and more versatile approach to the forward calibration. However, it shows  that the instrument matrix provides an accurate prediction of the observational spectra as well, and that the complications caused by the ill-posedness of the calibration are absent in the forward direction.

   \begin{figure}
   \centering
   \includegraphics[width=0.48\textwidth]{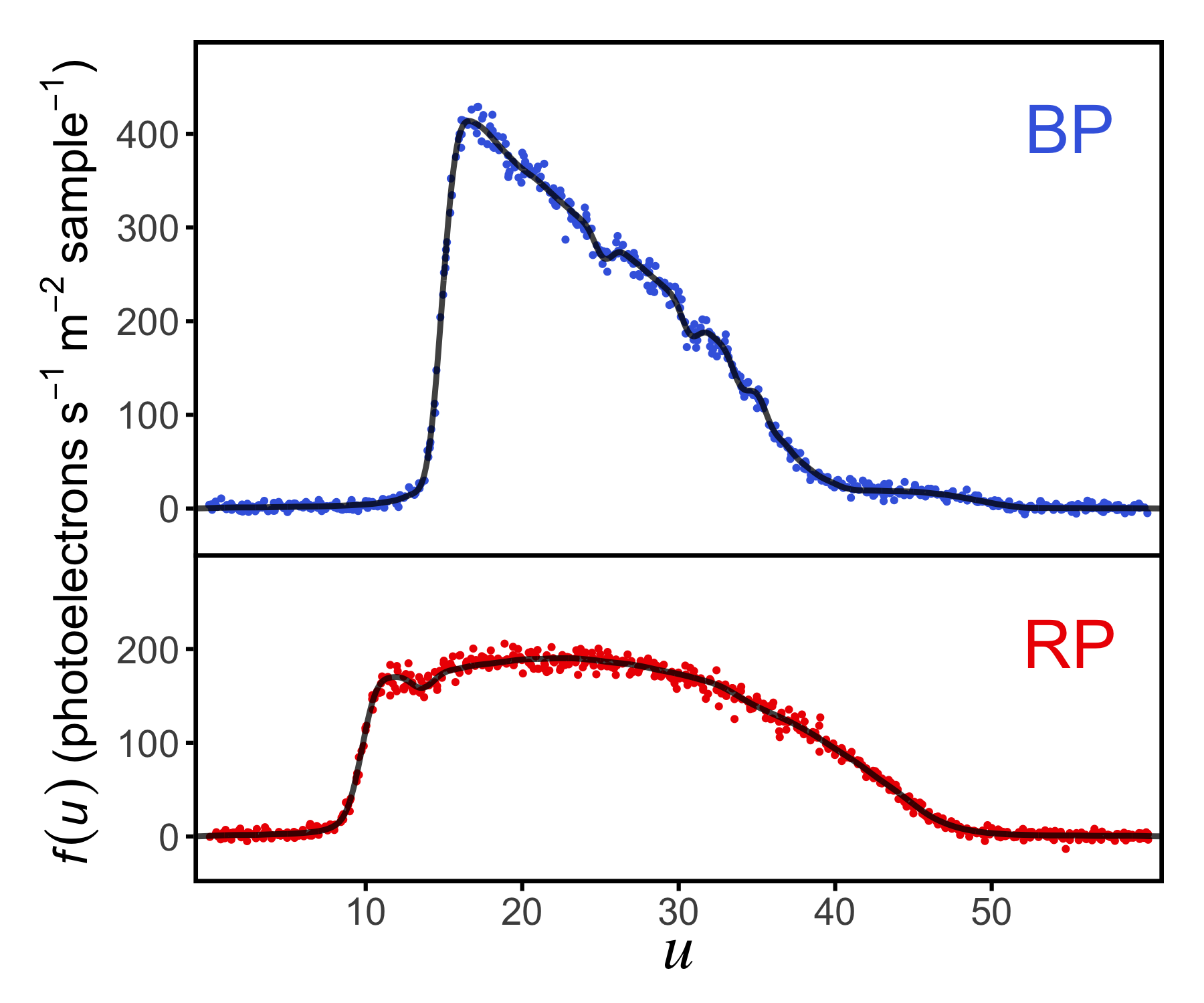}
   \caption{Result for the forward calibration of the same star as in Fig.~\ref{fig:performance1}, with $G=$ 16\m. Dots show simulated observations for BP (top panel) and RP (bottom panel). Lines show predicted observational spectra.}
              \label{fig:forwardCalibration}
    \end{figure}

\subsubsection{Performance for a set of stars \label{sec:performanceSet}}

The behaviour of the calibration illustrated in the previous section, with systematically overly low errors for bright sources with $G = $ 13\m~ and a good calibration of the coefficients within their errors is typical for most stars in the test set. In the following, we consider the performance of the calibration for the bulk of test stars, using the standard deviation of the normalised residuals as discussed in the previous section as an indicator of the goodness of the calibration of an individual star, and analyse the distribution of this parameter as a function of effective temperature, for the reddened and unreddened stars, for the three different magnitudes, and for the three different approaches for the expansion of $V$ and $W$. As a reference for the scatter of the standard deviation of the normalised residuals, we use the standard error (s.e.) of this quantity, which is given by
\begin{equation}
s.e. = \frac{1}{\sqrt{2(N - 1)}} \quad .
\end{equation}
The upper part of Fig.~\ref{fig:performance2} shows the standard deviation of the normalised residuals for all stars in the test set, in three columns for the $G$ band magnitudes 13\m, 16\m, and 19\m, respectively. The top two panels in each column show the results for the vector space expansion with stellar model spectra, the central two panels show the results for the expansion with B-spline basis functions, and the bottom two panels show the results for the expansion with the combination of stellar model spectra and B-spline basis functions. For each method of expansion, the results for the unreddened ($E(B-V)=0$) and reddened ($E(B-V) = 1$) case are shown separately.\par
Starting with the brightest sources of $G=$ 13\m, one can see the increase of the scatter of the normalised residuals shown for one individual source in Sect.~\ref{sec:individualPerformance} and for all sources and all methods of expanding $V$ and $W$. With the effective temperature of 7750~K, the example source was among the best cases at this magnitude, with decreasing performance for cooler and significantly hotter stars. Particularly for M-type sources, the performance of the calibration becomes rather poor.\par
We identify the two fundamental reasons for the systematic errors, as already discussed in Sect.~\ref{sec:uncertainties}: the error on the instrument matrix; and a breakdown of the fundamental assumption for the calibration that the observational spectra can be linked to the principal components of their corresponding SPDs. Using the true instrument matrix in the calibration, we can separate the two effects. For a schematic presentation, we do a polynomial fit to the standard deviations of the normalised residuals for the true instrument matrix to estimate the component produced by the breakdown. We then fit a polynomial to the standard deviations of the normalised residuals and subtract the previous fit to estimate the contribution from the instrument matrix. The contributions of the breakdown, the instrument matrix, and the overall systematic error are indicated as the green, the blue, and the red lines in the upper part of Fig.~\ref{fig:performance2} for $G = $ 13\m~ and the unreddened cases.\par
The contribution from the breakdown is very low for stars with effective temperatures larger than about 4000~K. Here, the effect of errors in the instrument matrix is fully dominant. For M-type stars, the contribution from the breakdown strongly increases with decreasing effective temperature, where it dominates at temperatures below about 3500~K. This increase of the breakdown contribution is caused by the increasingly complex shapes of the SPDs in the test set.\par
Towards higher magnitudes, and therefore lower signal-to-noise ratios, the systematic errors  quickly become covered by random noise. For the $G = $ 16\m~ case, the contribution of the instrument matrix is already entirely covered by random noise, and only the systematic effects due to the breakdown are visible at low effective temperatures. For $G = $ 19\m, all systematic errors are covered by random noise at all effective temperatures.\par
The low signal-to-noise ratio in the $G = $ 19\m~case also covers all differences between the three different approaches to expanding $V$ and $W$. Comparing the intermediate and bright sources, systematic effects can be spotted. The expansion with the stellar model spectra performs much better for unreddened stars than for reddened ones. This is caused by the lack of reddened SPDs in the model spectra. However, the B-spline expansion has no problem with reddened sources, as the effect of interstellar extinction is assumed to be a smooth variation with wavelength, and such smooth variations are easily represented with B-spline basis functions in wavelength. However, this approach shows slightly poorer performance than the expansion with stellar model spectra for unreddended sources, as it lacks typical stellar components in the basis functions. The best representation is therefore obtained with an expansion using a combination of stellar model spectra and B-spline basis functions. With this expansion, the calibration is very good for all effective temperatures for $G = $ 19\m, and also for all effective temperatures larger than about 4000~K at $G=$ 16\m, both for unreddened and reddened stars.\par
However, the poor standard deviations of the normalised residuals for bright sources do not imply an overly poor calibration after all.  As illustrated for the example source in Sect.~\ref{sec:individualPerformance}, the deviation of the coefficients from the calibration and the true coefficients show little systematic error and can be mostly compensated by allowing larger errors. This reflects the fact that for stars with signal-to-noise ratios close to or above the ones for the calibration sources, systematic errors prevent the full exploitation of the high signal-to-noise ratio. Furthermore, the systematic errors in the coefficients become particularly large for the coefficients of the higher-order basis functions. Restricting the standard deviation of the normalised residuals to the nine coefficients with the smallest absolute errors for each source, systematic errors are strongly reduced. This is illustrated in Fig.~\ref{fig:performance3}, which shows the same plots as Fig.~\ref{fig:performance2} but includes only the nine coefficients with the smallest absolute errors for each source. In this plot, the smaller number of coefficients per source taken into account results in a larger scatter. The standard error on the standard deviation of the normalised residuals is much larger in this case, as indicated by the shaded regions in Figs.~\ref{fig:performance2} and \ref{fig:performance3}. However, for the best case of an expansion with a combination of stellar model spectra and B-splines, all systematic effects are irrelevant even at $G =$ 16\m~ among the first nine coefficients and are strongly reduced for $G=$ 13\m. For the example source from Sect.~\ref{sec:individualPerformance}, the performance for the first nine coefficients is among the worst of all test sources. Therefore, the calibration with the matrix approach may still be useful for sources affected by systematic errors resulting from both the breakdown and the instrument matrix if the analysis of the calibration result is limited to the best constrained coefficients and/or larger errors on the coefficients are assumed.

   \begin{figure*}
   \centering
   \includegraphics[width=0.8\textwidth]{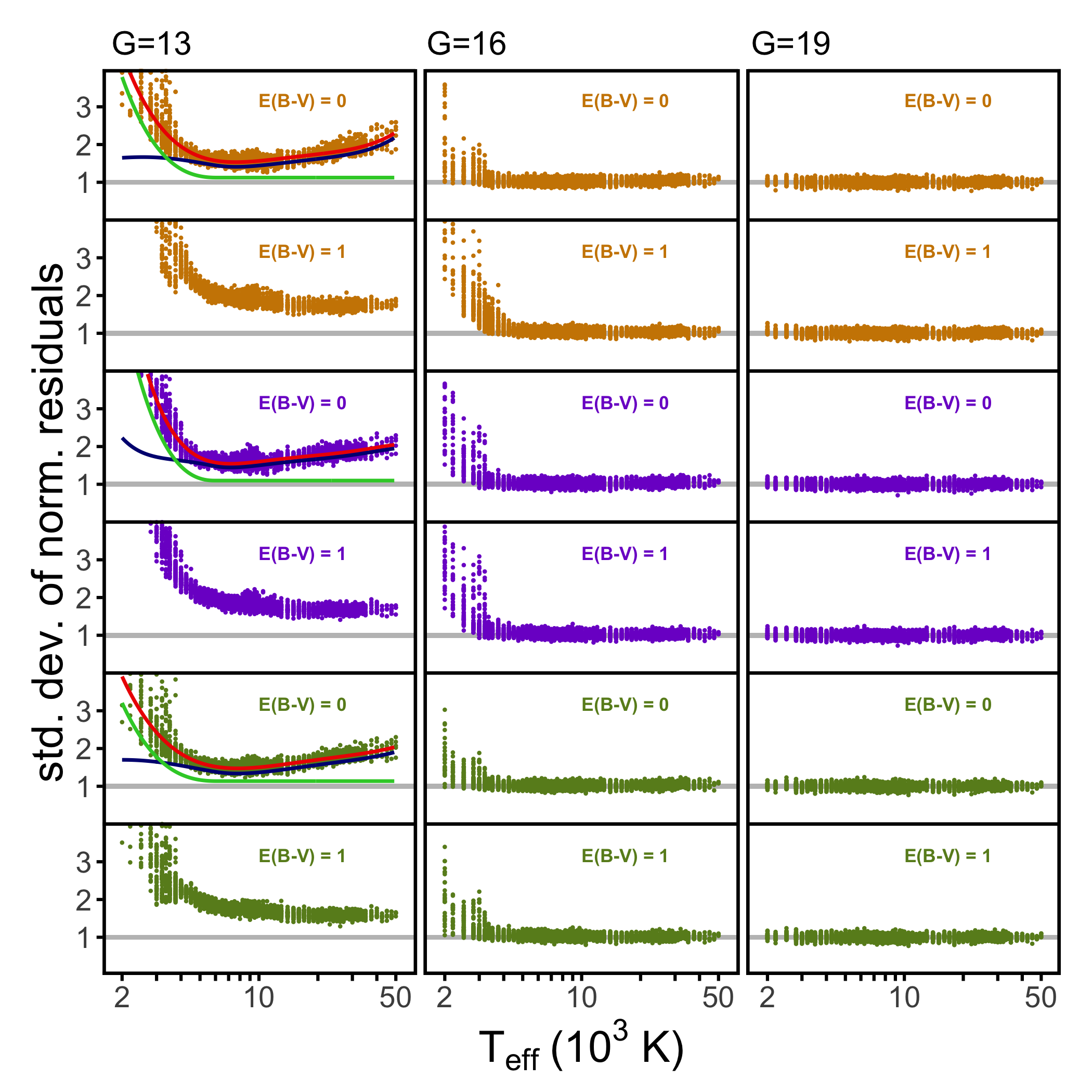}
    \includegraphics[width=0.8\textwidth]{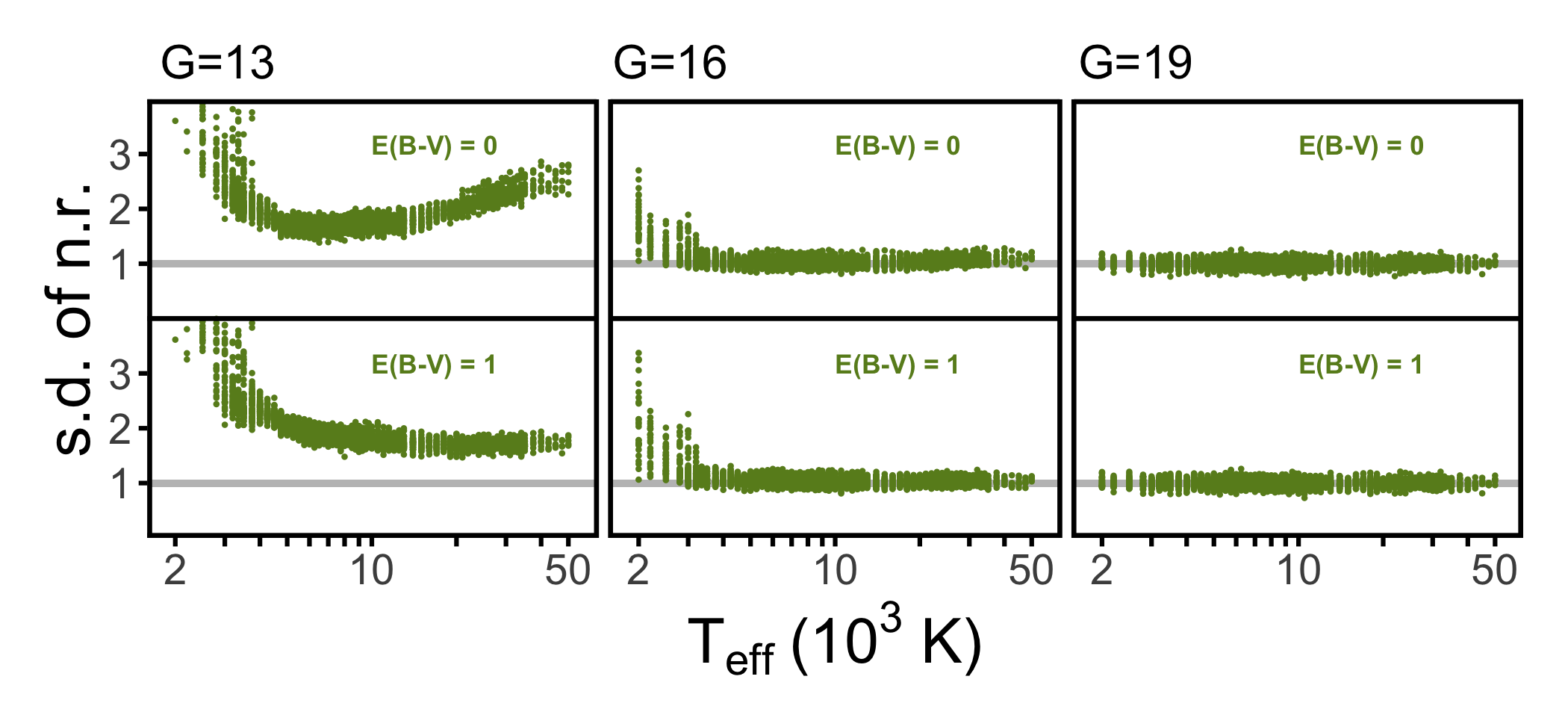}
   \caption{{\bf Upper panel:} Standard deviations of the normalised residuals for all stars in the test set. The three columns correspond to $G$ magnitudes of 13, 16, and 19. The top two panels show results for a basis expansion using stellar model spectra, the central two panels using B-splines, and the bottom two using a combination of stellar model spectra and B-splines. For each approach for the basis expansion, the results for unreddened stars ($E(B-V) = 0$) and reddened stars (using $E(B-V) = 1$) are shown.
   The grey shaded region indicates the standard error of the standard deviation. For the case of $G$ = 13\m~and $E(B-V) = 0$, lines indicate the approximate contributions of the  limitations by the breakdown (green lines) and the error in the instrument matrix (blue lines) to the total systematic error (red lines). {\bf Lower panel:} Same as the bottom two rows of the upper panel, but for the defocus test.}
              \label{fig:performance2}
    \end{figure*}
    
   \begin{figure*}
   \centering
   \includegraphics[width=0.8\textwidth]{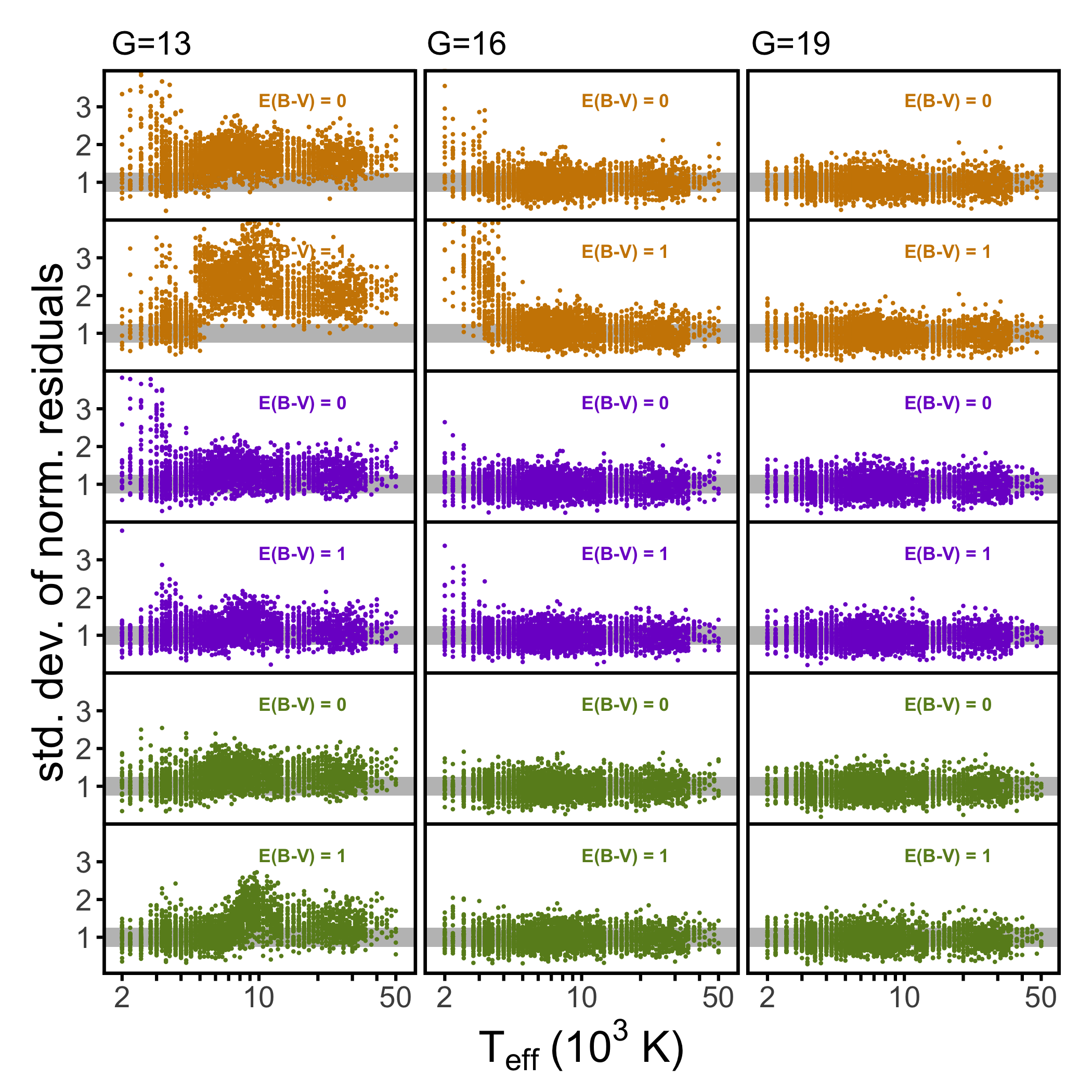}
   \caption{Same as Fig.~\ref{fig:performance2}, but taking only the nine best constrained coefficients into account.}
              \label{fig:performance3}
    \end{figure*}

\subsubsection{Performance for low-redshift QSOs \label{sec:QSOs}}

Up to now, tests for the performance of the calibration were carried out using stellar SPDs. As a test case for very non-stellar SPDs we consider QSOs. As a model spectrum we use the composite spectrum by \cite{Selsing2016}. This QSO spectrum covers rest wavelengths from about 100~nm to about 1135~nm. We are therefore able to apply redshifts $z$ from zero to two to the model spectrum while still covering the entire BP/RP wavelength range. For the calibration tests we generate 2001 QSO spectra with equally spaced redshifts between zero and two. We find that the limit in magnitude for which a good calibration can be obtained over the entire range of $z$ is at about $G=$ 17\m. The resulting standard deviations of the normalised residuals are shown in Fig.~\ref{fig:performance4}, again for the three cases of expanding $V$ with basis functions from stellar model spectra, B-spline basis functions, and a combination of both. For the case of an expansion with stellar model spectra, clear systematic errors can be seen, in particular for the largest and smallest redshifts in the range covered. This reflects the non-stellar SPDs of the QSOs, which are not well represented by basis functions solely based on stellar SPDs. For an expansion with B-splines, the systematic errors are strongly reduced. The combination of stellar model spectra and B-splines results in only a very slight improvement as compared to the B-splines alone. For the expansion of $V$ with the combination of stellar model spectra and B-splines, systematic errors are only tentatively visible at $G=$ 17\m~for redshifts close to zero and two. For QSOs brighter than 17\m, the systematic errors become more and more prominent. The systematic errors in the calibration of the QSO spectra result from the breakdown, occurring when the basis functions that span $V^\ast_{\rm BP/RP}$ cannot represent certain narrow spectral features. We discuss this issue in more detail and in a more general context in Sect.~\ref{sec:narrowFeatures}.\par
The matrix approach with an expansion of $V$ based on a combination of stellar model spectra and B-spline basis functions is therefore in principle capable of providing a good calibration also for non-stellar SPDs for  objects below a certain brightness. For QSOs, which are mostly fainter than 17\m, a good calibration can be achieved.

   \begin{figure}
   \centering
   \includegraphics[width=0.49\textwidth]{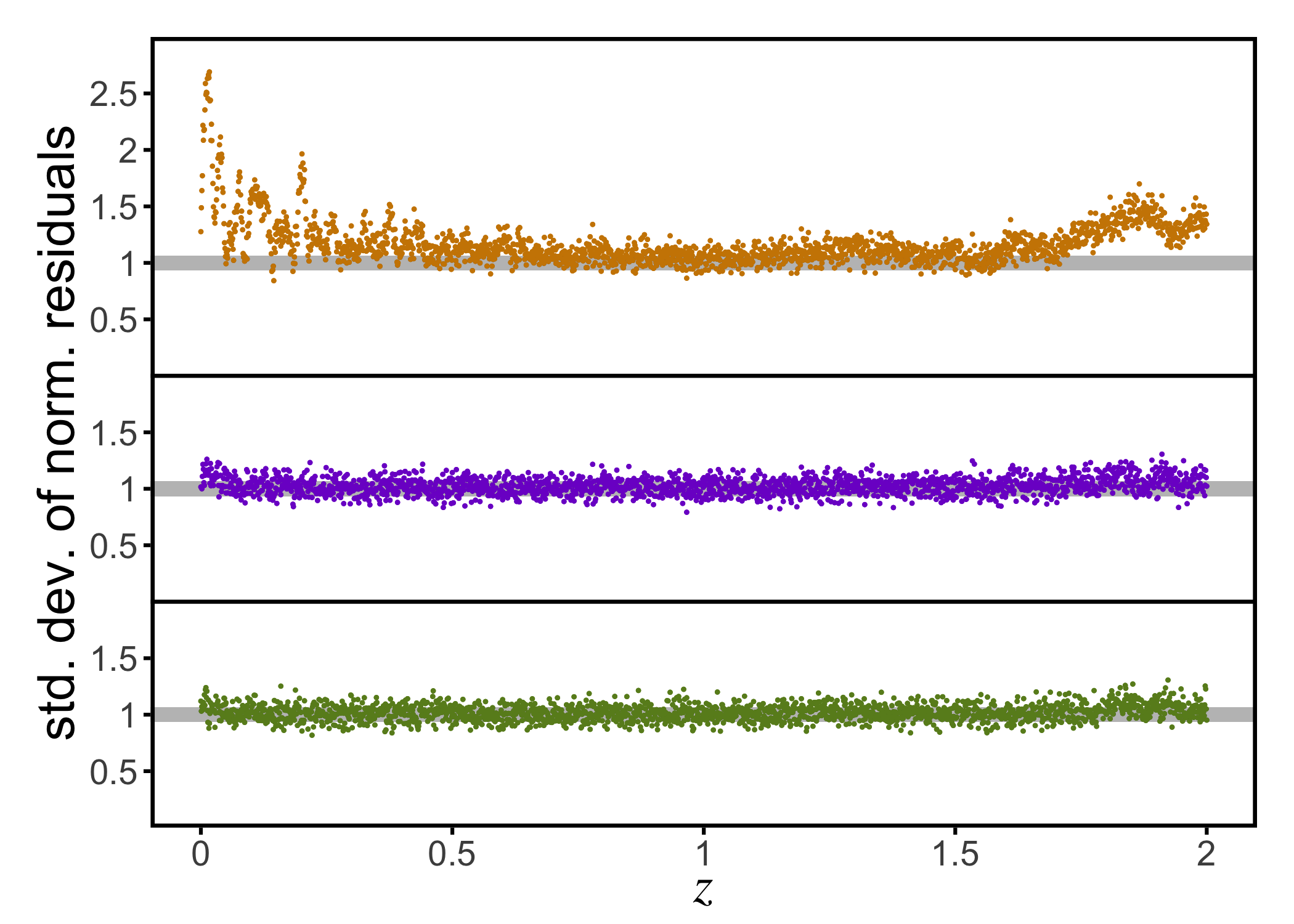}
    \includegraphics[width=0.49\textwidth]{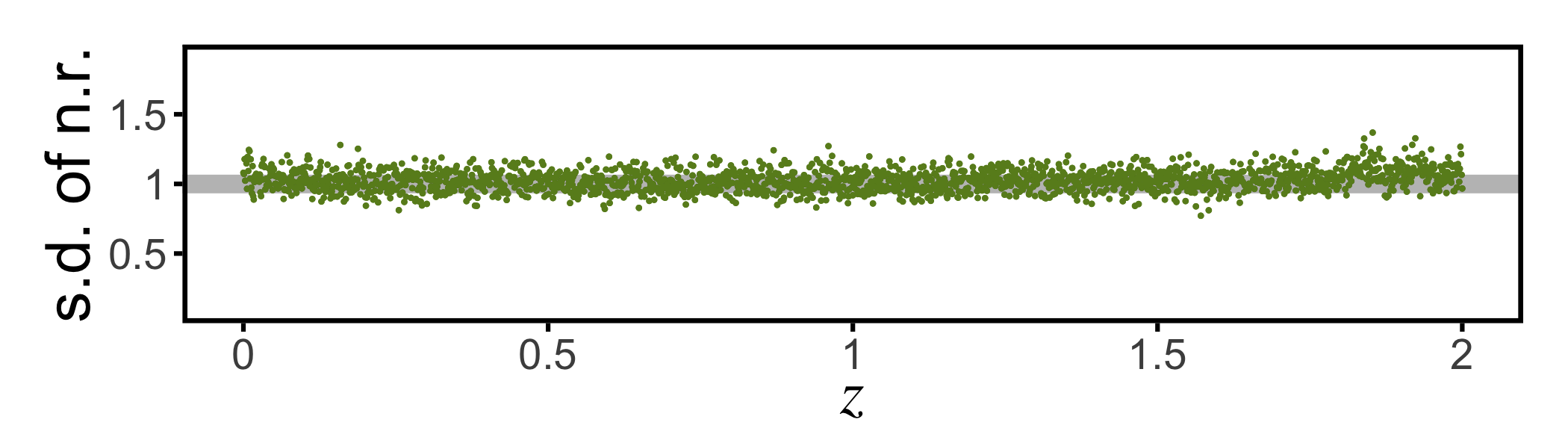}
   \caption{{\bf Upper part:} Standard deviation of the normalised residuals for a model QSO spectrum with $G$ = 17\m~and redshifts $z$ between 0 and 2. Top panel: Basis extension with stellar model spectra. Middle panel: Basis extension with B-splines. Bottom panel: Basis extension with both model spectra and B-splines. The grey shaded region indicates the standard error of the standard deviation. {\bf Lower part:} Same as the bottom panel of the upper part, but for the defocus test.}
              \label{fig:performance4}
    \end{figure}

\subsection{Calibration performance for a de-focused telescope \label{sec:performanceDefocusedTelescope}}

In the following we investigate the performance of the calibration of \gaia's spectrophotometry with the matrix approach in the test case of a defocused telescope. This test case, in which we try to absorb a convolution of the LSF with a rectangular function with a convolution with a Gaussian, is a test case for the presence of systematic errors introduced by an imperfect LSF model. The LSFs derived in the calibration process of BP, using Eq.~(\ref{eq:OTF2}) and adjusting $\sigma_u(u)$ with a power dependency in $\lambda$, are shown as solid lines in Fig.~\ref{fig:defocusedLSFs} for three different wavelengths, in comparison to the true LSFs. The differences between the true LSFs and the model LSFs are largest at short wavelengths, as here the LSFs are most affected by the defocusing. At the long-wavelength cut-off of BP, the true and the modelled LSFs almost agree.\par
In the performance test, we consider the same test objects as before, that is, unreddened stars and stars with a colour excess of $E(B-V) =$~1\m, and the model QSO with redshifts between zero and two. For brevity, we only present the results for the expansion of the $V$ and $W$ by a combination of stellar model and B-spline basis functions, which has been the best case. For the stars, the results of the defocusing test are shown in the lower part of Fig.~\ref{fig:performance2}, and can be directly compared with the previous test case. For the QSOs, the results are shown in the lower part of Fig.~\ref{fig:performance4}.\par
For the stars at $G = $ 19\m, the calibration result in the defocused test situation is basically the same as when using the correct LSF. At $G=$ 16\m, a slight decrease of performance for very hot stars is visible for the unreddened case. For the reddened stars, no difference occurs. For the brightest test stars, that is those with $G = $ 13\m, the performance decreases more strongly, again in particular for hot stars and more for unreddened than for reddened stars. For the test QSOs, the performance in the defocus scenario is almost the same as in the case of using the exact LSFs in the calibration tests, except for a tiny decrease of performance for redshifts between about 1.8 and 2.\par
The differences visible between the two test scenarios using the exact LSF and the defocused telescope with an inaccurate LSF model result from the fact that the differences between the modelled and the true LSFs in the latter case are strongest for short wavelengths. At longer wavelengths, the intrinsic difference between a focused and a defocused LSF are smaller, and the adjustment of $\sigma_u(\lambda)$ in the calibration model can almost entirely absorb the difference. Therefore, stars with large fluxes at short wavelengths, that is, hot stars, are more affected by the systematic errors in the calibration model. However, even with the relatively large differences between the true and the modelled LSFs at short wavelengths, as can be seen in Fig.~\ref{fig:defocusedLSFs}, a good overall calibration can be obtained. Moreover, the defocusing test case with a defocusing by 0.3~mm represents about the maximum deviation one may allow without further improving the LSF model, and therefore stronger decreases in the calibration performance than visible in the lower parts of Figs.~\ref{fig:performance2} and \ref{fig:performance4} may not pass unnoticed and uncorrected. We therefore conclude that the matrix approach is not particularly sensitive to the modelling of the LSF. A very accurate reproduction of the true LSF may not be possible in practice, but is also not required for obtaining a reliable calibration.

\subsection{Narrow spectral features \label{sec:narrowFeatures}}

   \begin{figure*}[h!]
   \centering
   \includegraphics[width=0.49\textwidth]{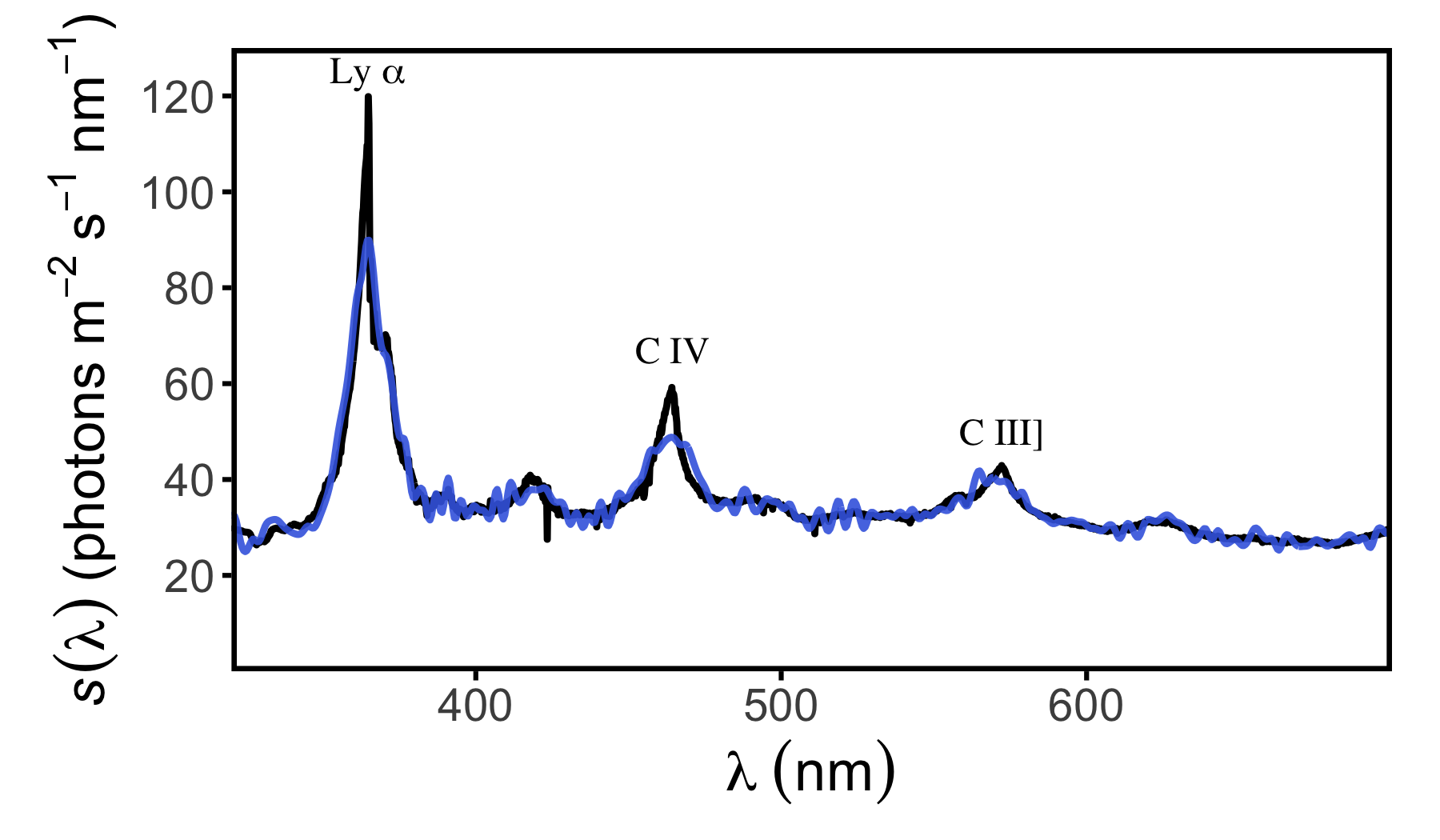}
   \includegraphics[width=0.49\textwidth]{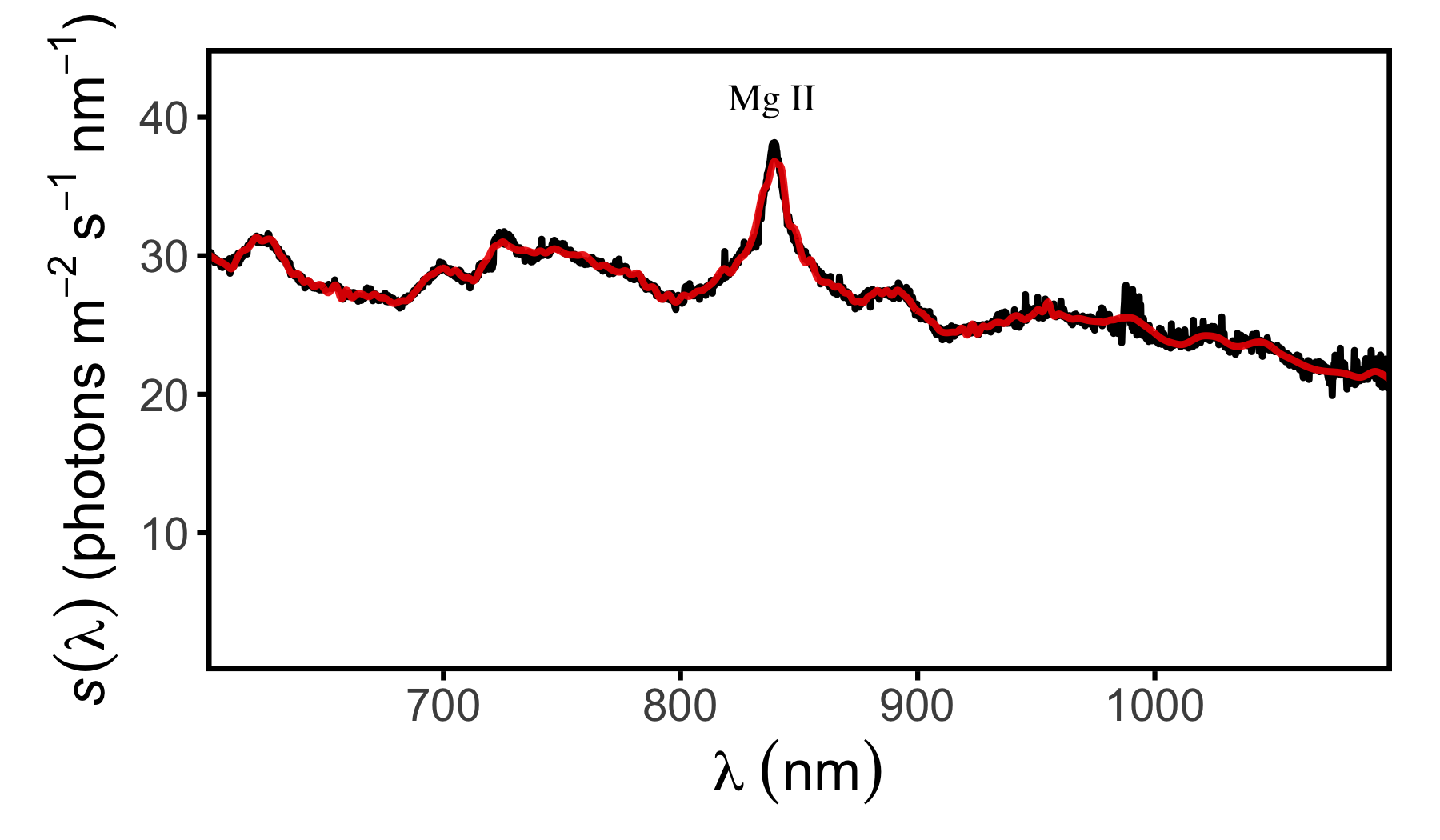}
   \includegraphics[width=0.49\textwidth]{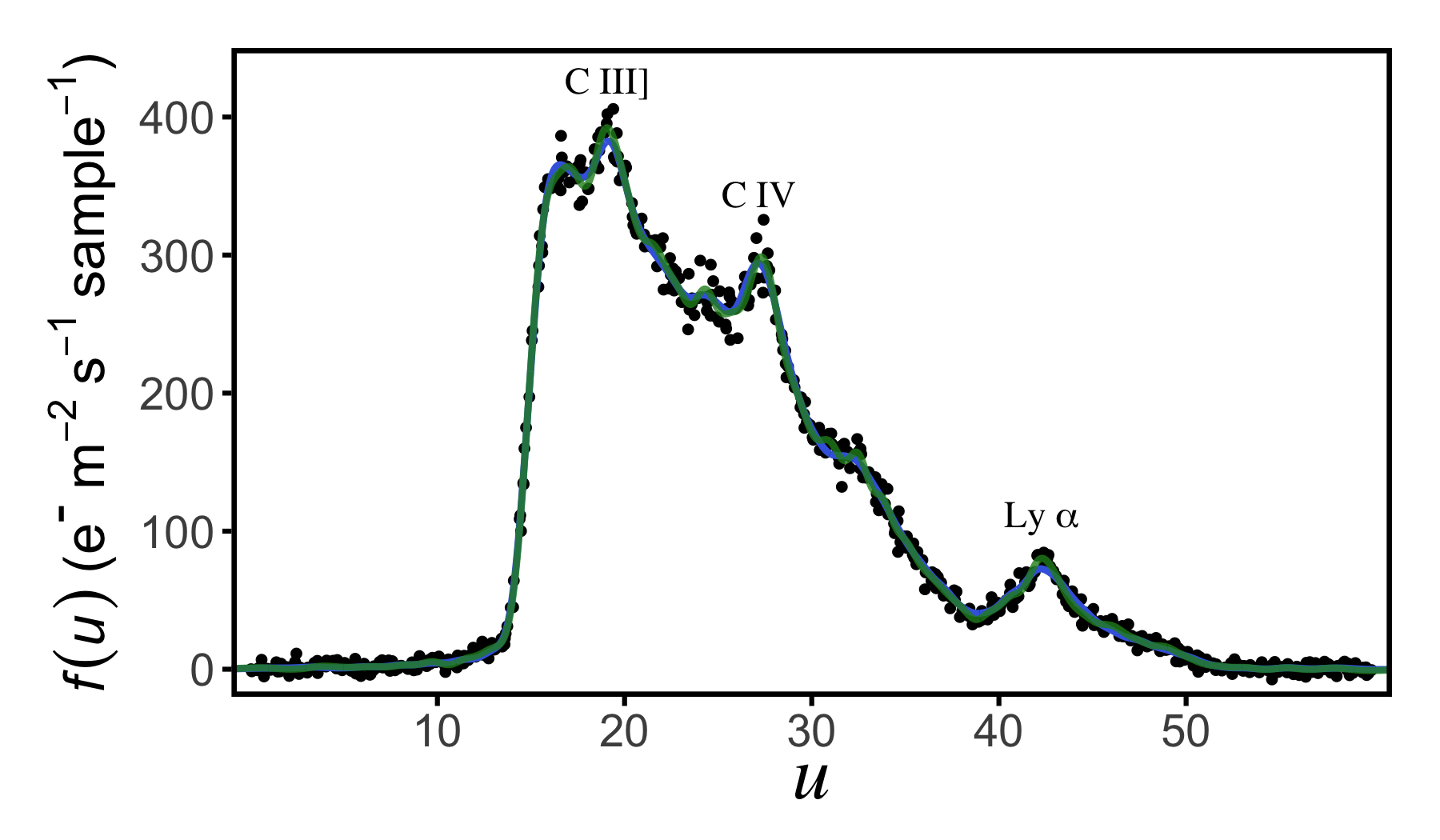}
   \includegraphics[width=0.49\textwidth]{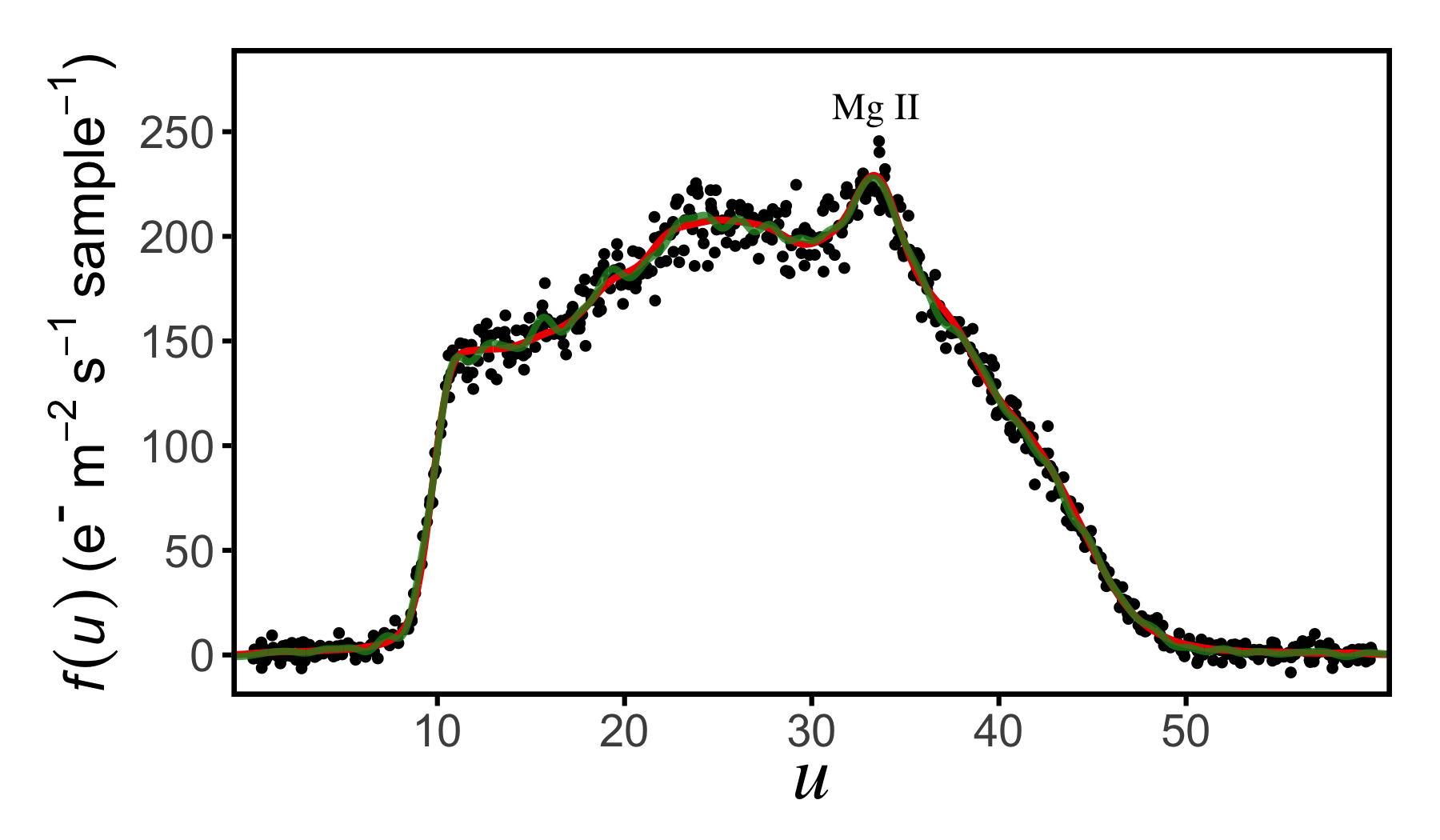}
   \includegraphics[width=0.49\textwidth]{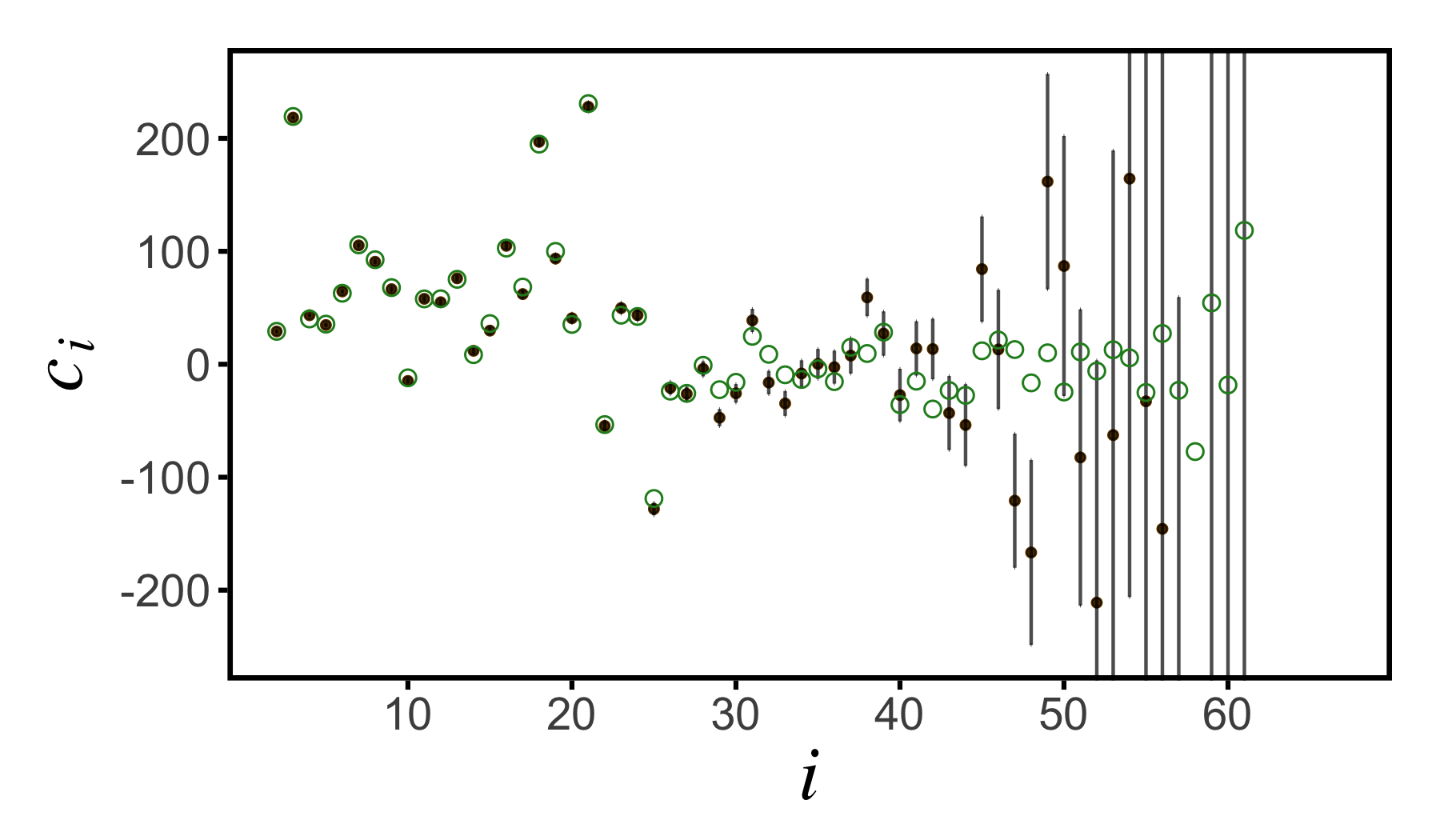}
   \includegraphics[width=0.49\textwidth]{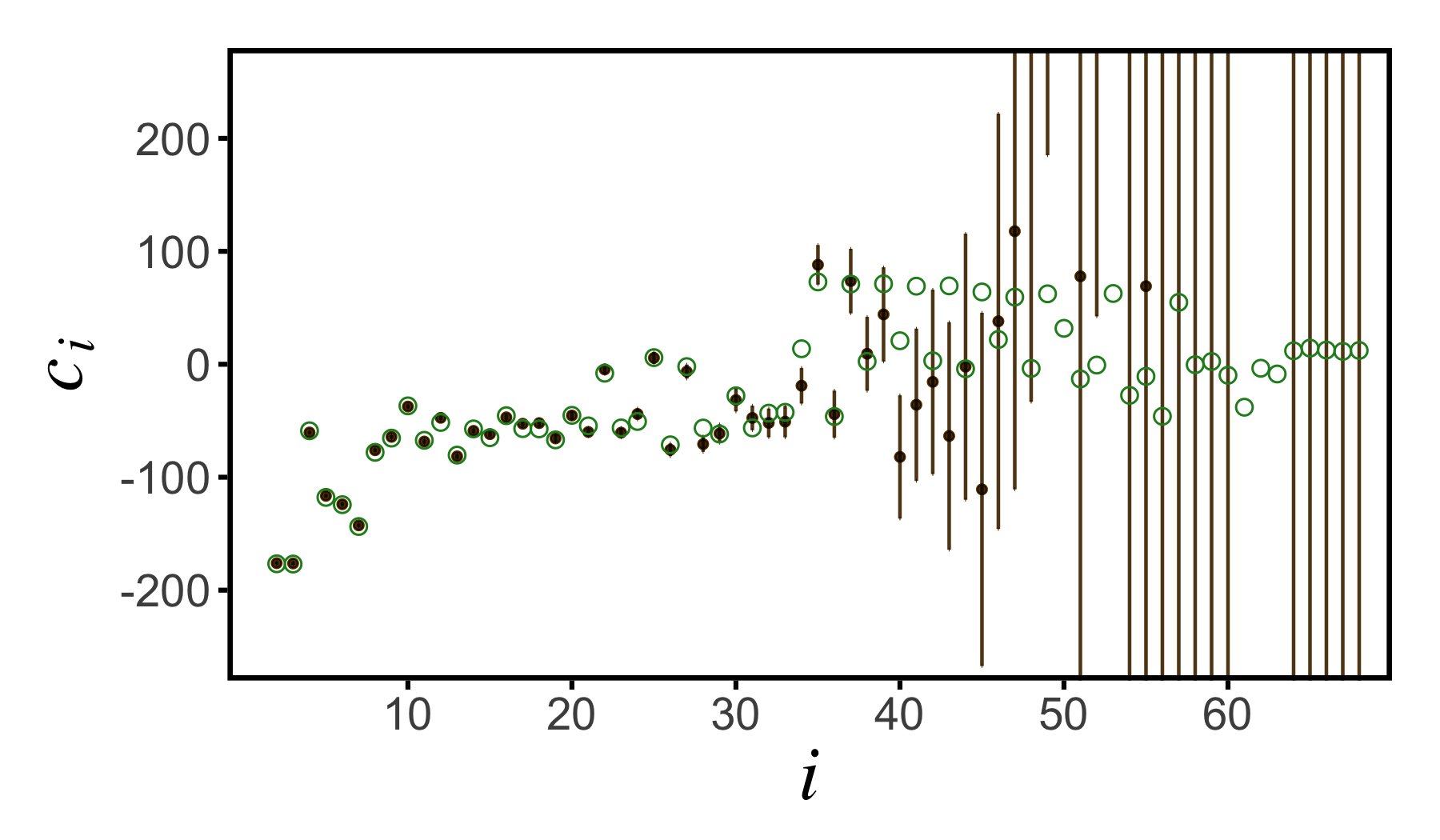}
   \includegraphics[width=0.49\textwidth]{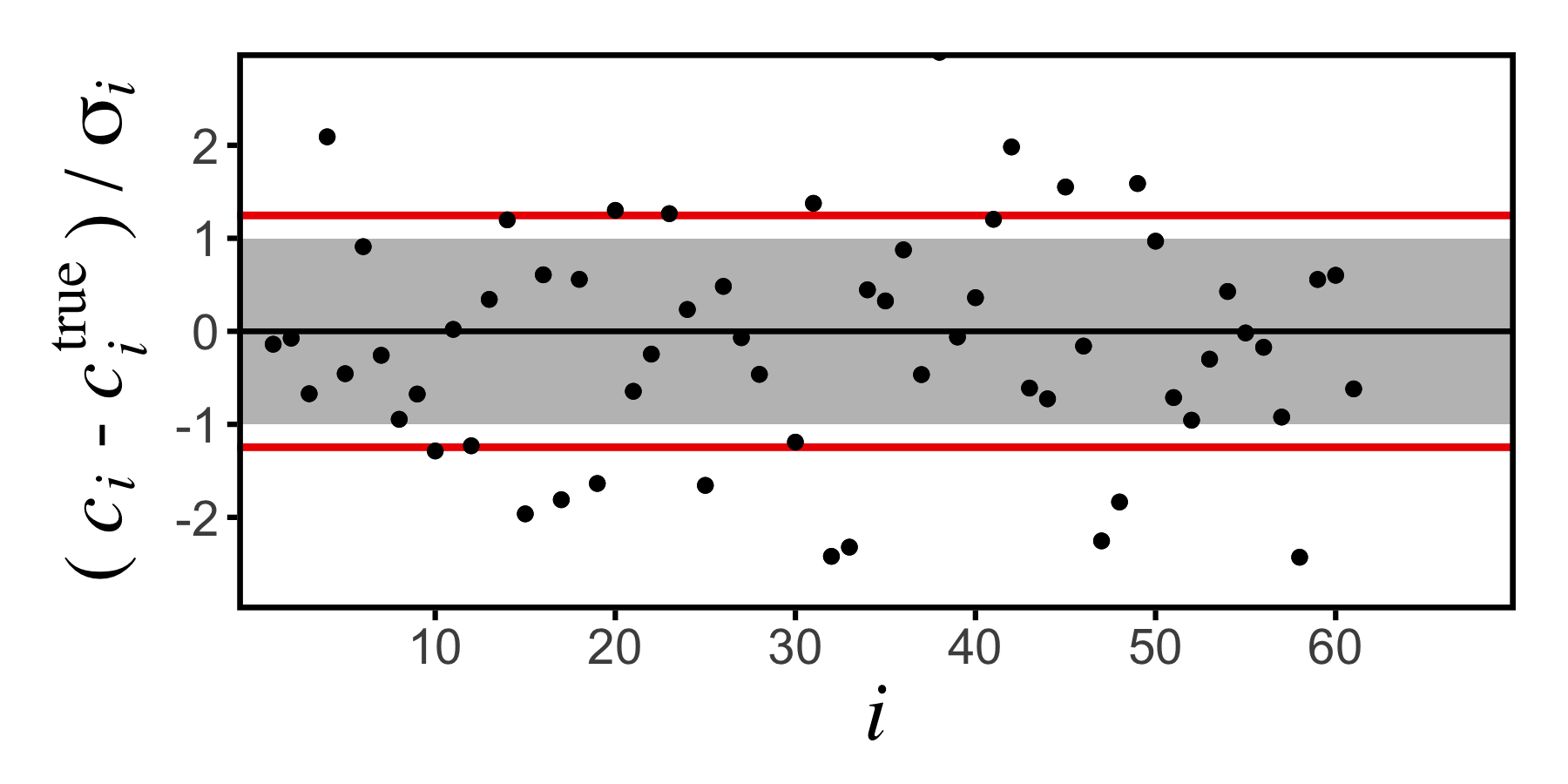}
   \includegraphics[width=0.49\textwidth]{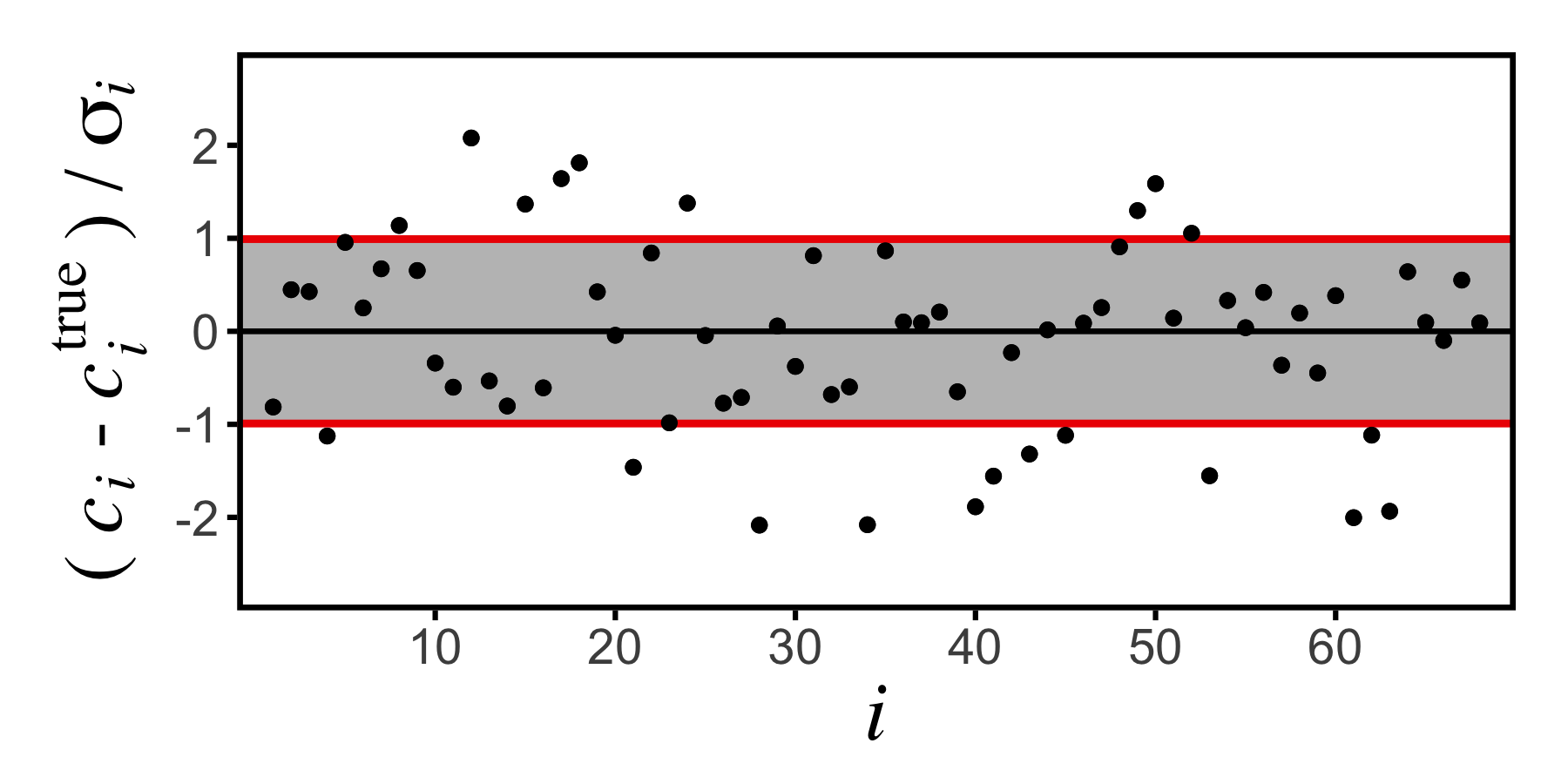}
   \caption{Spectral photon distributions and calibration results for a QSO with $G =$ 16\m~ and $z = 2$ for BP (left column) and RP (right column). Top row: SPDs (black lines) and their projections onto the bases for $V^\ast_{\rm BP}$ (blue line) and $V^\ast_{\rm RP}$ (red line). Second row: Simulated BP and RP observations (black dots), the result for the observational spectra from the calibration (green lines), and the results for the forward calibration (blue and red lines). Third row: Coefficients resulting from the calibration (black symbols) and true coefficients (green symbols). Bottom row: Normalised residuals of the coefficients. The red lines give the sample standard deviation.}
              \label{fig:QSOTest}
    \end{figure*}

We conclude the example calibrations by considering the case of narrow spectral features in the SPDs. The term 'narrow' refers to both isolated emission lines with a small intrinsic width and rather 'cusp-like' features that are present in the QSO model spectra considered above. The common aspect of these narrow features is that they cannot be well represented by the B-spline basis functions used to expand $V$, resulting in breakdown systematic errors. To illustrate these effects, we first consider the calibration of one QSO with a redshift of $z = 2$ (the worst case in the tests presented in Sect.~\ref{sec:QSOs}) and $G = $~16\m. As the effects are prominent in BP but not in RP, we consider the calibrations for the two instruments separately.\par
The uppermost panel of Fig.~\ref{fig:QSOTest} shows the model SPDs over the BP and RP wavelength ranges, together with the projections of the SPDs onto the bases for $V^\ast_{\rm BP}$ and $V^\ast_{\rm RP}$, respectively. For BP, the peaks of the emission lines cannot be well represented in $V^\ast_{\rm BP}$. For RP, the situation is better, as the emission line by Mg II present in the RP wavelength range is broader than the emission lines in the BP wavelength range, and the B-spline basis functions used in the construction of $V^\ast_{\rm RP}$ are a slightly narrower than in BP. The SPD of the QSO in RP is therefore relatively well represented in $V^\ast_{\rm RP}$, while this is not the case in BP, where the peaks of the three emission lines are 'missing' in $V^\ast_{\rm BP}$.\par
The other panels in Fig.~\ref{fig:QSOTest} show the simulated BP and RP spectra, in which the emission lines are clearly visible. In this approach, only the projection of the SPD is used to generate the observational spectrum. The poor representation of the emission lines in BP is therefore transferred to the observation spectrum (in addition to the higher computational cost, this is one reason to prefer the use of the instrument kernel $I(u,\lambda)$ over the matrix approach for the forward calibration). For the Ly~$\alpha$ line, although very poorly represented in $V^\ast_{\rm BP}$, the low response of the BP instrument at short wavelengths results in a weak, noisy feature in the observational spectrum. However, for the other two lines in BP, the poor representation is clear, with the C IV line being predicted to be overly faint in the observational spectrum, and the C III] line slightly too bright. In RP, the Mg II line, which is fairly well represented in $V^\ast_{\rm RP}$, is well predicted in the observational spectrum in the forward calibration.\par
The green lines in the second row of Fig.~\ref{fig:QSOTest} show the solution for $f(u)$ in BP and RP, respectively, in the inverse calibration. For RP, the difference is not significant. For BP however, a better representation of the C III] and the C IV emission lines is obtained. The basis of $W^\ast_{\rm BP}$ is thus able to accurately represent the observational BP spectrum. However, it is not possible to link the representation to the true basis functions for the SPD, as they are not included in $V^\ast_{\rm BP}$. As a consequence, the representation of the observational spectrum in $W^\ast_{\rm BP}$ has to be attributed to incorrect basis functions in $V^\ast_{\rm BP}$, which results in an error referred to as the breakdown systematic error.\par
The breakdown error manifests in the coefficients $\bf c$ obtained from solving Eq.~(\ref{eq:fullSolution}). The coefficients are shown for BP and RP in the third row of Fig.~\ref{fig:QSOTest} (black symbols), together with the true coefficients (green symbols). The reference system is again chosen such that the covariance matrices diagonalise. The same general behaviour as for the example star in Fig.~\ref{fig:elimination} in the combined basis can be seen in the BP and RP bases here as well. The effect of the breakdown is relatively weak in this example, and the poorer calibration for BP than for RP becomes visible when regarding the normalised residuals of the coefficients. These normalised residuals are shown in the bottom row of Fig.~\ref{fig:QSOTest}. The scatter of the BP coefficients is slightly larger than for the RP coefficients. The standard deviations of the normalised residuals are shows as red lines. For RP, the standard deviation is 1.001, in close agreement with the theoretical value of $1 \pm 0.086$. For BP, the standard deviation is 1.201, and is thus larger than the theoretical value of $1 \pm 0.091$. Repeating the calibration of the QSO spectrum with different random noise realisations confirms that the increase of the standard deviation is indeed real.\par
The effect of the breakdown is rather small in this case. The increase of the scatter in the normalised residuals still results in a good representation of the observed spectrum (second row of Fig.~\ref{fig:QSOTest}) and the corresponding coefficients (third row of Fig.~\ref{fig:QSOTest}), without systematic deviations of particular coefficients. However, it may not be possible to detect such errors. The representation of the observational BP spectrum in $W^\ast_{\rm BP}$ is perfectly adequate (green line in the second row of Fig.~\ref{fig:QSOTest}), and therefore no indications of problems in the calibration occur. This means that even in cases where an excellent reproduction of an observational spectrum is obtained, caution is in order when interpreting the calibration result, in particular if narrow spectral features may be involved.\par
A similar situation to that for the QSOs may be encountered with emission line stars. In this case, the emission lines can be relatively narrow, and also in this case the basis functions spanning $V^\ast$ may not be able to accurately represent the emission line. To illustrate this effect, we construct a test spectrum that mimics a Be star. We use the SPD of a B-type main sequence star (with an effective temperature of 20.000~K, $log(g)=4$, $[M/H] = 0$, and $E(B-V)=0$), and add a strong, narrow H$\alpha$ line with a Gaussian profile with a  full width at half maximum of 1~nm. The resulting RP calibration is shown in Fig.~\ref{fig:ELS} for $G =$ 16\m.\par
The top panel of Fig.~\ref{fig:ELS} shows the model SPD (black line), together with its projection onto $V^\ast_{\rm RP}$ (red line). As was the case for the QSO emission lines, the H$\alpha$ line cannot be accurately represented in $V^\ast_{\rm RP}$, as the B-spline basis functions do not provide the necessary resolution. The projection of the line is therefore broader and flatter than the true line. The forward calibration result nevertheless shows that the imperfect representation of the H$\alpha$ line in $V^\ast_{\rm RP}$ is well below the sensitivity limit introduced by the noise. The RP simulations are shown in the middle panel of Fig.~\ref{fig:ELS} (black dots) together with the forward calibration result (red line). The bottom panel of Fig.~\ref{fig:ELS} shows the coefficients for RP as black dots, as compared to the true coefficients (green symbols). Close agreement is obtained in the calibrations. The orange dots in the bottom panel of Fig.~\ref{fig:ELS} show the coefficients obtained for a calibration without introducing the H$\alpha$ line. These coefficients are clearly different from the ones with the H$\alpha$ line when considering the errors, illustrating the detection of the emission line.\par
A good determination of the coefficients $\bf c$ may therefore be possible even in the presence of narrow spectral features. However, the analysis of the emission lines may be affected by the fact that although the projection of the SPD on the vector space $V^\ast$ is reliable, this projection may not provide a good representation of narrow spectral features. For emission line stars, an alternative approach may be feasible. This approach requires that the emission line be narrow enough to be approximated by a delta distribution $\delta(\lambda - \lambda_l)$, with the central wavelength of the emission line, $\lambda_l$, being known. If the SPD is, apart from the emission line, a stellar one, one may fit the observational spectrum of an emission line star with a linear combination of observational spectra of stars, excluding the region of the observational spectrum affected by the emission line, and subtract this fit to separate the emission line from the underlying stellar spectrum. The underlying stellar spectrum can then be calibrated using the method  outlined in this work. The emission line is then introduced as a delta distribution centred on $\lambda_l$ with a scaling corresponding to the integrated line flux in the observational spectrum after subtraction of the fitted stellar background, divided by $R(\lambda_l)$. This approach may allow for accurate determination of the line flux. For more complex SPDs involving narrow spectral features such as QSOs, this alternative approach may not be feasible, and one is left with the interpretation of $\bf c$ only.

   \begin{figure}
   \centering
   \includegraphics[width=0.49\textwidth]{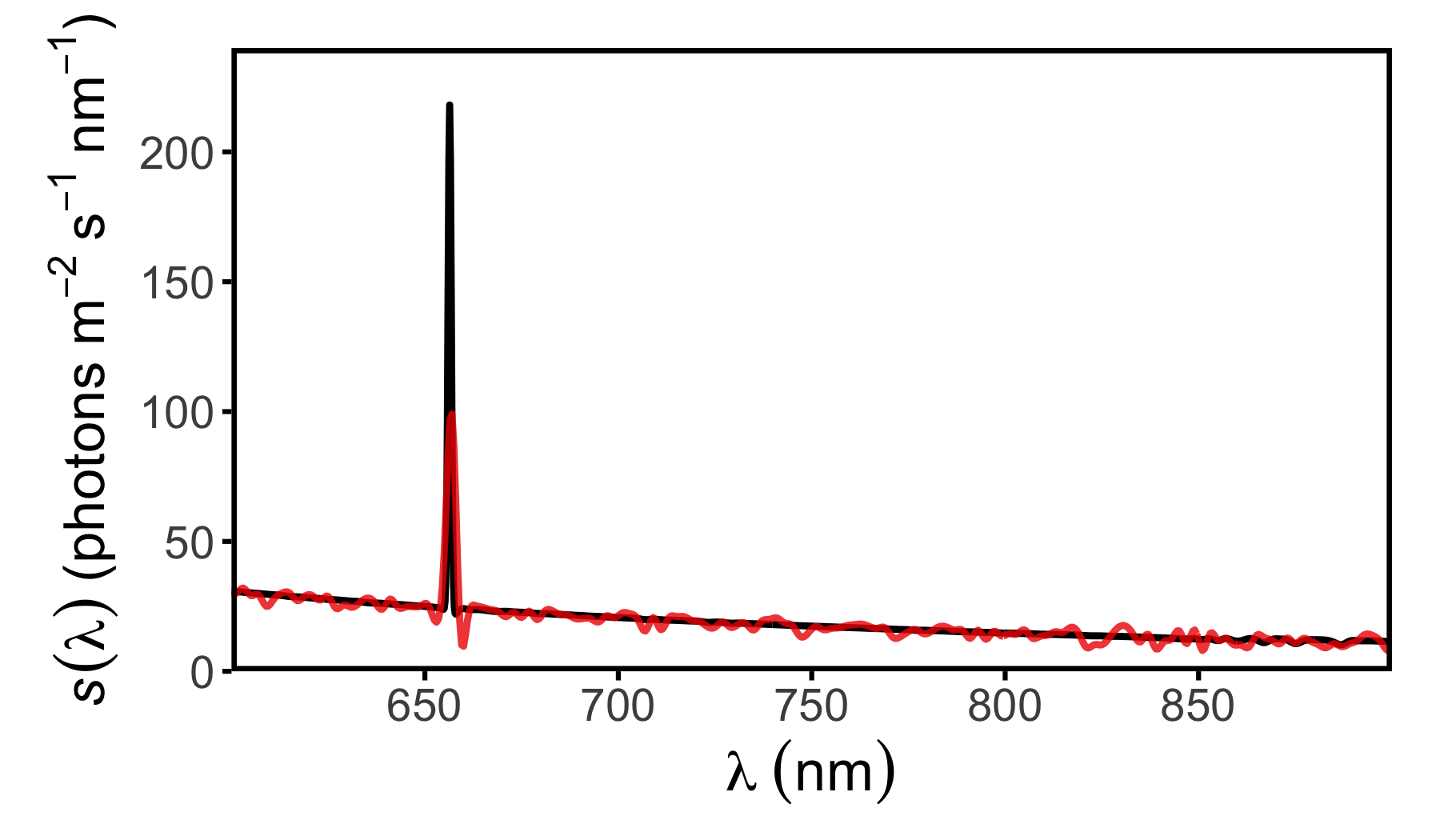}
    \includegraphics[width=0.49\textwidth]{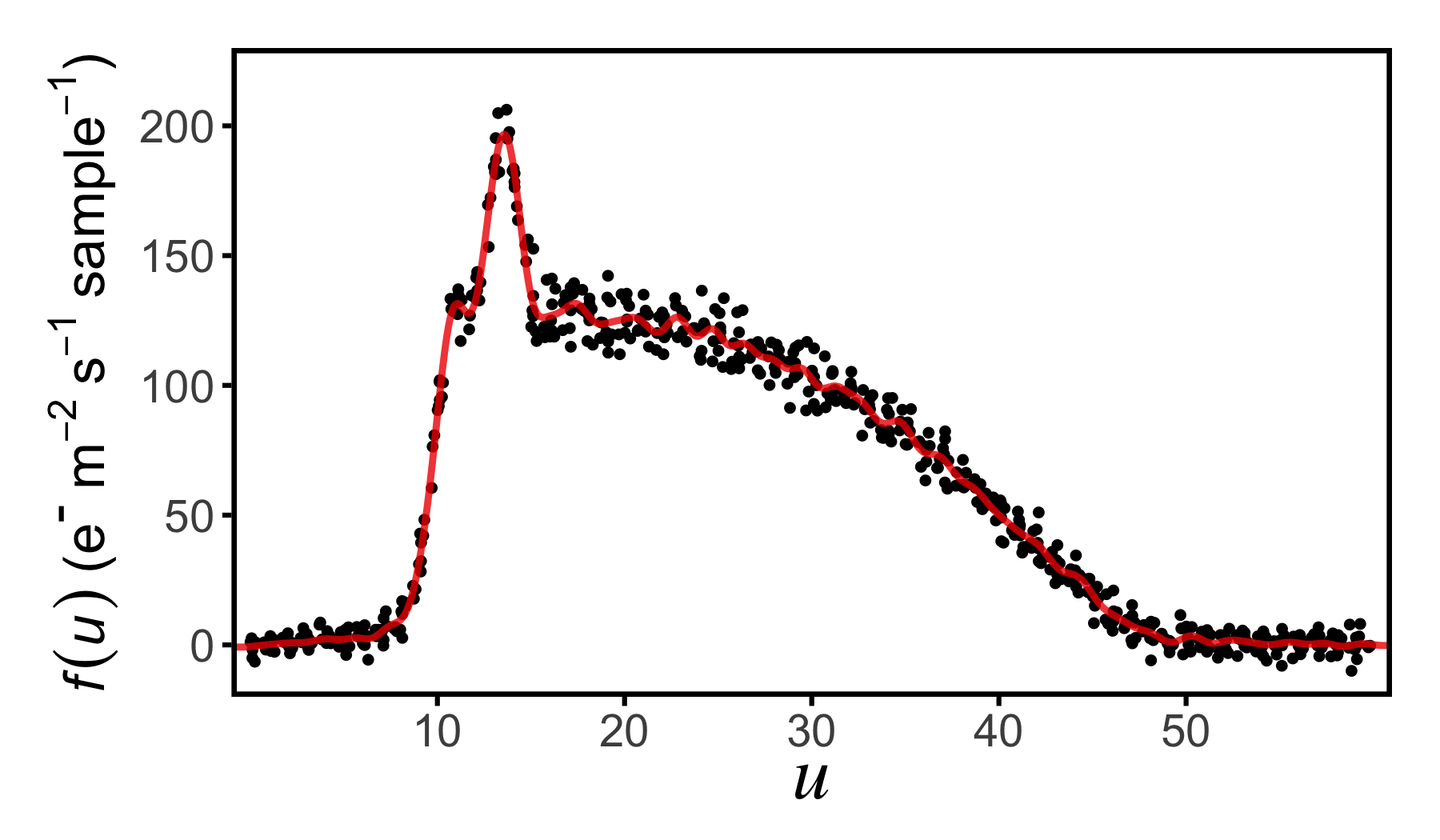}
    \includegraphics[width=0.49\textwidth]{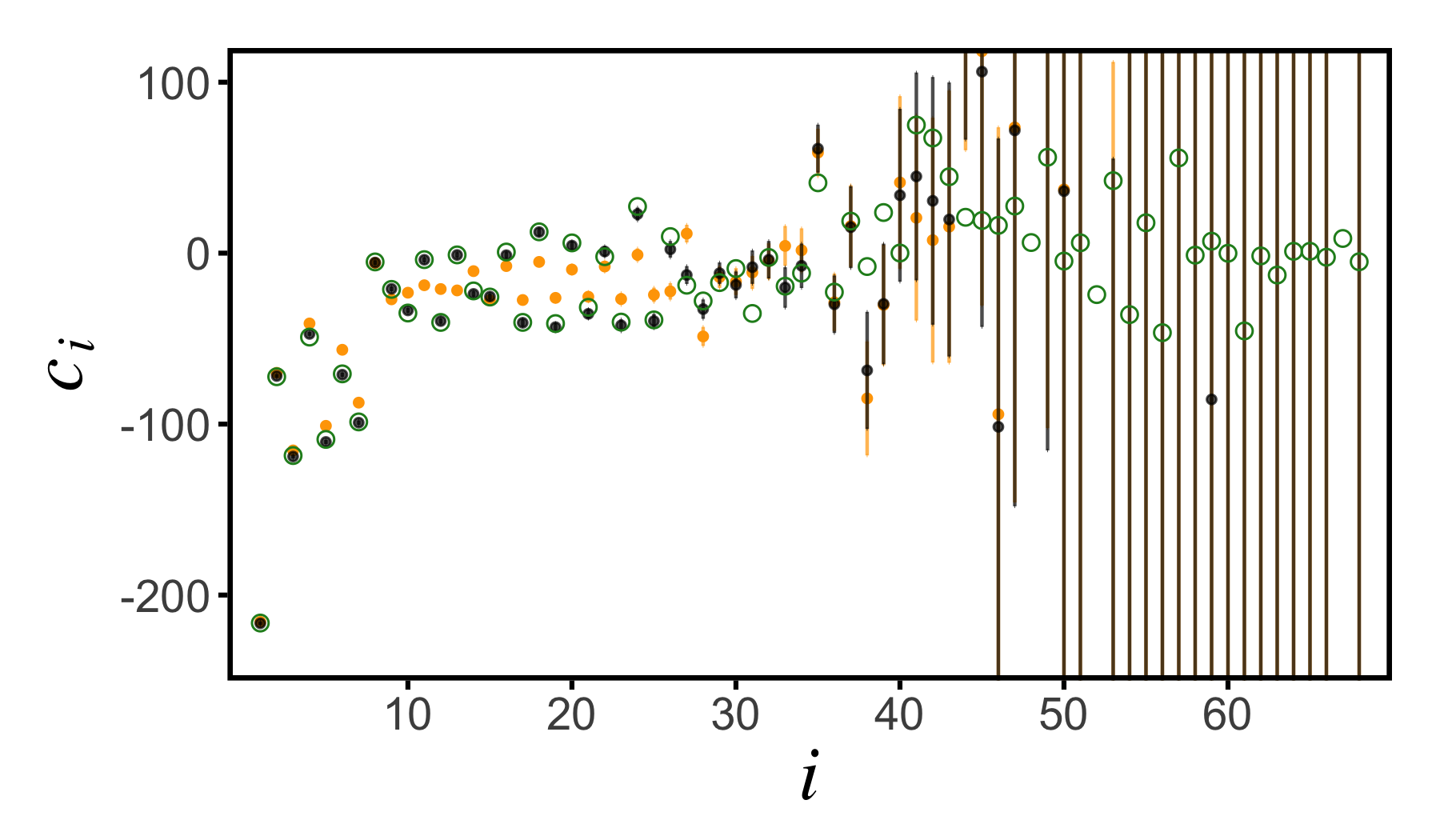}
   \caption{Upper panel: SPD of a B-type star with the added H$\alpha$ emission line (black line). The projection of this SPD onto the basis of $V^\ast_{\rm RP}$ is shown by the red line. Middle panel: Simulated RP observations (black dots) and the forward calibration result (red line). The H$\alpha$ line is the peak centred at $u = 13.6$. Bottom panel: Coefficients in $V^\ast_{\rm RP}$ in a reference system in which the covariance matrix diagonalises (black dots). The true coefficients are shown as green circles. The orange dots show the coefficients in the absence of the H$\alpha$ line.}
              \label{fig:ELS}
    \end{figure}

\section{Summary and Conclusions \label{sec:summary}}

Here, we investigate the effects of a non-negligible LSF width on the spectrophotometric calibration of spectra. We find the conventional approach, consisting of dividing the observational spectrum of a calibration star by its SPD to estimate the response function, to be a relatively poor approximation if the response function changes on scales that are not wide compared to the LSF. Here, we refer to cases in which this approach is insufficient  as 'low-resolution spectroscopy'. In this case, the spectrophotometric calibration becomes an intrinsically ill-posed problem. Aspects of relevance here are wavelength cut-on/off filters that introduce rather abrupt changes in the response function with wavelength. Moreover, electronic resonances in the mirror coatings may result in rather abrupt changes in the refractive indices of the materials with wavelength, and with it in the response function.\par
For low-resolution spectroscopy, the spectrophotometric calibration procedure needs to be fundamentally revised if a quantitative relation between the observational spectrum and the SPD is to be obtained. Instead of deriving a solution for the SPD of the observed object directly, the representation of the SPD in a suitably chosen basis with the corresponding coefficients can be derived. This is achieved by constructing bases of function spaces describing the observational spectra and the SPDs, and linking them via a linear transformation described by a transformation matrix. This matrix represents the spectroscopic instrument, and is restricted to the constructed function spaces;  we refer to this method as the matrix approach. The ill-posed nature of the problem is then absorbed by large uncertainties on some of the coefficients in this representation, while other coefficients are well constrained. An analysis of the calibration result should therefore be based on the interpretation of the coefficients describing the SPDs, rather than the SPDs themselves. For the coefficients, unbiased solutions with reliable covariance matrices can in principle be determined. Although the interpretation of the coefficients of an SPD rather than the SPD itself is unfamiliar, most processes typically done with SPDs, such as comparison of different spectra or spectra and model spectra, or synthetic photometry, can also be done using the coefficients. This can be done in an even more efficient and convenient way due to the compression of the SPD in a rather small number of coefficients, a number much smaller than the numbers of samples typical for sampled spectra over a wide wavelength range.\par
The matrix approach also implies different requirements for the calibration spectra. While in the conventional approach, in principle, one known SPD is sufficient for deriving the response function, the matrix approach requires a wide variety of calibration sources, such that the space spanned by the SPDs of the calibration sources has a dimension that is as high as possible. By using such a high-dimensional space, the instrument matrix becomes more constrained by the calibration sources and less dependent on the estimate on the instrument kernel based on the expansion by model spectra. This requirement on the calibration spectra is more demanding than for a calibration with the conventional approach; however it fully coincides with the requirement for calibration sources in passband determination \citep{WeilerEtAl2018}.\par
While the conventional approach is sufficient for the majority of spectroscopic instruments, a noteworthy exception are the spectrophotometric instruments on board the \gaia~spacecraft. \gaia's BP and RP spectrophotometric instruments are clearly examples for low-resolution spectroscopy as defined above. Taking full advantage of the large number of observations of this mission under excellent observing conditions in space requires the stringent treatment of the low-resolution regime. Test calibrations with simulated \gaia~XP spectra show that excellent spectrophotometric calibration of XP spectra is feasible with the methods presented in this work for astronomical sources covering a wide range of magnitudes and astrophysical parameters. The result of this work may therefore be of help in validating the release of \gaia's spectrophotometry in DR3, and in obtaining the best calibration possible.

\FloatBarrier

\begin{acknowledgements}
      This work has been developed in the context of the Gaia Data Processing and Analysis Consortium and presented and discussed in meetings, which resulted in positive inputs. The authors also thank F. De Angeli and D. W. Evans for their comments to the manuscript.\\
      This work was supported by the MINECO (Spanish Ministry of Economy) through grant ESP2016-80079-C2-1-R and RTI2018-095076-B-C21 (MINECO/FEDER, UE) and MDM-2014-0369 of ICCUB (Unidad de Excelencia ''Mar{\'i}a de Maeztu'').
\end{acknowledgements}

%
   \bibliographystyle{aa} 
   \bibliography{LowResolutionSpectrophotometry} 

\begin{thebibliography}{33}
\expandafter\ifx\csname natexlab\endcsname\relax\def\natexlab#1{#1}\fi

\bibitem[{{Akeson} {et~al.}(2019){Akeson}, {Armus}, {Bachelet}, {Bailey},
  {Bartusek}, {Bellini}, {Benford}, {Bennett}, {Bhattacharya}, {Bohlin},
  {Boyer}, {Bozza}, {Bryden}, {Calchi Novati}, {Carpenter}, {Casertano},
  {Choi}, {Content}, {Dayal}, {Dressler}, {Dor{\'e}}, {Fall}, {Fan}, {Fang},
  {Filippenko}, {Finkelstein}, {Foley}, {Furlanetto}, {Kalirai}, {Gaudi},
  {Gilbert}, {Girard}, {Grady}, {Greene}, {Guhathakurta}, {Heinrich},
  {Hemmati}, {Hendel}, {Henderson}, {Henning}, {Hirata}, {Ho}, {Huff},
  {Hutter}, {Jansen}, {Jha}, {Johnson}, {Jones}, {Kasdin}, {Kelly}, {Kirshner},
  {Koekemoer}, {Kruk}, {Lewis}, {Macintosh}, {Madau}, {Malhotra}, {Mand el},
  {Massara}, {Masters}, {McEnery}, {McQuinn}, {Melchior}, {Melton},
  {Mennesson}, {Peeples}, {Penny}, {Perlmutter}, {Pisani}, {Plazas}, {Poleski},
  {Postman}, {Ranc}, {Rauscher}, {Rest}, {Roberge}, {Robertson}, {Rodney},
  {Rhoads}, {Rhodes}, {Ryan}, {Sahu}, {Sand}, {Scolnic}, {Seth}, {Shvartzvald},
  {Siellez}, {Smith}, {Spergel}, {Stassun}, {Street}, {Strolger}, {Szalay},
  {Trauger}, {Troxel}, {Turnbull}, {van der Marel}, {von der Linden}, {Wang},
  {Weinberg}, {Williams}, {Windhorst}, {Wollack}, {Wu}, {Yee}, \&
  {Zimmerman}}]{Akeson2019}
{Akeson}, R., {Armus}, L., {Bachelet}, E., {et~al.} 2019, arXiv e-prints,
  arXiv:1902.05569

\bibitem[{Babusiaux {et~al.}(2004)Babusiaux, Ch{\'e}reau, \&
  de~Bruijne}]{Babusiaux2004}
Babusiaux, C., Ch{\'e}reau, F., \& de~Bruijne, J. 2004, Simulation of the Gaia
  point spread functions for GIBIS, Tech. Rep. GAIA-CB-02, Observatoire de
  Paris-Meudon, GEPI,
  (\url{http://ulisse.pd.astro.it/Gaia_doc/Simulation_of_the_Gaia_point_spread_functions_for_GIBIS.pdf})

\bibitem[{Bailey \& Swarztrauber(1994)}]{Bailey1994}
Bailey, D.~H. \& Swarztrauber, P.~N. 1994, SIAM J. in Scientific Computing, 15,
  1105

\bibitem[{Bessell(2000)}]{Bessell2000}
Bessell, M. 2000, PASP, 112, 961

\bibitem[{Bessell \& Murphy(2012)}]{Bessell2012}
Bessell, M. \& Murphy, S. 2012, PASP, 124, 140

\bibitem[{Bohlin {et~al.}(1980)Bohlin, Holm, Savage, Snijders, \&
  Sparks}]{Bohlin1980}
Bohlin, R.~C., Holm, A.~V., Savage, B.~D., Snijders, M. A.~J., \& Sparks, W.~M.
  1980, A\&A, 85, 1

\bibitem[{Cacciari(2010)}]{Cacciari2010}
Cacciari, C. 2010, in EAS Publication Series, Vol.~45, Gaia: At the frontiers
  of astrometry, ed. C.~{Turon}, F.~{Meynadier}, \& F.~{Arenou}, 155--160

\bibitem[{{Cardelli} {et~al.}(1989){Cardelli}, {Clayton}, \&
  {Mathis}}]{Cardelli1989}
{Cardelli}, J.~A., {Clayton}, G.~C., \& {Mathis}, J.~S. 1989, \apj, 345, 245

\bibitem[{Carrasco {et~al.}(2016)Carrasco, Evans, Montegriffo, Dummy1, Dummy2,
  \& Dummy3}]{Carrasco2016}
Carrasco, J.~M., Evans, D.~W., Montegriffo, P., {et~al.} 2016, A\&A, 595, A7

\bibitem[{{Carrasco} {et~al.}(2017){Carrasco}, Weiler, Jordi, \&
  Fabricius}]{Carrasco2017}
{Carrasco}, J.~M., Weiler, M., Jordi, C., \& Fabricius, C. 2017, in Highlights
  on Spanish Astrophysics IX, ed. A.~{Alonso-Herrero}, F.~Figueras,
  C.~Hern{\'a}ndez-Monteagudo, A.~S{\'a}nchez-Lavega, \& S.~P{\'e}rez-Hoyos,
  622--627

\bibitem[{Chen \& Jiang(2017)}]{Chen2017}
Chen, L.-H. \& Jiang, C.-R. 2017, Statistics and Computation, 27, 1181

\bibitem[{{Costille} {et~al.}(2016){Costille}, {Caillat}, {Rossin}, {Pascal},
  {Sanchez}, {Foulon}, \& {Vives}}]{Costille2016}
{Costille}, A., {Caillat}, A., {Rossin}, C., {et~al.} 2016, Society of
  Photo-Optical Instrumentation Engineers (SPIE) Conference Series, Vol. 9912,
  {Final design and choices for EUCLID NISP grism}, 99122C

\bibitem[{Evans {et~al.}(2018)Evans, Riello, De~Angeli, Carrasco, Montegriffo,
  Fabricius, Jordi, Palaversa, Diener, Busso, Cacciari, \& van
  Leeuwen}]{Evans2018}
Evans, D.~W., Riello, M., De~Angeli, F., {et~al.} 2018, A\&A, 616, A4

\bibitem[{{\it Gaia}~Collaboration~Brown {et~al.}(2018){\it
  Gaia}~Collaboration~Brown, Vallenari, Prusti, Dummy1, Dummy2, \&
  Dummy3}]{Gaia2018a}
{\it Gaia}~Collaboration~Brown, A. G.~A., Vallenari, A., Prusti, T., {et~al.}
  2018, A\&A, 616, A1

\bibitem[{Goodman(1996)}]{Goodman1996}
Goodman, J.~W. 1996, Introduction to Fourier Optics, 2nd edn. (McGraw--Hill)

\bibitem[{Henrici(1988)}]{Henrici1988}
Henrici, P. 1988, Applied and Computational Complex Analysis, Vol.~2 (Wiley)

\bibitem[{H{\"o}rmander(2003)}]{Hoermander2003}
H{\"o}rmander, L. 2003, The Analysis of Linear Partial Differential Operators
  I, 2nd edn. (Springer)

\bibitem[{{{\it Gaia} Collaboration,}~Brown {et~al.}(2016){{\it Gaia}
  Collaboration,}~Brown, Vallenari, Prusti, Dummy1, Dummy2, \&
  Dummy3}]{Gaia2016b}
{{\it Gaia} Collaboration,}~Brown, A. G.~A., Vallenari, A., Prusti, T.,
  {et~al.} 2016, A\&A, 595, A2

\bibitem[{{{\it Gaia} Collaboration,}~Prusti {et~al.}(2016){{\it Gaia}
  Collaboration,}~Prusti, de~Bruijne, Brown, Dummy1, Dummy2, \&
  Dummy3}]{Gaia2016a}
{{\it Gaia} Collaboration,}~Prusti, T., de~Bruijne, J. H.~J., Brown, A. G.~A.,
  {et~al.} 2016, A\&A, 595, A1

\bibitem[{Jordi {et~al.}(2010)Jordi, Gebran, Carrasco, de~Bruijne, Voss,
  Fabricius, Vallenari, Kohley, \& Mora}]{Jordi2010}
Jordi, C., Gebran, M., Carrasco, J.~M., {et~al.} 2010, A\&A, 523, A48

\bibitem[{Lena {et~al.}(1998)Lena, Lebrun, \& Mignard}]{Lena1998}
Lena, P., Lebrun, F., \& Mignard, F. 1998, Observational Astrophysics, 2nd
  edn., Astronomy and Astrophysics Library (Springer)

\bibitem[{Ma{\'i}z~Apell{\'a}niz(2005)}]{JMA2005}
Ma{\'i}z~Apell{\'a}niz, J. 2005, A new flux calibration for the STIS objective
  prism, Tech. rep., Space Telescope Science Institute

\bibitem[{Ma{\'i}z~Apell{\'a}niz \& Weiler(2018)}]{MAW2018}
Ma{\'i}z~Apell{\'a}niz, J. \& Weiler, M. 2018, A\&A, 619, A180

\bibitem[{Pancino(2010)}]{Pancino2010}
Pancino, E. 2010, in Hubble after SM4. Preparing JWST, ed. S.~{Deustua} \&
  C.~{Oliveira}

\bibitem[{Pickering(1890)}]{Pickering1890}
Pickering, E.~C. 1890, Annals of the Harvard College Observatory, 27, 1

\bibitem[{Press {et~al.}(2007)Press, Teukolsky, Vetterling, \&
  Flannery}]{Press2007}
Press, W.~H., Teukolsky, S.~A., Vetterling, W.~T., \& Flannery, B.~P. 2007,
  Numerical Recipes, 3rd edn. (Cambridge University Press)

\bibitem[{Schmidt(1987)}]{Schmidt1987}
Schmidt, R.~C. 1987, PhD thesis, Iowa State University

\bibitem[{Selsing {et~al.}(2016)Selsing, Fynbo, Christensen, \&
  Krogager}]{Selsing2016}
Selsing, J., Fynbo, J. P.~U., Christensen, L., \& Krogager, J.-K. 2016, A\&A,
  585, A87

\bibitem[{Vacca {et~al.}(2003)Vacca, Cushing, \& Rayner}]{Vacca2003}
Vacca, W.~D., Cushing, C., \& Rayner, J.~T. 2003, PASP, 115, 389

\bibitem[{Weiler(2018)}]{Weiler2018}
Weiler, M. 2018, A\&A, 617, A138

\bibitem[{Weiler {et~al.}(2018)Weiler, Jordi, Fabricius, \&
  Carrasco}]{WeilerEtAl2018}
Weiler, M., Jordi, C., Fabricius, C., \& Carrasco, J.~M. 2018, A\&A, 615, A24

\bibitem[{{Westera} {et~al.}(2002){Westera}, {Lejeune}, {Buser}, {Cuisinier},
  \& {Bruzual}}]{Westera2002}
{Westera}, P., {Lejeune}, T., {Buser}, R., {Cuisinier}, F., \& {Bruzual}, G.
  2002, \aap, 381, 524

\bibitem[{Young(1994)}]{Young1994}
Young, A.~T. 1994, A\&A, 288, 683

\end{thebibliography}
%

\end{document}